\documentclass[journal=jacsat,manuscript=article]{achemso}

\usepackage[version=3]{mhchem} 

\newcommand{\VO}{V$_\text{O}$}
\newcommand{\NO}{N$_\text{O}$}

\author{Akitaka Nakanishi}
\affiliation{The Institute for Solid State Physics, The University of Tokyo, \
5-1-5 Kashiwanoha, Kashiwa-shi, Chiba 277-8581, Japan}
\email{nakanishi@issp.u-tokyo.ac.jp}
\author{Shusuke Kasamatsu}
\affiliation{Faculty of Science, Yamagata University, \
1-4-12 Kojirakawa, Yamagata-shi, Yamagata 990-8560, Japan}
\author{Jun Haruyama}
\affiliation{The Institute for Solid State Physics, The University of Tokyo, \
5-1-5 Kashiwanoha, Kashiwa-shi, Chiba 277-8581, Japan}
\author{Osamu Sugino}
\affiliation{The Institute for Solid State Physics, The University of Tokyo, \
5-1-5 Kashiwanoha, Kashiwa-shi, Chiba 277-8581, Japan}

\title{Theoretical analysis of zirconium oxynitride/water interface 
    using neural network potential}

\begin{document}

\begin{abstract}
Zr oxides and oxynitrides are promising candidates to replace precious metal cathodes in polymer electrolyte fuel cells. Oxygen 
reduction reaction activity in this class of materials has been correlated with the amount of oxygen vacancies, but a microscopic 
understanding of this correlation is still lacking. To address this, we simulate a defective Zr$_7$O$_8$N$_4$/H$_2$O interface model 
and compare it with a pristine ZrO$_2$/H$_2$O interface model. First, ab initio replica exchange Monte Carlo sampling was performed 
to determine defect segregation at the surface in the oxynitride slab model, then molecular dynamics accelerated by neural 
network potentials was used to perform 1000 of 500 ps-long simulations to attain sufficient statistical accuracy of the solid/liquid interface structure. 
The presence of oxygen vacancies on the surface was found to clearly modify the local adsorption 
structure: water molecules were found to adsorb preferentially on Zr atoms surrounding oxygen vacancies, but not on the oxygen 
vacancies themselves. The fact that oxygen vacancy sites are free from poisoning by water molecules may explain the activity enhancement in defective systems. 
The layering of water molecules was also modified considerably, which should influence the proton and O$_2$ transport near the 
interfaces which is another parameter that determines the overall activity.
\end{abstract}

\section{Introduction}
Polymer electrolyte fuel cells (PEFCs) are attracting attention 
as energy conversion devices, and their performance depends on 
the catalytic activity of the oxygen reduction reaction (ORR) taking place 
on the cathode side. 
Currently, Pt or its alloys are used as the ORR catalyst owing to their high
catalytic activity, although  durability and cost issues are hindering the
wide-spread production and use of PEFCs.
\cite{Debe2012}

Zr oxides 
\cite{Doi2007,Chisaka2017,Ishihara2019}
and Ti oxides
\cite{Arashi2014,Ishihara2020a,Ishihara2016,Ishihara2018}
are promising candidates to replace Pt as electrocatalysts 
due to their high durability, low cost, and high catalytic activity obtained 
by introducing defects such as 
oxygen vacancies (\VO s) and oxygen-nitrogen substitutions (\NO s).
\cite{Ishihara2020b,Ota2012}
Previous experimental studies have shown that the catalytic activity correlates with 
the amount of \VO s\ in ZrO$_2$, 
\cite{Doi2007,Ukita2011,Yin2013} suggesting that \VO s\ are the active centers
for the ORR. The presence of oxynitride phases such as Zr$_2$ON$_2$ ($\gamma$ phase)
and Zr$_7$O$_8$N$_4$ ($\beta$ phase), which can be regarded as fluorite-structured
ZrO$_2$ with vacancy ordering  \cite{Bredow2004,Bredow2007},
have also been reported in catalytically active
samples. \cite{Maekawa2008}
However, the reaction pathways, rate-determining steps, 
and various other aspects of the ORR remain unknown. 

To the best of our knowledge, 
there have not been many theoretical studies on the ORR activity 
of defective oxide surfaces compared to pristine ones.
\cite{Yamamoto2019,Muhammady2022a,Muhammady2022b}
One such work is by 
Muhammady et al., who performed an exhaustive search for ORR intermediate structures
and calculated the corresponding ab initio free energy diagrams on 
tetragonal ZrO$_2$(101) surfaces with \VO\,and \NO\ 
based on the computational hydrogen electrode (CHE) model. 
\cite{Muhammady2022b, norskov_origin_2004}
The results suggest that Zr atoms and \VO\ on the surface
do not differ significantly in terms of local site activity, in
apparent contrast to the experimental literature.
However, the influence of the solvent was only considered through
solvation correction energies, and neither solvent nor adsorbed water molecules
were considered explicitly. This means that possibility of poisoning of active sites
by the solvent is ignored.
In addition, the CHE model does not contain contributions from finite-temperature dynamics of the intermediates.

Thus, the natural way to resolve the abovementioned discrepancies between theory and experiment 
would be to accurately consider explicit solvent effects and dynamics using
ab initio molecular dynamics (AIMD).
However, the high computational cost is a major problem,
especially for models with a large number of atoms in the unit cell
such as catalysts with defects, 
or when dynamic properties such as proton diffusion are studied 
which requires simulations of 1\,ns or longer. 
In recent years, this issue has been solved to some extent through the
advent of machine learning potentials (MLPs),
which enable more efficient molecular dynamics calculations
while maintaining the same level of accuracy as ab initio calculations.\cite{Behler2021}
Numerous works have been reported that employ MLPs to study adsorption structures and proton transfer at the interfaces of pristine solids and water.
\cite{Natarajan2016,Kondati2017,Quaranta2017,Quaranta2018,
Quaranta2019,Artrith2019,Andrade2020,Ghorbanfekr2020,
Eckhoff2021,Schran2021,Mikkelsen2021,
Schienbein2022,Mikkelsen2022,
Fan2023,Wen2023a,Zeng2023}
However, application to complex and defective oxides are still lacking, partially due to difficulties in
obtaining a stable MLP that can treat complex solid/liquid interfaces with a large number of distinct chemical
environments. It is often the case that simulations `blow up' when the system wanders into regions of space
that the MLP has not been trained on.

Another issue in modeling of complex oxides is in determining the ion ordering. In the case of zirconium oxynitrides, two stoichiometric 
phases are known with vacancy ordering in the bulk as mentioned above, but 
the anion ordering is still under debate because x-ray diffraction cannot be used to separate oxygen and nitrogen sites. The 
situation is even more complicated when considering surfaces, as there is no experimental information on segregation of \NO \ or \VO 
\ and whether vacancy ordering persists at the surface. Such segregation would be decisive in determining the surface 
activity and stability, but theoretical prediction is challenging due to the combinatorial explosion in the number of possible configurations.
Molecular dynamics cannot solve this issue because ion and defect arrangements in complex oxides are very slow to relax and are determined by 
firing or sintering processes whose time scales are beyond the reach of atomistic simulations.

In this work, we apply recently developed methodologies to solve the above issues
and investigate the structure of water molecules at the interface 
between water and defective oxide catalysts, 
as a first step towards understanding 
the effects of defects and solvents on ORR activity.
Examples of such effects include the influence of water layering
on proton transfer and the poisoning of O$_2$ adsorption sites by water.
We compare zirconium oxynitride with Zr$_7$O$_8$N$_4$ composition
to ZrO$_2$, where the former can be considered as ZrO$_2$ with a large
number of \NO\ and \VO\ defects, with one \VO\ for every two \NO\ to
satisfy formal ion charge neutrality.
First, ab initio replica exchange Monte Carlo simulation accelerated by
active learning of a neural network configuration energy model \cite{Artrith2017,Kasamatsu2022,Hoshino2023} was carried out to
obtain a Zr$_7$O$_8$N$_4$ slab model with thermodynamically stable \NO\ and \VO\ segregation profiles.
Then, neural network potentials (NNPs) \cite{Behler2007,Behler2015} that reproduce ab initio calculations
for interfaces of water and ZrO$_2$/Zr$_7$O$_8$N$_4$ were constructed using this surface slab model
through an iterative active learning approach detailed in the next section.
Once a stable NNP was obtained, hundreds of nanosecond-long trajectories were
generated using NNP molecular dynamics (NNPMD) to study water molecule adsorption on the surfaces of Zr$_7$O$_8$N$_4$ and ZrO$_2$.
The possibility of the \VO\,
as catalytically active sites is discussed by comparing the two systems.

\section{Model and Methods}
\subsection{Surface slab models}
\begin{figure}
\includegraphics[width=16cm]{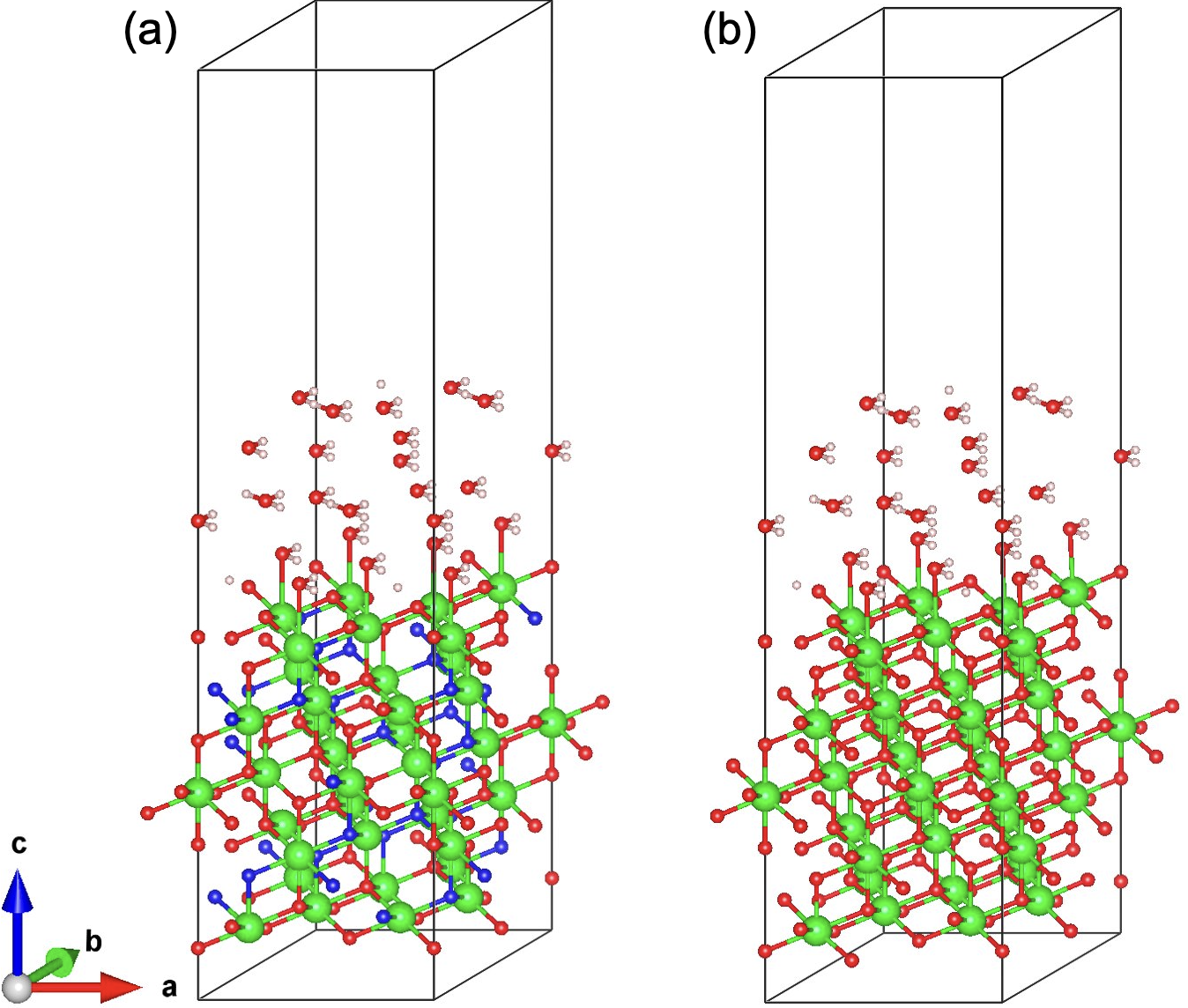}
\caption{
The initial structures of molecular dynamics
for interfaces between H$_2$O and (a) Zr$_7$O$_8$N$_4$ or (b) ZrO$_2$.
Green, blue, red and white indicate Zr, N, O and H atoms respectively.
The Figures have been generated using the VESTA software package.
\cite{Momma2011}
}
\label{fgr:interface}
\end{figure}

In this work, we model the tetragonal zirconia and zirconium oxynitride 
surfaces which correspond to the (111) surface in the cubic fluorite setting. This corresponds to 
the (101) orientation in tetragonal zirconia (whose structure is slightly distorted from the fluorite) and the (???) orientation of the orthorhombic
$\beta$ phase. This surface was chosen because the (111) facet has been predicted to be most stable 
in pure zirconia and is often observed in experiments. 
The (111) orientation in the fluorite setting consists of neutral
units of --(O--Zr--O)--(O--Zr--O)--$\cdots$; here, we consider ? repetitions of the (O-Zr-O) unit.
We employ the in-plane unit cell with $a = b = 9.542394$\,{\AA} shown in Fig.~\ref{fgr:interface} (a), which results in ?? 
Zr sites and ?? O sites, where the latter can be occupied by O, \NO\, or \VO in the oxynitride.
For the oxynitride slab, replica exchange Monte Carlo (RXMC) sampling using a neural network configuration 
energy model is first performed without water to simulate nitrogen and oxygen vacancy segregation near the surface as detailed in Sec.~??. One of the lowest energy configurations found is used in subsequent NNPMD simulations.
Figure~\ref{fgr:interface} (b) shows the interface models simulated by NNPMD where water molecules were placed on top of the surface of the pristine tetragonal ZrO$_2$ surface and the oxynitride surface from RXMC sampling. 
The interface model contains a random arrangement of water molecules 
in a unit cell with an additional vacuum layer of 15\,{\AA}.
The number of water molecules was set to 21 per calculation cell, 
which corresponds to about three layers of water.

To improve the thermodynamic sampling of water structures, 
we performed numerous NNPMD runs with 
different initial water configurations. The following procedure was
employed to prepare initial configurations that are not correlated with each other.
First, the xy-coordinates of the water molecules were set so that 
the oxygen of the water molecules was above one of the atoms of the slab
and the dipole moment of the water molecules was parallel to the x-axis. 
Then, the 21 water molecules were randomly assigned to 
three different groups of seven water molecules each, and
depending on the group, the z-component of the atomic coordinates was 
set to a value of 3, 6, and 9\,{\AA} higher than 
the z-coordinate of outermost surface layer of the oxide and oxynitride. 
We also performed NNPMD on a supercell expanded by a factor of 2 in the $a$ and $b$ directions,
which we will refer to as the $2\times2\times1$ supercell. In this case, 84 water molecules
are arranged randomly with in a similar fashion as detailed above.

\subsection{Iterative training of on-lattice neural network model for replica exchange Monte Carlo sampling}

\subsection{Iterative training of neural network potential for molecular dynamics}

Active learning is now recognized as an effective and often necessary procedure for 
obtaining reliable NNPs in complex systems such as solid/liquid interfaces.
\cite{Andrade2020,Wen2023a,Schienbein2022,Schran2021}
In this work, the following active learning procedure was used 
to generate the NNPs and analyze the structures of adsorbed water:
\begin{enumerate}
    \item The total energy, interatomic forces, and stress tensor data are obtained
    for atomic configurations from 4 independent 1 ps AIMD trajectories 
    for the interface model (Fig.~\ref{fgr:interface}).
    \item An NNP is trained on this data set and used to perform 4 independent 1 ns NNPMD simulations.
    \item 400 structures are sampled from the NNPMD trajectories and recalculated using ab initio calculations
    to obtain the total energy, interatomic forces and stress tensor data. Structures with NNP errors $\geq 1$ meV/atom are added
    to the training set.
    \item Repeat from 2 until the RMSE of the NNP prediction vs. ab initio calculations in step 3 is smaller than 10 meV/atom.
\end{enumerate}

We then carefully examined the convergence of the resulting NNP model when applied to the $2\times2\times1$ expanded supercell.
NNPMD was performed for 1 ns in the $2\times2\times1$ supercell, then 40 configurations from the NNPMD trajectory was recalculated 
using ab initio calculations to evaluate the NNP energy errors. If the RMSE $\geq 10$ meV/atom, we return to the active learning procedure
outlined above. After passing this energy error test, we performed 1000 independent NNPMD simulations to calculate
average adsorption statistics. In the first pass, we return to the active learning in the smaller cell for one more iteration.
In the subsequent iterations, convergence is examined for the adsorption statistics; if the convergence is judged to be
insufficient, we perform additional active learning iterations in the original supercell of Fig.~\ref{fgr:interface}.
We judge the convergence to be sufficient when the errors are small enough to discuss the difference between the oxide and the oxynitride. The full procedure including details of splitting of training, validation, and test data is outlined in Fig.~\ref{fig:alprocedure}.
%

The calculation software used for each procedure
and the detailed calculation conditions are as follows.
ab initio calculations were performed 
using the VASP code,
\cite{Kresse1996,VASP}
which is based on the plane-wave pseudopotential method.
Projector augmented wave pseudopotentials
\cite{Blochl1994}
and the exchange correlation functional 
based on the generalised gradient approximation 
of the Perdew-Burke-Ernzerhof type
\cite{Perdew1996}
were used.
The projection operator was evaluated in real space, 
including the non-spherical contribution to 
the gradient of the density of potential spheres.
Only the $\Gamma$ point was sampled for the Brillouin zone 
and partial occupancies were
calculated by Gaussian smearing with a width of 0.03\,eV.
The cutoff energy for the plane-wave-basis set is 520\,eV.
Self-consistent calculations were performed for up to 100 loops and 
stopped when the energy difference was less than 0.1\,meV.
The initial magnetic moment was set to 0.6\,$\mu_\text{B}$ for all atoms 
to account for spin polarisation.
The mass of the hydrogen atoms was replaced by the mass of deuterium,
allowing a larger time step in the integration of the equation of motion. 
AIMD was performed in the NVT ensemble whose temperature was controlled at 300\,K 
by the Langevin thermostat with a friction coefficient of 10\,ps$^{-1}$.

The NNPs were generated using SIMPLE-NN (ver. 2),
\cite{Lee2019,SIMPLENNv2}
which is based on the Behler-Parinello type symmetry function.
\cite{Behler2007}
The parameters of the radial symmetry function G$_2$ are 
$R_\text{cut}$ = 6.0, 
$\eta$ = 0.003214, 0.035711, 0.071421, 0.124987, 0.214264, 0.357106, 0.714213, 1.428426,
$R_\text{s}$ = 0.0.
Those of the angular symmetry function G$_3$ are 
$R_\text{cut}$ = 6.0,
$\eta$ = 0.000357, 0.028569, 0.089277, 
$\lambda$ = -1, 1, 
$\zeta$ = 1, 2, 4.
A neural network is constructed with these symmetry functions as input 
and the total energy, interatomic forces and stress tensor as output.
The neural network contains two hidden layers with 30 nodes each and employs the tanh activation function.
Its weights are optimized up to 1000 epochs 
using the Adam method with a learning rate of 0.001.

The NNPMD calculations were performed using LAMMPS.
\cite{Plimpton1995,LAMMPS}
The NVT ensemble was used with the temperature controlled to 300\,K 
by the Langevin thermostat with the damping factor of 0.1\,ps. 
The mass of the hydrogen atom was replaced by the mass of the deuterium atom. 
Four initial structures were created using the method described above, 
each with a different initial velocity and 
a 1\,ns NNPMDs was performed for each with a 1\,fs time step. 

Self-consistent field calculations were performed in VASP for 400 structures 
every 10\,ps in four 1\,ns trajectories,
and the ab initio and NNP energy errors were checked.
Samples with an energy error greater than 1\,meV/atom were added to 
the training and validation data, but not to the test data.

When the energy error was less than 10\,meV/atom 
for both the unit cell and the $2\times2\times1$ supercell, 
the structural properties used for the convergence decision condition 
of active learning were calculated.
For the $2\times2\times1$ supercell, 1000 new initial structures were generated 
and NNPMD was performed on each of them for 500\,ps with a time step of 1\,fs.
The mean and standard deviation of 
the number of molecular and dissociative adsorptions
were calculated using 1000 structures at 500\,ps.
The conditions for O adsorption on the Zr site and OH covalent bonds 
were set to 3.09651\,{\AA} and 1.2\,{\AA} respectively. 
When it was determined that these properties had sufficiently converged, 
active learning was stopped and the other properties 
- number density distribution for  the surface perpendicular and parallel, 
adsorption distance d, adsorption angle $\alpha$, 
molecular orientation $\beta$, 
OH orientation $\gamma$, number of hydrogen bonds and 
radial distribution function - were also calculated. 
Definitions of angles and hydrogen bonds will be explained later.
In summary, 
if the RMSE of the unit cell is below 10 meV/atom, 
the RMSE of the $2\times2\times1$ supercell is calculated; 
if it is also below 10 meV/atom, the number of adsorptions is calculated; 
if it converges, active learning is stopped and 
the other properties are calculated.

\subsection{Proton transfer}
We calculated the free energy in proton transfer (PT) 
using different collective variables (CVs).
The PT reactions treated in this study are summarised in the Table~\ref{tbl:PT_list}.
The CV = $\delta_\text{min}$ was used for PT between 
donor oxygen O$_\text{d}$ and acceptor oxygen O$_\text{a}$ in hydrogen bonding.
\cite{Tuckerman2002}
This CV is the minimum value of the difference $\delta$ 
between the O$_\text{a}$-H and O$_\text{d}$-H distances for each time step arrangement of the MD,
and used for surface-PT, adlayer-PT, and PT involving solvent 1 to 4
in the Table~\ref{tbl:PT_list}.
For the PT between O$_\text{d}$ and O$_\text{a}$ mediated by H$_2$O,
we used the CV = $(\delta_1+\delta_2)/2$.
This CV is the average value of $\delta_1,\delta_2$
in the hydrogen bond between O$_\text{d}$ or O$_\text{a}$ and the intermediate oxygen H$_2$O,
and used for solvent-assisted PT in the Table~\ref{tbl:PT_list}.
In addition, we used CV = S$_\text{O-H}$ for the PT between surface oxygen atoms and any oxygen.
This CV is the minimum distance between a given oxygen and any hydrogen on the surface.
When using this CV, 
it is not clear which PT is between the surface oxygen and which oxygen.
We calculated the histogram $W(CV)$ with a bin width of 0.1\,{\AA} for each CV and
evaluated the Helmholtz free energy $\Delta F(CV) = -k_\text{B}T \ln W(CV)$.
In the free energy calculation, 
structures at 401,402,...,500 ps were extracted 
from the 1000 trajectories of the 500 ps MD described above, 
and a total of 100000 structures were used.

\begin{table}
\caption{
The PT reaction formulas and names treated in this study.
Adsorbed oxygen, oxygen of the slab surface and oxygen in the solvent are expressed
as O$^*$, O$_\text{s}$ and O$^\sim$ respectively.
}
  \label{tbl:PT_list}
  \begin{tabular}{llllllll}
    \hline
    O$_\text{s}$H$^-$ & + O$^*$H$^-$ &        & $\rightleftharpoons$ & O$_\text{s}^{2-}$ & + H$_2$O$^*$ &        & surface-PT \\
    H$_2$O$^*$ & + O$^*$H$^-$ &        & $\rightleftharpoons$ & O$^*$H$^-$ & + H$_2$O$^*$ &        & adlayer-PT \\
    O$_\text{s}$H$^-$ & + O$^\sim$H$^-$ &        & $\rightleftharpoons$ & O$_\text{s}^{2-}$ & + H$_2$O$^\sim$ &        & PT involving solvent 1 \\
    O$_\text{s}$H$^-$ & + H$_2$O$^\sim$ &        & $\rightleftharpoons$ & O$_\text{s}^{2-}$ & + O$^\sim$H$_3^+$ &        & PT involving solvent 2 \\
    H$_2$O$^*$ & + O$^\sim$H$^-$ &        & $\rightleftharpoons$ & O$^*$H$^-$ & + H$_2$O$^\sim$ &        & PT involving solvent 3 \\
    O$^*$H$^-$ & + O$^\sim$H$_3^+$ &        & $\rightleftharpoons$ & H$_2$O$^*$ & + H$_2$O$^\sim$ &        & PT involving solvent 4 \\
    O$_\text{s}$H$^-$ & + H$_2$O$^\sim$ & + O$^*$H$^-$ & $\rightleftharpoons$ & O$_\text{s}^{2-}$ & + H$_2$O$^\sim$ & + H$_2$O$^*$ & solvent-assisted PT 1\\
    O$_\text{s}$H$^-$ & + H$_2$O$^\sim$ & + O$_\text{s}^{2-}$ & $\rightleftharpoons$ & O$_\text{s}^{2-}$ & + H$_2$O$^\sim$ & + O$_\text{s}$H$^-$ & solvent-assisted PT 2\\
    H$_2$O$^*$ & + H$_2$O$^\sim$ & + O$^*$H$^-$ & $\rightleftharpoons$ & O$^*$H$^-$ & + H$_2$O$^\sim$ & + H$_2$O$^*$ & solvent-assisted PT 3\\
    \hline
  \end{tabular}
\end{table}

The lifetime $\tau$ was evaluated from the time correlation function $C(t)$ 
based on the “stable states picture” (SSP).
\cite{Laage2008}
The $C(t)$ represents the probability that if a molecule is a "stable" reactant at time $t_0$, 
it is not yet a "stable" product at time $t_0+t$.
The SSP assumes that an intermediate region exists between stable reactants and stable products.
The definition of stable state in PT is as follows:
\begin{enumerate}
    \item In O$^*$H$^-$ and O$_\text{s}$H$^-$, 
    one H atom is coordinated within an oxygen sphere with radius of 1.1 {\AA},
    and no other H atom is coordinated within a maximum distance of 1.4 {\AA}.
    \item In H$_2$O$^*$, 
    two H atoms are coordinated within an oxygen sphere of radius 1.1 {\AA}.
    \item In O$_\text{s}^{2-}$, H atoms are not coordinated 
    within an oxygen sphere of radius 1.5 {\AA}.  
    In the same way, 
    we also calculated the lifetime of the water adsorption/desorption.  
    In this case, the definition of stable state is as follows.
    \item In the adsorbed state, 
    one Zr atom is coordinated 
    within the 2.2 {\AA} radius shpere around the oxygen of the water molecule. 
    \item In the desorption state, 
    the Zr atom is not coordinated 
    within the 3.2 {\AA} radius sphere around the oxygen of the water molecule.
\end{enumerate}
In order to calculate the time correlation function, 
50\,ps MD was newly executed using 100 initial structures selected 
from the 1000 final structures of 500\,ps MD described above.
We took the average of $C(t)$ calculated by varying the time origin from 0 to 25 ps, 
and averaged them over 100 trajectories.
We fitted $C(t)$ with the double exponential function 
$A\,\exp(-t/\tau_1)+(1-A)\,\exp(-t/\tau_2)$ 
and calculate the lifetime as a weighted average $\tau = A\,\tau_1 + (1-A)\,\tau_2$,
which is equivalent to integrating $C(t)$ from time zero to infinity.

To investigate the effective long-range diffusion of protons, 
we calculated the mean squared displacement (MSD) and 
the diffusion coefficient of the proton hole centers (PHCs). 
\cite{Hellstrom2019}
The PHC is defined as a combination of typical acceptors, O$_\text{s}^{2-}$ and O$^*$H$^-$.
For this purpose,
we performed a new 50\,ns MD with an initial structure 
selected from the 1000 final 500\,ps MD structures mentioned above. 
The MD trajectory were recorded at 100 ps intervals.

To investigate the types of PT 
that contribute to the long-range diffusion of protons, 
we have visualised the pathway along which PT occurs.
If the oxygen binding to H changes from O$_\text{a}$ to O$_\text{b}$ at time t=i,i+1, 
draw a line segment connecting the coordinates O$_\text{a}$$(i)$ and O$_\text{b}$$(i+1)$ 
at each time and the PT path.
In addition, the PT paths were classified according to their oxygen states.
In this calculation, 
we again used the 100 trajectories of 50\,ps MD
used in the calculation of the time correlation function.

\subsection{Anharmonic OH vibrational spectrum}
In addition, we have calculated anharmonic OH stretching vibration frequencies 
based on instantaneous MD simulation snapshots.
\cite{Mitev2015,Pejov2010,Quaranta2018}
For more information on this method, see reference.
All anharmonic stretching frequencies were calculated as follows.
First, we selected a snapshot of the MD simulation 
and generated configurations with different O and H coordinates for each OH group contained in it.
In these configurations, 
the centroid coordinates of OH remain unchanged 
and the distance $r_\text{OH}$\,({\AA}) expands and contracts from 0.68 to 1.70 in steps of 0.02.
The potential energy of each configuration generated in this way was calculated 
using the NNP generated by the method described above, 
and the one-dimensional potential energy $V(r_\text{OH})$ was obtained.
We fitted $V(r_\text{OH})$ with an 11th order polynomial $V_{11}(r_\text{OH})$ 
 and expanded it around the minimum equilibrium bond length $r_\text{OH}$(eq).
\begin{equation}
    V_{11}(r_{OH}) = V_0 + \sum_{n=1}^{11}\frac{k_n}{n!} (r_{OH} - r_{OH}(eq))^n
\end{equation}
The one-dimensional Schr\"{o}dinger equation for $V_{11}(r)$ is solved 
using the discrete variable representation method,
\cite{Lill1982,Light1985,Bacic1989}
and the anharmonic OH frequency $\nu$ is calculated 
from the energy difference between the ground state and the first excited state. 
A histogram with a bin width of 100 ($cm^{-1}$) was calculated 
from the $\nu$ data set for all oscillators, 
and the OH vibrational spectral bands were obtained 
by fitting to an asymmetric Gaussian function.
\cite{Stancik2008}
\begin{eqnarray}
    G(\nu) & = & \frac{A}{\gamma} \sqrt{\frac{4\ln 2}{\pi}} \exp\left[-4\ln 2\left(\frac{\nu-\nu_{max}}{\gamma}\right)^2\right] \\
    \gamma & = & \frac{2\gamma_{0}}{1+\exp[\alpha(\nu-\nu_{max})]}
\end{eqnarray}
where $A$ is the peak area, $\nu$ is the wavenumber frequency, 
$\nu_{max}$ is the position of the peak maximum, 
and $\gamma_0$ is the full width at half maximum.
In the free energy calculation, 
structures at 500 ps were extracted 
from the 1000 trajectories of the 500 ps MD described above, 
and a total of 1000 structures were used.

\section{Results and discussion}

\subsection{Validation of NNP}
\begin{table}
\caption{
The root mean squared error (RMSE) of the existing data set
for the total energy, the interatomic force and the stress tensor 
for the NNP of each loop in Zr$_7$O$_8$N$_4$ and ZrO$_2$ 
for training, validation and test data.
}
  \label{tbl:rmse_train}
  \begin{tabular}{rrrrrrrrrr}
    \hline
Zr$_7$O$_8$N$_4$ & \multicolumn{3}{l}{Energy\,(meV/atom)} & \multicolumn{3}{l}{Force\,(eV/{\AA})} & \multicolumn{3}{l}{Stress\,(kbar)} \\
   loop & train & valid &  test & train & valid &  test & train & valid &  test \\
    \hline
      0 & 1.119 & 1.184 & 1.729 & 0.088 & 0.118 & 0.157 & 1.752 & 1.987 & 2.022 \\
      1 & 1.294 & 1.521 & 1.548 & 0.109 & 0.133 & 0.160 & 2.298 & 2.263 & 2.178 \\
      2 & 1.998 & 1.695 & 2.347 & 0.113 & 0.128 & 0.148 & 2.023 & 1.980 & 2.103 \\
      3 & 2.122 & 2.587 & 2.302 & 0.133 & 0.162 & 0.158 & 2.144 & 2.241 & 2.292 \\
      4 & 1.530 & 1.866 & 1.818 & 0.133 & 0.141 & 0.159 & 2.488 & 2.558 & 2.795 \\
    \hline
   ZrO$_2$ & \multicolumn{3}{l}{Energy\,(meV/atom)} & \multicolumn{3}{l}{Force\,(eV/{\AA})} & \multicolumn{3}{l}{Stress\,(kbar)} \\
   loop & train & valid &  test & train & valid &  test & train & valid &  test \\
    \hline
      0 & 0.826 & 0.819 & 1.061 & 0.083 & 0.103 & 0.130 & 1.822 & 1.781 & 1.938 \\
      1 & 3.865 & 2.135 & 2.117 & 0.137 & 0.164 & 0.147 & 2.866 & 4.953 & 1.927 \\
      2 & 3.565 & 2.088 & 2.177 & 0.151 & 0.163 & 0.141 & 3.232 & 4.531 & 2.647 \\
      3 & 3.043 & 4.622 & 1.826 & 0.153 & 0.155 & 0.150 & 3.011 & 3.020 & 2.092 \\
      4 & 3.267 & 1.834 & 1.295 & 0.151 & 0.163 & 0.138 & 2.919 & 2.508 & 1.783 \\
    \hline
  \end{tabular}
\end{table}
Table~\ref{tbl:rmse_train} shows the root mean squared error (RMSE) 
for the Zr$_7$O$_8$N$_4$ and ZrO$_2$ NNPs against existing data sets. 
The energy error is less than 10\,meV/atom in each loop. 
In most cases the relative difference between loops is less than 10\,\%. 
Although the NNP appears to converge sufficiently without going to loop 4, 
the NNP accuracy should not be judged 
solely on the basis of the RMSEs of the existing data.
This is because, as discussed below, 
the RMSEs of the newly acquired data
in the molecular dynamics trajectories performed 
with these NNPs are not necessarily 10\,meV/atom.

\begin{table}
  \caption{
RMSE in the total energy of the newly acquired data,
which are included in the NNPMD trajectories of Zr$_7$O$_8$N$_4$ and ZrO$_2$. 
For loops 0 and 1,
RMSEs for $2\times2\times1$ supercells were not calculated
due to large RMSEs for unit cell and 
the non-convergence for self-consistent calculation, respectively.
}
  \label{tbl:rmse_nnpmd}
  \begin{tabular}{rrrrrrrrr}
    \hline
Zr$_7$O$_8$N$_4$ & \multicolumn{4}{l}{unit cell} & \multicolumn{4}{l}{$2\times2\times1$ supercell} \\
   loop & run = 1 &      2 &      3 &      4 & run = 1 &     2 &     3 &     4  \\
    \hline
      0 &  10.572 & 17.765 & 11.313 & 10.633 &         &       &       &        \\
      1 &   4.833 &  7.105 &  5.323 &  5.113 &         &       &       &        \\
      2 &   2.317 &  2.140 &  8.145 &  2.143 &   2.279 & 2.353 & 5.398 & 2.299  \\
      3 &   1.626 &  1.541 &  1.603 &  1.442 &   1.931 & 6.112 & 6.553 & 6.647  \\
      4 &   1.202 &  1.178 &  1.392 &  1.384 &   1.310 & 1.199 & 1.428 & 1.345  \\
    \hline
ZrO$_2$    & \multicolumn{4}{l}{unit cell} & \multicolumn{4}{l}{$2\times2\times1$ supercell} \\
   loop & run = 1 &      2 &      3 &      4 & run = 1 &     2 &     3 &     4  \\
    \hline
      0 &  28.576 & 28.416 & 28.955 & 25.323 &         &       &       &        \\
      1 &   1.900 &  1.939 &  1.899 &  2.100 &         &       &       &        \\
      2 &   1.628 &  2.553 &  1.514 &  1.573 &   6.632 & 6.231 & 5.308 & 5.153  \\
      3 &   1.469 &  1.094 &  1.088 &  1.152 &   4.936 & 5.024 & 5.159 & 5.116  \\
      4 &   2.644 &  1.160 &  1.625 &  2.123 &   3.021 & 3.902 & 3.884 & 3.543  \\
    \hline
  \end{tabular}
\end{table}

Table~\ref{tbl:rmse_nnpmd} shows the errors 
in the ab initio self-consistent calculations 
for the structures included in the NNPMD trajectories of Zr$_7$O$_8$N$_4$ and ZrO$_2$. 
Ab initio self-consistent calculations were also performed 
for $2\times2\times1$ supercells only in loops 1-4, 
where the RMSE in the unit cell is less than 10\,meV/atom. 
In loop 1, 
the RMSEs and structural properties were not calculated
because the self-consistent calculations of $2\times2\times1$ supercells 
did not converge for some structures.
The RMSEs of loops 2, 3, and 4 were $<$ 10\,meV/atom in both cells, 
so structural properties were calculated.
Note that the total energies in 1\,ns NNPMDs 
reach equilibrium after about 0.1\,ns
and are conserved without significant increase or decrease thereafter.

\begin{figure}
\begin{tabular}{cc}
\includegraphics[width=3.2in]{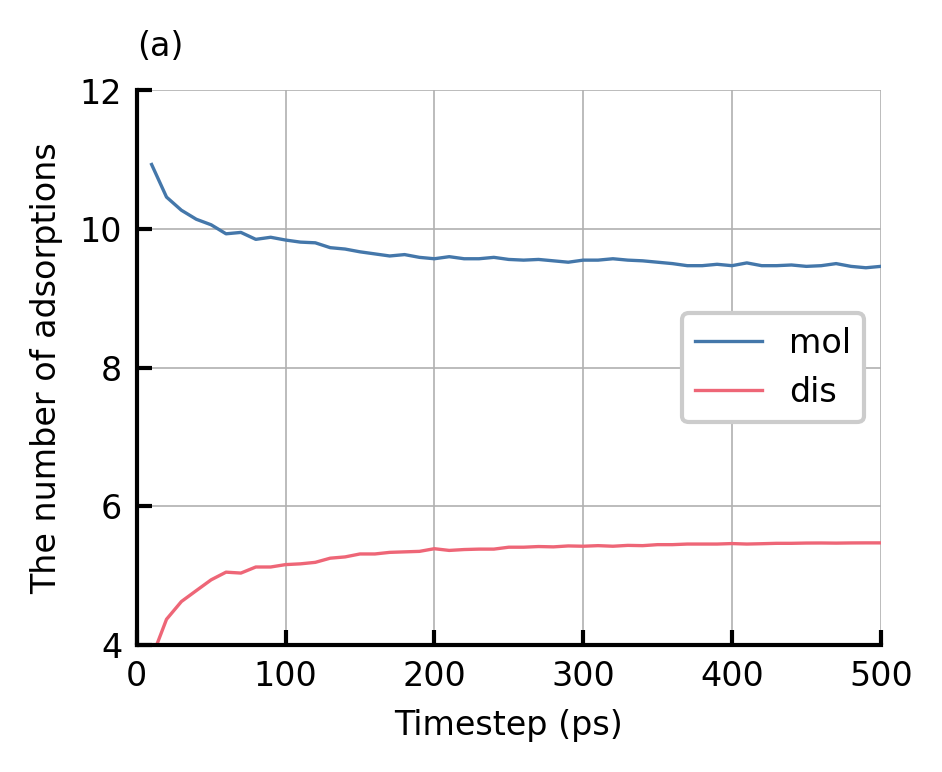} 
\includegraphics[width=3.2in]{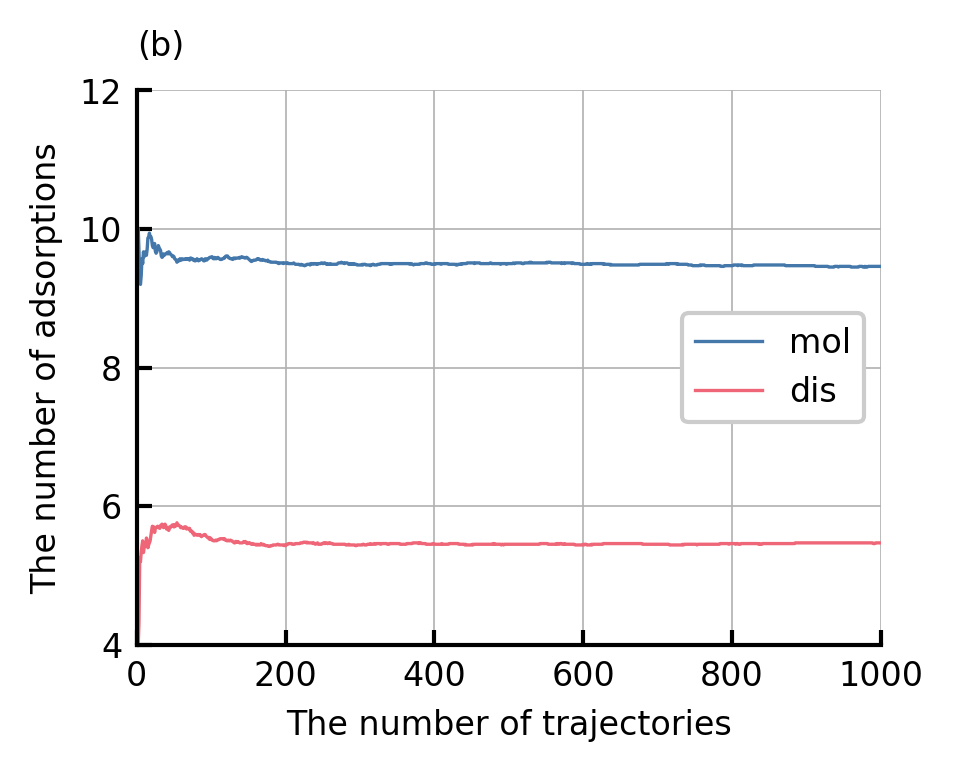}
\end{tabular}
\caption{
Dependence of the number of adsorptions on 
(a) timestep and
(b) the number of trajectories.
Blue and red represent molecular (mol) and dissociative (dis) adsorption, 
respectively.
These were calculated from loop 4 NNPMD trajectories of Zr$_7$O$_8$N$_4$.
}
\label{fgr:depend}
\end{figure}

Whether sufficient time and the number of trajectories were used
to calculate the structural properties was verified by 
the number of adsorptions calculated from 
the loop 4 NNPMD trajectories of Zr$_7$O$_8$N$_4$.
Figure~\ref{fgr:depend} (a) shows
the time dependence of the number of adsorptions
calculated from the 10, 20, ..., 500\,ps structure of 1000 trajectories.
It can be seen that it converges to 0.1 at 400\,ps.
Figure~\ref{fgr:depend} (b) shows
the dependence of the number of adsorptions on the number of trajectories
calculated from 
the 500\,ps structure of 2, 3, ..., 1000 trajectories.
It can be seen that the number converges to 0.1 for 200 trajectories.
Thus the number of adsorptions calculated from 
the 500\,ps structure of 1000 trajectories can be trusted to the order of 0.1.

\begin{table}
  \caption{
Mean values of the number of molecular (mol) and dissociative (dis) adsorptions
in Zr$_7$O$_8$N$_4$ and ZrO$_2$.
The values in brackets are standard deviations.
For both materials,
these were calculated from loop 2, 3 and 4 NNPMD trajectories.
N = the number of adsorptions in a $2\times2\times1$ supercell.
M = N/4 = the number of adsorptions per unit cell.
C = coverage (\%) = N/28(= the number of Zr sites in $2\times2\times1$ supercell surface)*100.
}
  \label{tbl:nads}
  \begin{tabular}{rrrrrrr}
    \hline
\multicolumn{7}{l}{Zr$_7$O$_8$N$_4$} \\
   loop &  N$_\text{mol}$ & N$_\text{dis}$ & M$_\text{mol}$ & M$_\text{dis}$ & C$_\text{mol}$ & C$_\text{dis}$ \\
    \hline
      2 &  9.6 (1.2) & 5.8 (1.1) & 2.4 (0.3) & 1.5 (0.3) & 34 (4) & 21 (4) \\
      3 & 10.2 (1.2) & 5.5 (1.0) & 2.5 (0.3) & 1.4 (0.2) & 36 (4) & 20 (3) \\
      4 &  9.5 (1.1) & 5.5 (1.0) & 2.4 (0.3) & 1.4 (0.3) & 34 (4) & 20 (4) \\
    \hline
\multicolumn{7}{l}{ZrO$_2$} \\
   loop & N$_\text{mol}$ & N$_\text{dis}$ & M$_\text{mol}$ & M$_\text{dis}$ & C$_\text{mol}$ & C$_\text{dis}$ \\
    \hline
      2 &  4.1 (1.4) & 9.3 (1.3) & 1.0 (0.4) & 2.3 (0.3) & 15 (5) & 33 (5) \\
      3 &  6.2 (1.4) & 9.1 (1.2) & 1.6 (0.4) & 2.3 (0.3) & 22 (5) & 33 (4) \\
      4 &  5.6 (1.4) & 9.6 (1.2) & 1.4 (0.4) & 2.4 (0.3) & 20 (5) & 34 (4) \\
    \hline
  \end{tabular}
\end{table}

Table~\ref{tbl:nads} shows 
the number of molecular and dissociative adsorptions (N)
of Zr$_7$O$_8$N$_4$ and ZrO$_2$ calculated by each loop NNP.
For reference, the number of adsorptions per unit cell (M), 
converted from the number of adsorptions per $2\times2\times1$ supercell,
and the coverage (C) are also given.
For the number of adsorptions N, 
the mean difference between loops 3 and 4 is up to 0.7 
and the standard deviation between the materials is up to 2.5.
Both are small compared to the difference between materials in loop 4, 2.9,
so we consider that the convergence is not a problem 
for comparing these materials.

\subsection{Structural analysis}
The relationship between the number of molecular and dissociative adsorptions
is reversed between Zr$_7$O$_8$N$_4$ and ZrO$_2$.
Because Zr$_7$O$_8$N$_4$ has less oxygen for the \VO,
the dissociated excess H$^+$ is less likely to be adsorbed on oxygen
and more likely to re-bind to dissociatively adsorbed OH$^-$, 
which is the cause. 
This idea is also supported by the fact that 
the total number of adsorptions is almost the same for Zr$_7$O$_8$N$_4$ and ZrO$_2$.
In addition, Zr$_7$O$_8$N$_4$ is charge neutral 
due to the presence of the \VO\,and \NO\,defects, 
so it is unlikely that the \VO\,is responsible for attracting the O atom.
According to a study by Muhammady et al.,
\cite{Muhammady2022b}
where one \VO\,and two \NO\,defects were introduced 
into the tetragonal-ZrO$_2$ (101) surface,
the \VO\,is not more easily adsorbed by the O$_2$ molecule than the Zr atom. 
Therefore, it is consistent with the present results to consider 
that the \VO\,does not promote dissociation 
by accepting the O of water molecules.
The dissociative adsorption becomes molecular adsorption, 
which reduces the probability that intermediates and 
the H$^+$ that binds to them in the ORR
will have an additional reaction with nearby adsorbates, 
i.e. the ORR is more likely to 
proceed in the presence of oxygen defects.
Also, while the total number of adsorptions is almost the same and 
the number of vacant Zr sites is almost the same,
Zr$_7$O$_8$N$_4$ has two \VO s per unit cell, 
which could also be an active site for the ORR.
In other words, Zr$_7$O$_8$N$_4$ has more active sites 
and is more likely to promote the ORR than ZrO$_2$.

\begin{figure}
\begin{tabular}{cc}
\includegraphics[width=3.2in]{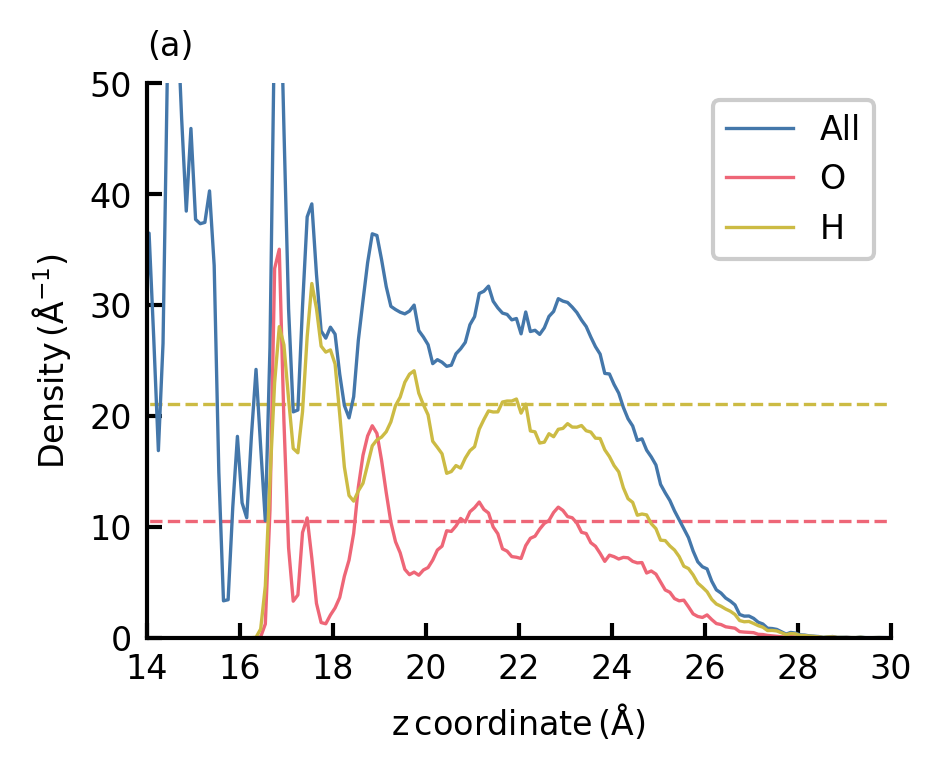} 
\includegraphics[width=3.2in]{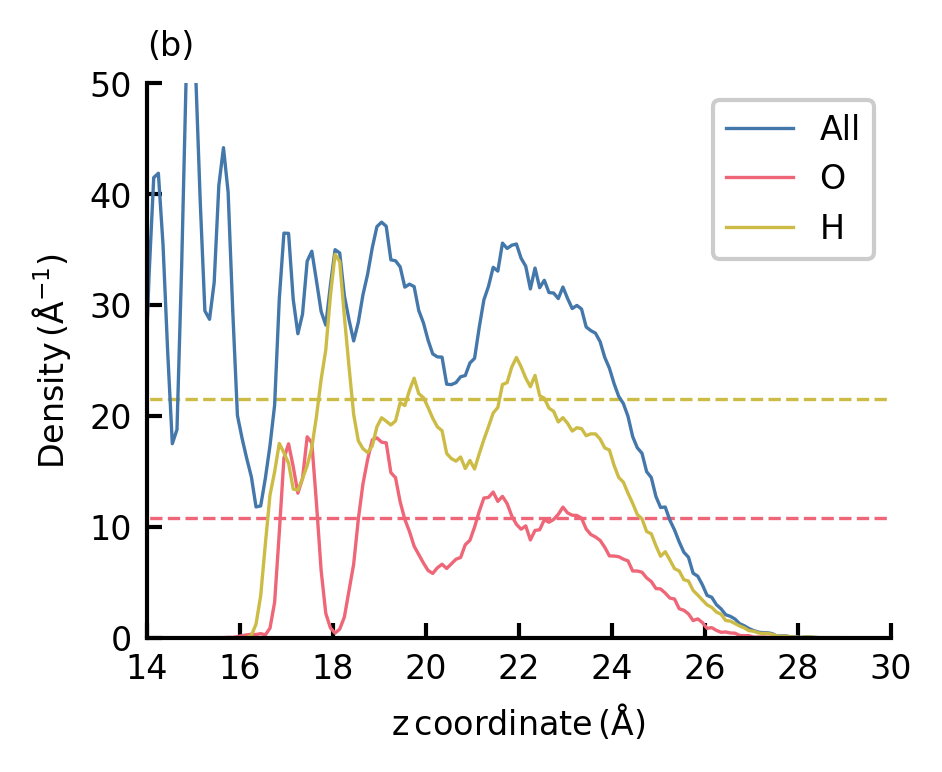}
\end{tabular}
\caption{
Surface perpendicular number density distributions\,({\AA}$^{-1}$)
of (a) Zr$_7$O$_8$N$_4$ and (b) ZrO$_2$.
The blue, red and yellow solid lines represent 
total, oxygen and hydrogen atoms in water at the interface.
The red and yellow dashed lines represent
oxygen and hydrogen atoms in bulk water.
These densities for both materials were calculated 
from loop 4 NNPMD trajectories.
}
\label{fgr:density_1d}
\end{figure}

Figure~\ref{fgr:density_1d} shows 
the surface perpendicular number density distributions of Zr$_7$O$_8$N$_4$ and ZrO$_2$. 
The densities of total, oxygen and hydrogen atoms in water at the interface
and hydrogen atoms in bulk water are shown respectively. 
The surface perpendicular number density distribution shows 
that the oxygen peak in the adsorption layer (16-18\,{\AA}) is one in ZrO$_2$, 
whereas it is divided into two in Zr$_7$O$_8$N$_4$,
which is due to the fact that the height of the Zr sites 
is almost the same in all ZrO$_2$ and is divided into two types in Zr$_7$O$_8$N$_4$.
Differences in the surface-derived adsorption structures 
will affect their function as catalysts for the ORR,
as well as the diffusion of molecules and protons.

Away from the surface, 
the number density oscillates and is not constant as in the bulk. 
In other words, the motion is limited compared to the bulk.
According to a previous study of the Cu/H$_2$O interface, 
\cite{Natarajan2016}
water molecules move like the bulk 
when the thickness of their layer is more than 30\,{\AA}.
In the present study,
the thickness of the water molecule layer is less than 15\,{\AA},
which is consistent with the results that it is not constant like the bulk.

\begin{figure}
\begin{tabular}{cc}
\includegraphics[width=3.2in]{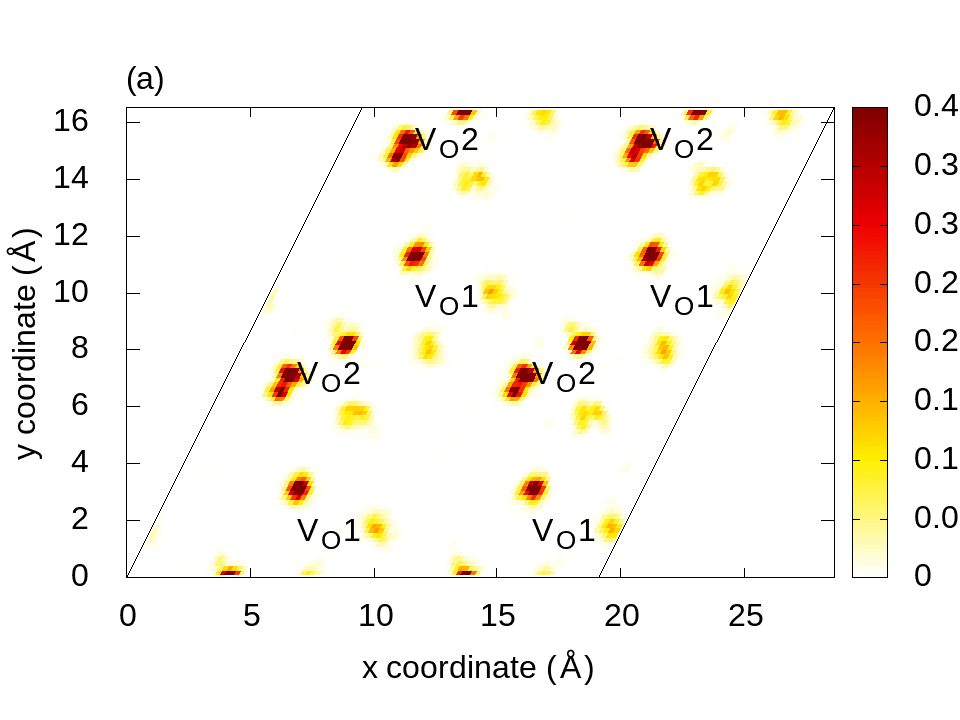}
\includegraphics[width=3.2in]{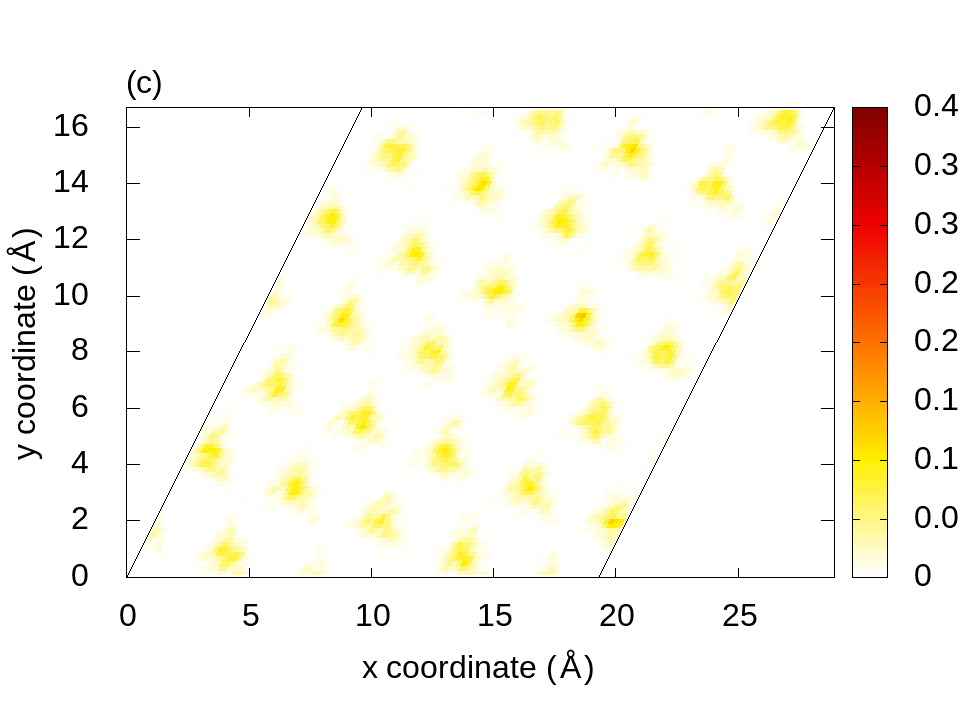} \\
\includegraphics[width=3.2in]{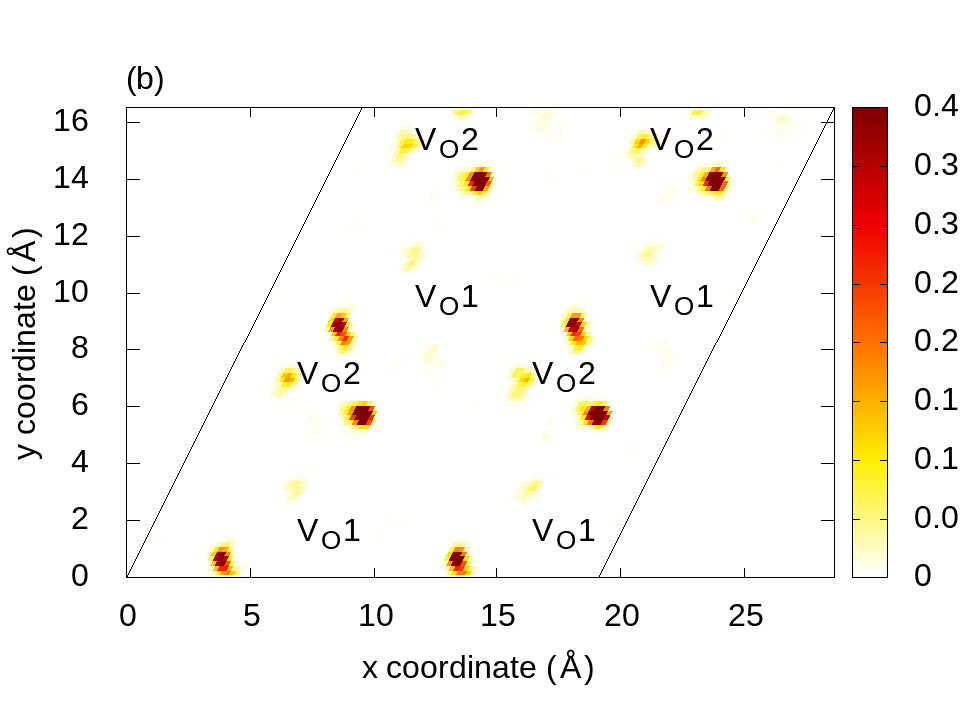}
\includegraphics[width=3.2in]{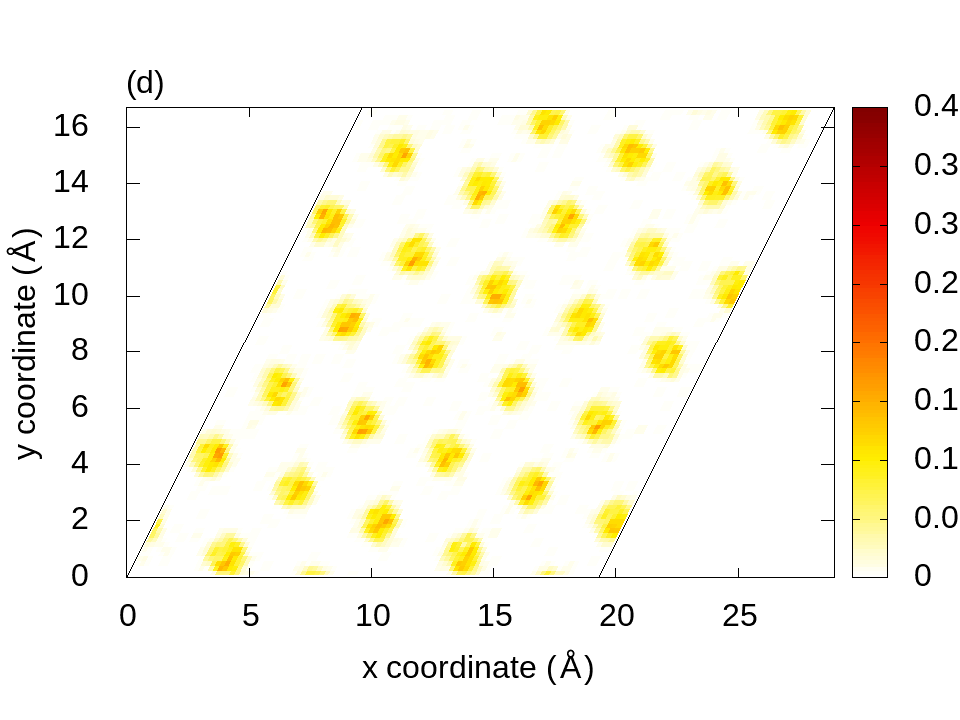}
\end{tabular}
\caption{
Surface parallel number density distributions\,({\AA}$^{-3}$) of O atoms in
molecular (top) and dissociative adsorption (bottom) 
in Zr$_7$O$_8$N$_4$ (left) and ZrO$_2$ (right). 
Two types of oxygen vacancies in a Zr$_7$O$_8$N$_4$ unit cell
are represented by \VO1 and \VO2.
}
\label{fgr:density_3d}
\end{figure}

Figure~\ref{fgr:density_3d} shows 
the surface parallel number density distributions of Zr$_7$O$_8$N$_4$ and ZrO$_2$.
Only the densities of the oxygen atoms 
in the H$_2$O or OH$^-$ adsorbed on the Zr sites are shown.
The distribution of ZrO$_2$ is uniform and translational symmetry is observed.
As mentioned above, this is not due to the initial structure,
as the translational symmetry of water is not guaranteed.
ZrO$_2$ was assumed to have a uniform adsorption distribution 
due to the equivalent Zr sites on its surface.
However, if the adsorption distribution does not change much in 500\,ps MD,
the adsorption distribution calculated from only one trajectory
may not be uniform.
In the present study, the adsorption distribution was 
calculated from 1000 independent trajectories,
so that a uniform distribution was obtained, 
regardless of whether the adsorption distribution 
was likely to change in 500\,ps MD. 
The distribution of Zr$_7$O$_8$N$_4$ is non-uniform, 
but translational symmetry is established.
This is also a result of obtaining a sufficiently large number of trajectories,
rather than the initial structure, 
and is due to the low computational cost of NNPMD, 
which allows so many MDs to be performed.

Particles were found to adsorb on the Zr, but not on the \VO.
Zr atoms around \VO\,2 are more likely to adsorb particles 
than those around \VO\,1
These results suggest 
that the \VO\,act as active sites for the ORR and 
that differences in adsorption around them affect their function.
In particular, 
we suggest that the \VO\,1 is more likely to adsorb O$_2$ molecules 
as there is less adsorption on the surrounding Zr atoms.

\begin{figure}
\includegraphics[width=16cm]{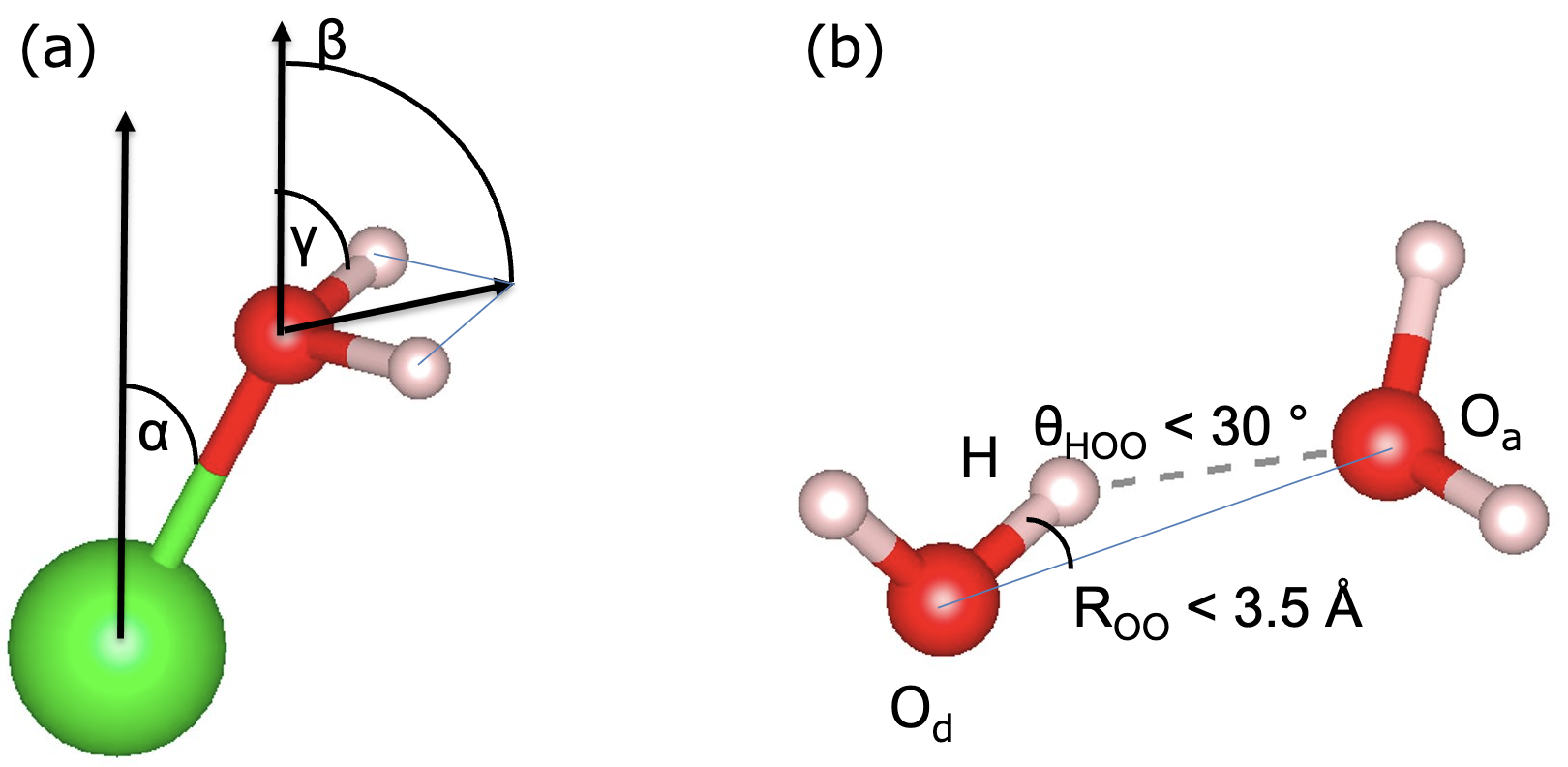}
\caption{
(a) Definition of adsorption angle $\alpha$, molecular orientation $\beta$ 
and OH orientation $\gamma$. 
(b) Definition of hydrogen bonding. 
Oxygens donating and accepting hydrogen are denoted O$_d$ and O$_a$. 
Green, red and white indicate Zr, O and H atoms respectively.
The Figures have been generated using the VESTA software package.
}
\label{fgr:anglebond}
\end{figure}

\begin{table}
  \caption{
Adsorption distance d\,({\AA}), adsorption angle $\alpha$, molecular orientation $\beta$, 
OH orientation $\gamma$ and number of hydrogen bonds N$_\text{HB}$ for Zr$_7$O$_8$N$_4$ and ZrO$_2$, 
for molecular (mol) and dissociative (dis) adsorption.
}
  \label{tbl:anglebond}
  \begin{tabular}{lllllllll}
    \hline
        & d$_\text{mol}$ & d$_\text{dis}$ & $\alpha_\text{mol}$ & $\alpha_\text{dis}$ & $\beta_\text{mol}$ & $\gamma_\text{mol}$ & $\gamma_\text{dis}$ & N$_\text{HB}$  \\
    \hline
Zr$_7$O$_8$N$_4$ &  2.39 &  2.20 & 18.28 & 17.74 & 51.68 & 68.40 & 53.35 & 68.70  \\
ZrO$_2$    &  2.29 &  2.11 & 12.81 & 20.23 & 35.03 & 62.14 & 62.38 & 67.56  \\
    \hline
  \end{tabular}
\end{table}

For other properties, 
(adsorption distance d, adsorption angle $\alpha$, 
molecular orientation $\beta$, OH orientation $\gamma$, and
number of hydrogen bonds N$_\text{HB}$)
their definitions are given before the results are presented.
Figure~\ref{fgr:anglebond} (a) shows the definition of 
the adsorption angle $\alpha$, molecular orientation $\beta$ 
and OH orientation $\gamma$.
The adsorption angle $\alpha$ is the angle between the position vector of O 
with respect to Zr and the surface normal.
The adsorption distance d is the absolute value of the vector.
The water molecule orientation $\beta$ is the angle 
between the dipole vector of the water molecule and the surface normal.
The OH orientation $\gamma$ is the angle 
between the position vector of H with respect to O and the surface normal.
Figure~\ref{fgr:anglebond} (b) shows the definition of a hydrogen bond.
A hydrogen bond is defined when the angle H-O$_d$-O$_a$ 
is less than 30$^\circ$, 
the distance O$_d$-H is less than 1.2\,{\AA} and 
the distance O$_d$-O$_a$ is less than 3.5\,{\AA}, 
where O$_d$ and O$_a$ are the oxygen donating and accepting hydrogen, 
respectively.
\cite{Luzar1996}

Table~\ref{tbl:anglebond} shows 
the adsorption distance d, adsorption angle $\alpha$,
molecular orientation $\beta$, OH orientation $\gamma$, and
the number of hydrogen bonds N$_\text{HB}$ 
calculated based on the above definition.
The adsorption distance d of ZrO$_2$ is 0.1\,{\AA} shorter 
in both molecular and dissociation.
The increase or decrease of the adsorption angle $\alpha$ is opposite 
for molecular and dissociation, with the adsorption angles becoming unequal,
whereas for Zr$_7$O$_8$N$_4$ they were almost equal at 18$^\circ$.
The molecular orientation $\beta$ of ZrO$_2$ decreases, i.e. it turns upwards.
The increase and decrease of the OH orientation $\gamma$ is 
opposite in molecular and dissociation, 
going from unequal in Zr$_7$O$_8$N$_4$ to aligned at almost 62$^\circ$.
The number of hydrogen bonds in ZrO$_2$ is reduced by one.
We consider this to be the effect of a distortion in the distribution 
of water molecules due to the defect.

\begin{figure}
\begin{tabular}{cc}
\includegraphics[width=3.2in]{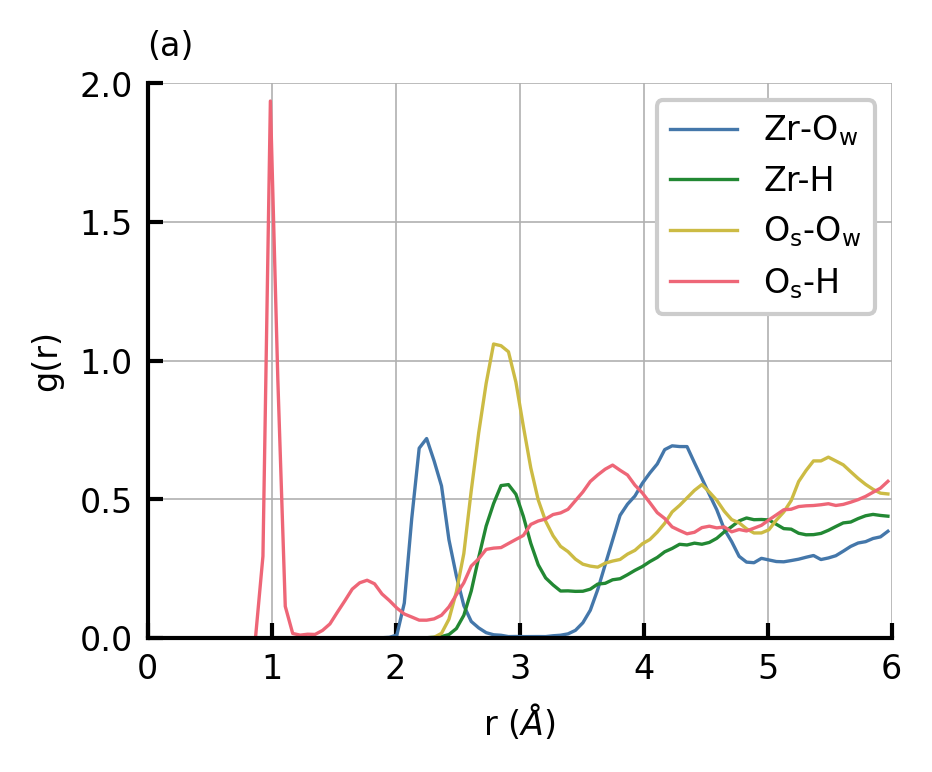} 
\includegraphics[width=3.2in]{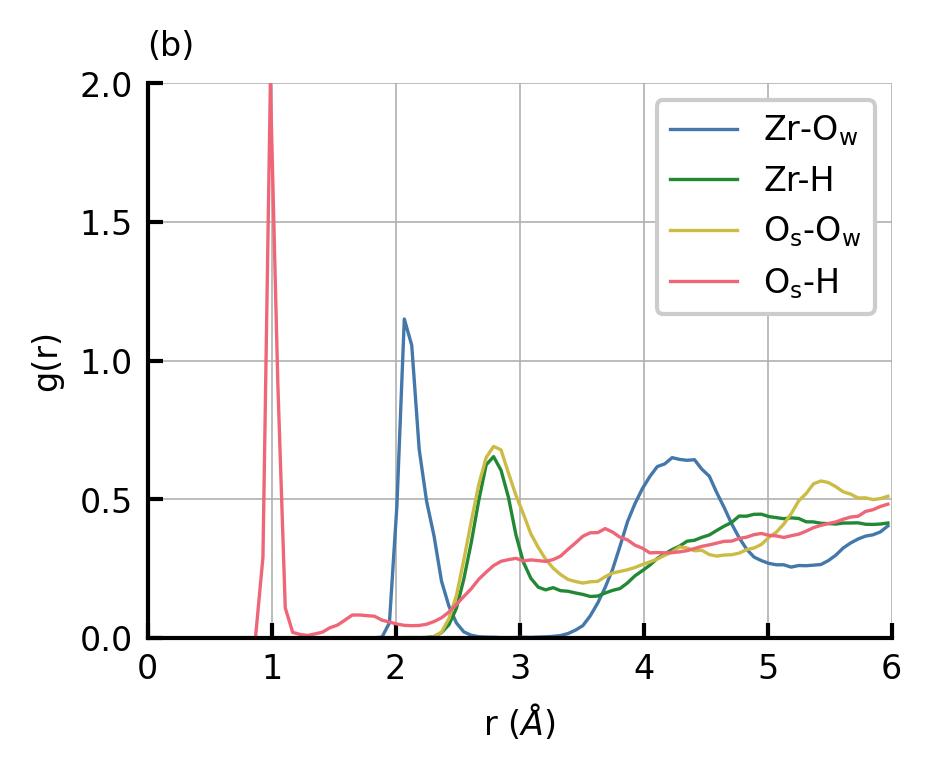}
\end{tabular}
\caption{
Radial distribution function $g(r)$ of (a) Zr$_7$O$_8$N$_4$ and (b) ZrO$_2$.
O$_\text{s}$ and O$_\text{w}$ represent oxygen in the slab and in the water molecule.
Blue, green, yellow and red represent 
Zr-O$_\text{w}$, Zr-H, O$_\text{s}$-O$_\text{w}$ and O$_\text{s}$-H, respectively.
}
\label{fgr:rdf}
\end{figure}

Figure~\ref{fgr:rdf} shows the Radial Distribution Function (RDF).
The oxygens in the slab and in the water is distinguished as O$_\text{s}$ and O$_\text{w}$. 
Compared to ZrO$_2$, 
the RDF of Zr-O$_\text{w}$ in Zr$_7$O$_8$N$_4$ 
has a lower peak at 2.1\,{\AA} and is shifted to the right.
This means that the average adsorption distance increases. 
Both molecular and dissociative adsorption distances are shorter for Zr$_7$O$_8$N$_4$,
but the increase in the proportion of molecular adsorption 
at longer adsorption distances is thought to be the cause.
The lower peak at 1\,{\AA} in the 
O$_\text{s}$-H RDF of Zr$_7$O$_8$N$_4$ is thought to be 
due to a decrease in H adsorbed 
on O$_\text{s}$ as dissociative adsorption decreases.
\newpage

\subsection{Proton transfer}

\begin{figure}
\begin{tabular}{cc}
\includegraphics[width=3.2in]{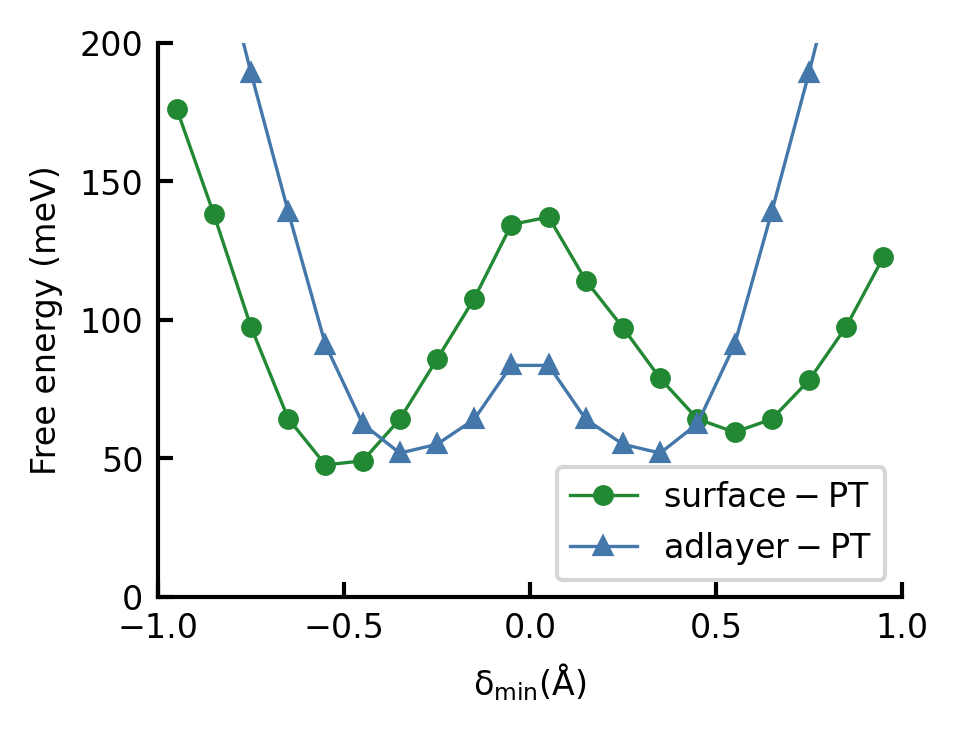} & 
\includegraphics[width=3.2in]{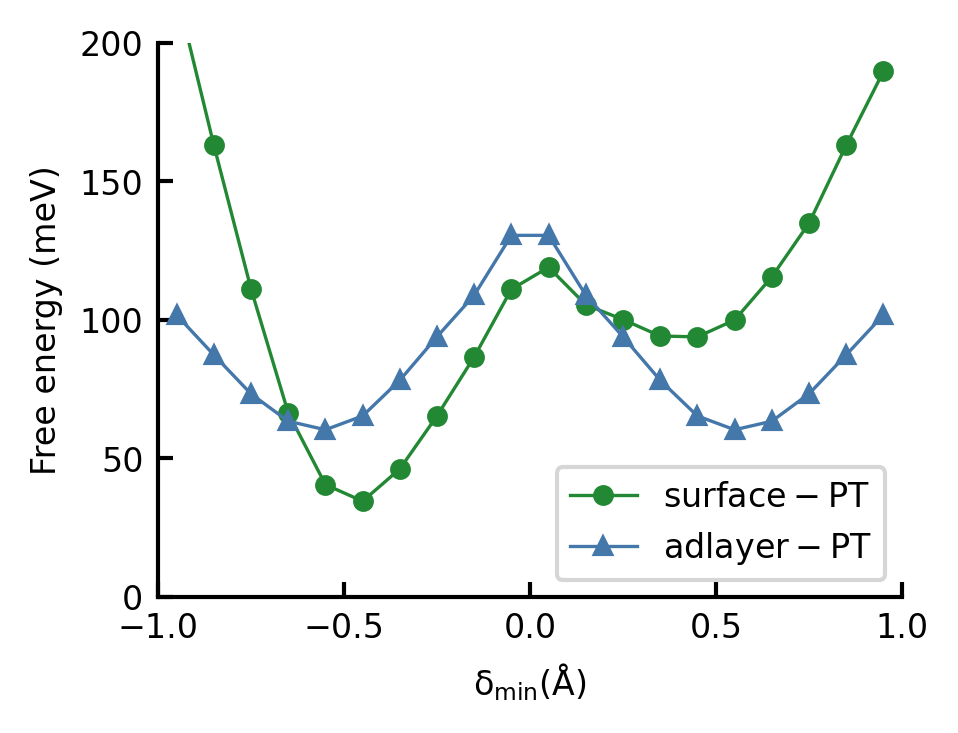} \\
\end{tabular}
\caption{
Free energy for CV = $\delta_\text{min}$ of surface-PT and adlayer-PT
in Zr$_7$O$_8$N$_4$ (left) and ZrO$_2$ (right).
The PT reaction formulas are summarised in the Table~\ref{tbl:PT_list}.
}
\label{fgr:fe_suad}
\end{figure}
\begin{table}
\caption{
Free energy barrier for surface-PT and adlayer PT reaction in Zr$_7$O$_8$N$_4$ and ZrO$_2$.
The PT reaction formulas are summarised in the Table~\ref{tbl:PT_list}.
}
  \label{tbl:fe_barrier}
  \begin{tabular}{rrrrr}
    \hline
      & \multicolumn{2}{r}{surface-PT} & \multicolumn{2}{r}{adlayer-PT} \\ 
      & $\rightarrow$ & $\leftarrow$ & $\rightarrow$ & $\leftarrow$ \\
    \hline
      Zr$_7$O$_8$N$_4$ & 89 & 78 & 34 & 34 \\
      ZrO$_2$    & 89 & 25 & 71 & 71 \\
    \hline
  \end{tabular}
\end{table}
Figure~\ref{fgr:fe_suad} represents the free energy of 
the surface PT and the adlayer PT.
The surface PT is the PT between adsorbed H$_2$O and surface oxygen,
The adlayer PT is the PT between adsorbed H$_2$O and adsorbed OH$^-$.
Table~\ref{tbl:fe_barrier} shows the barriers in these free energy.
The leftward barrier at the surface-PT represents the ease with which 
molecular adsorption turns into dissociative adsorption, 
and ZrO$_2$ has a much lower barrier than Zr$_7$O$_8$N$_4$.
This is consistent with the fact that 
the number of dissociative adsorptions of Zr$_7$O$_8$N$_4$ is smaller than that of ZrO$_2$, 
as mentioned above.
As Zr$_7$O$_8$N$_4$ has less O$_\text{s}$, adsorption of H onto O$_\text{s}$ is less likely to occur than on ZrO$_2$.
The barrier in the adlayer PT is lower for Zr$_7$O$_8$N$_4$ than for ZrO$_2$.
This is thought to be 
because in Zr$_7$O$_8$N$_4$ the adsorbates in Zr collect around \VO, 
so there is a high probability that H$_2$O$^*$ and O$^*$H$^-$ are adjacent.
However, the fact that PT is more likely to occur in Zr$_7$O$_8$N$_4$ 
does not necessarily mean that protons are more likely to diffuse over long distances.
In Zr$_7$O$_8$N$_4$ the adsorbates tend to be densely packed and 
even if an adlayer PT frequently occurs between them, 
protons will circulate in the same place and will not contribute to effective diffusion.
We will discuss later how easy it is for protons to actually diffuse over 
long distances by calculating mean squared displacement 
and diffusion coefficient.

\newpage

\begin{figure}
\begin{tabular}{cc}
\includegraphics[width=3.2in]{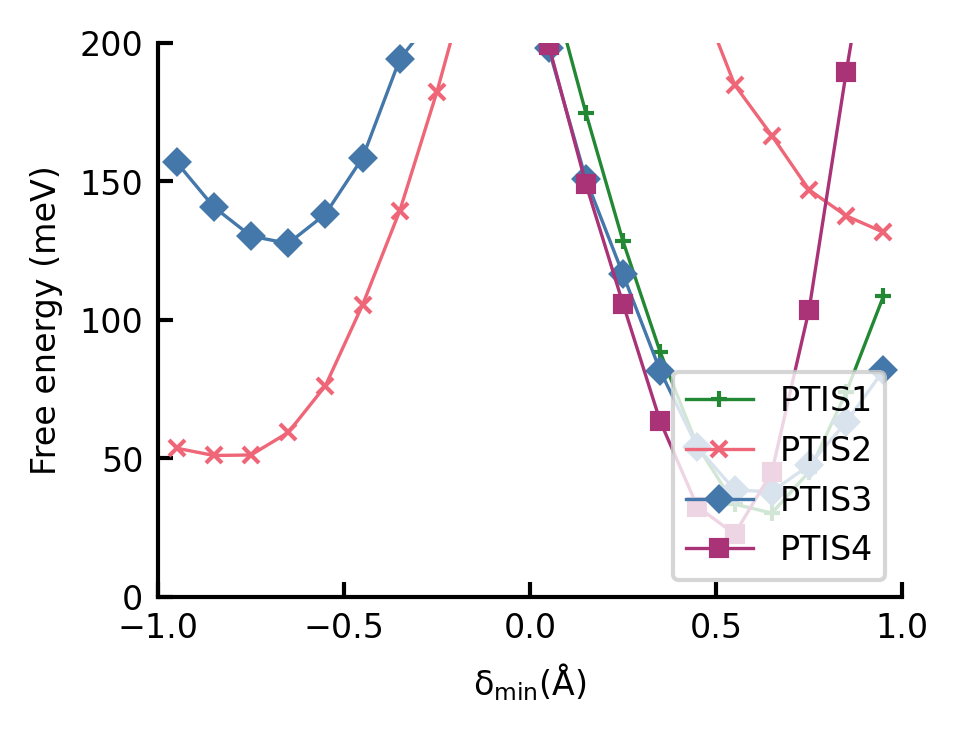} & 
\includegraphics[width=3.2in]{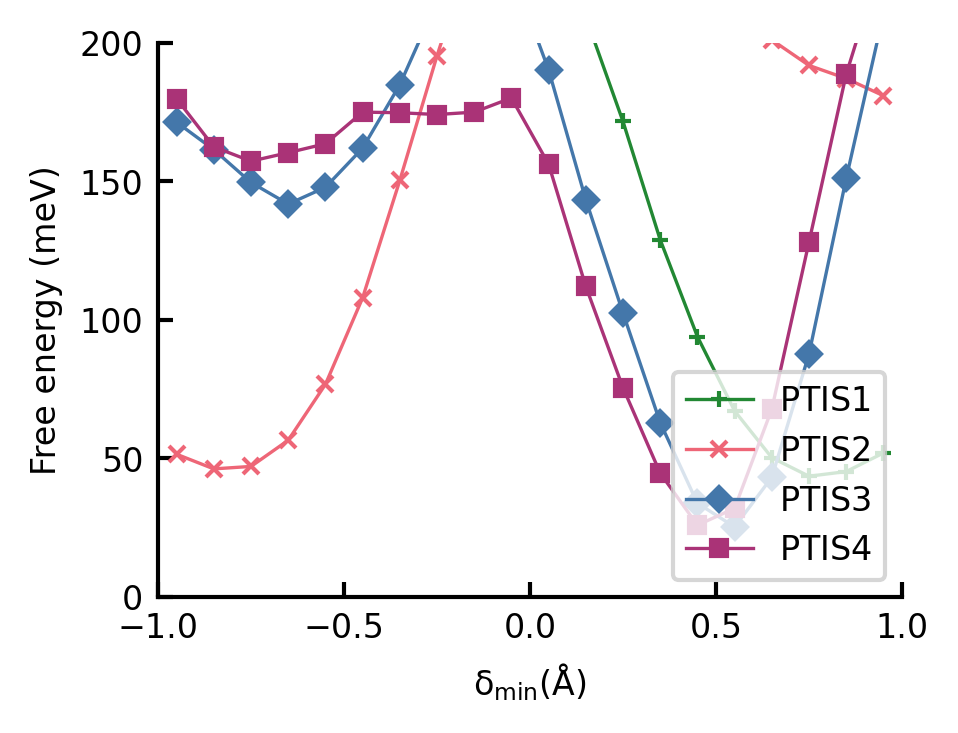} \\
\end{tabular}
\caption{
Free energy for CV = $\delta_\text{min}$ of PT involving the solvent (PTIS)
in Zr$_7$O$_8$N$_4$ (left) and ZrO$_2$ (right).
The PT reaction formulas are summarised in the Table~\ref{tbl:PT_list}.
}
\label{fgr:fe_ptis}
\end{figure}
Figure~\ref{fgr:fe_ptis}
shows the free energy for CV = $\delta_\text{min}$ of PT involving the solvent (PTIS).
For most PTs the barrier is so high 
that there is almost no chance of a reaction taking place.
Only PTIS4 of ZrO$_2$ has a relatively low barrier 
and a reaction can occur in rare cases.
\newpage

\begin{figure}
\begin{tabular}{cc}
\includegraphics[width=3.2in]{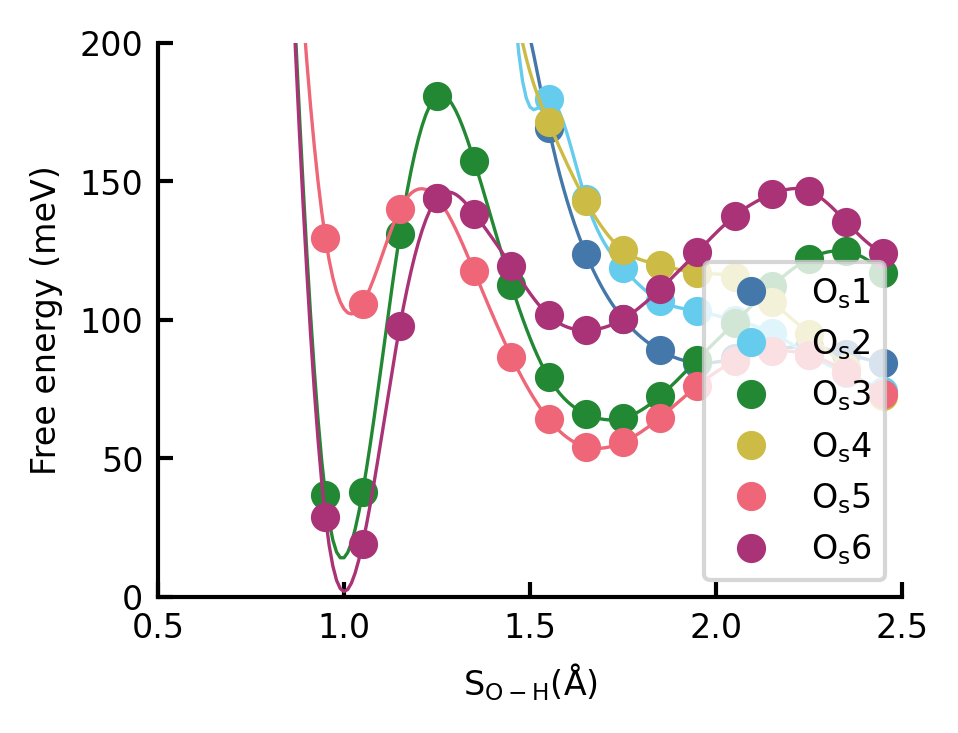} & 
\includegraphics[width=3.2in]{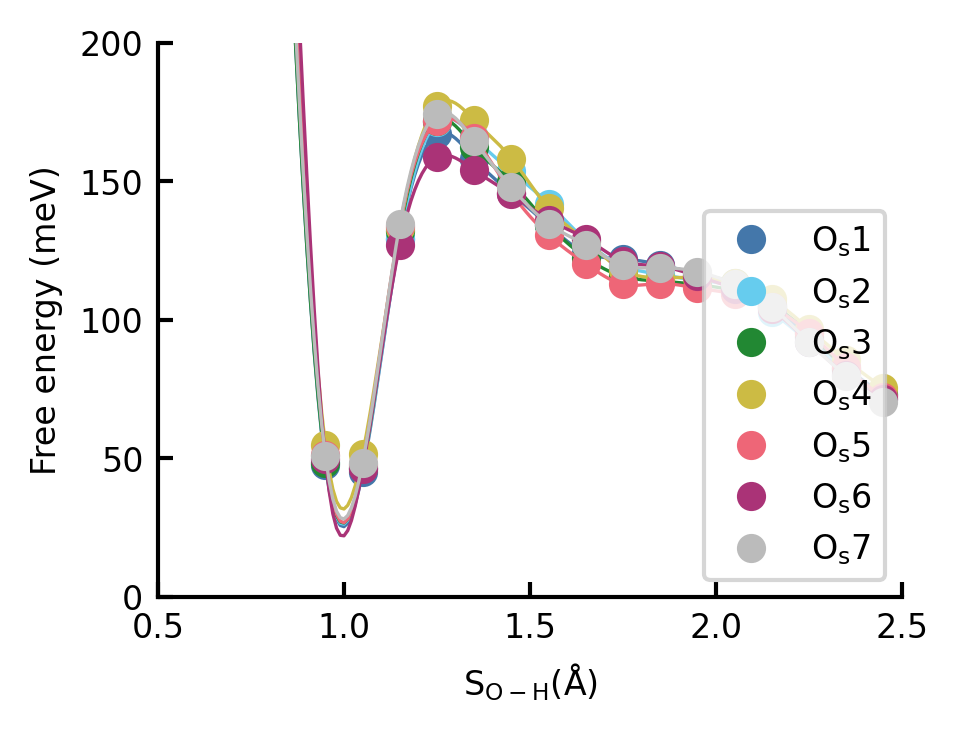} \\
\end{tabular}
\caption{
Free energy for CV = S$_\text{O-H}$ of PT involving on each O$_\text{s}$ 
in Zr$_7$O$_8$N$_4$ (left) and ZrO$_2$ (right).
Positions of each O$_\text{s}$ sites is shown in Figure~\ref{fgr:fe_os}.
}
\label{fgr:fe_soh}
\end{figure}
\begin{figure}
\begin{tabular}{cc}
\includegraphics[width=3.2in]{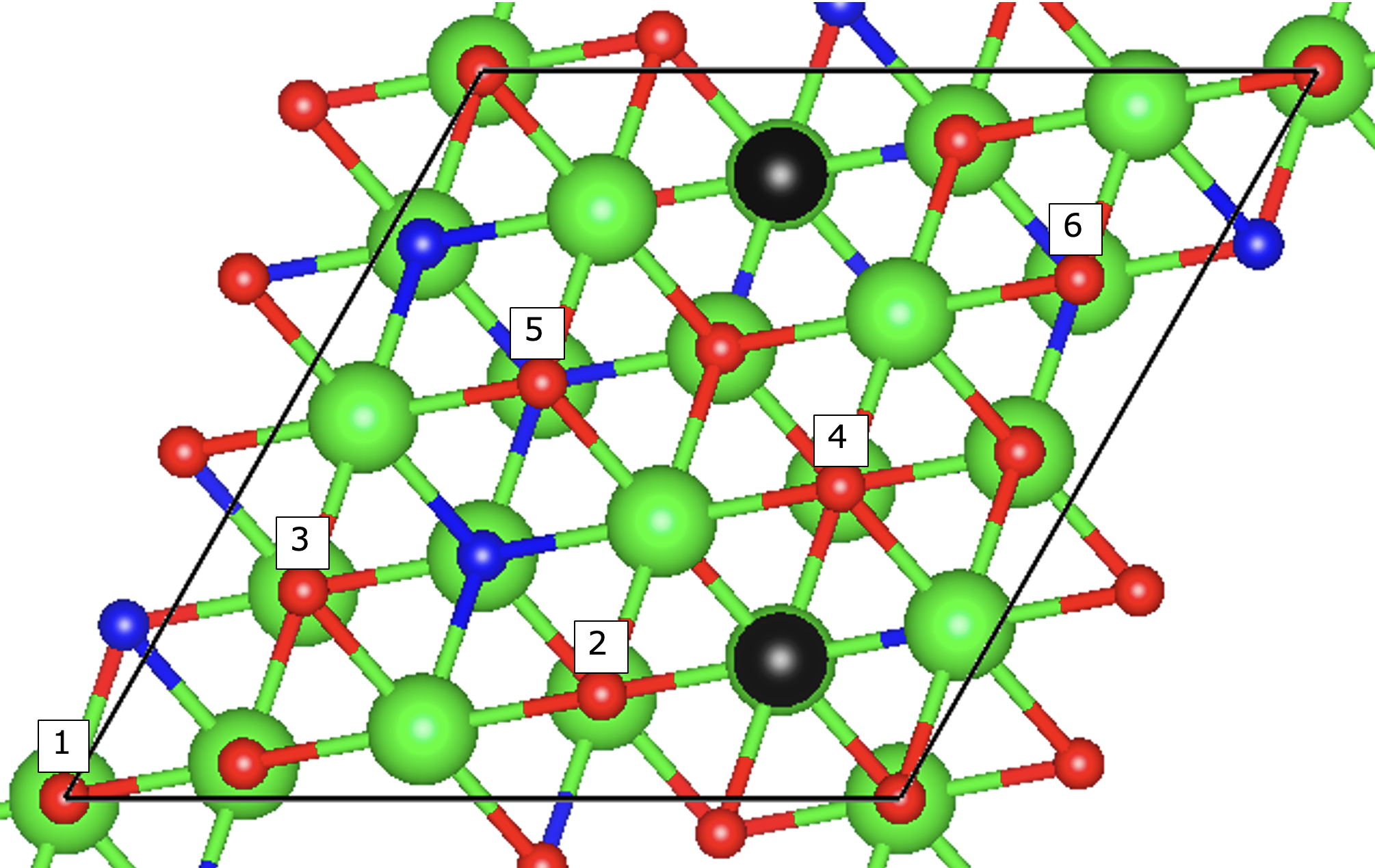} & 
\includegraphics[width=3.2in]{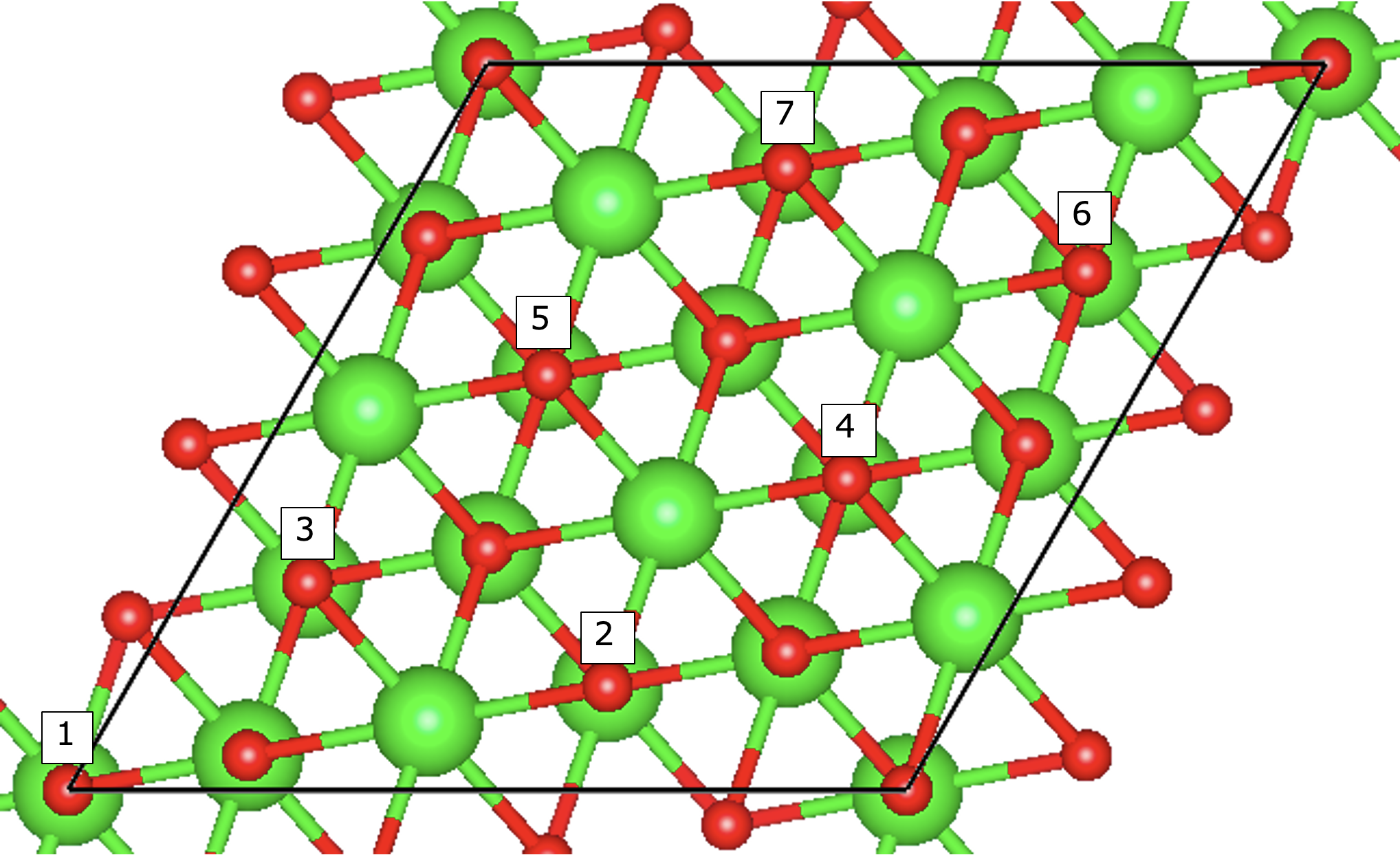} \\
\end{tabular}
\caption{
Positions of each O$_\text{s}$ sites 
in Zr$_7$O$_8$N$_4$ (left) and ZrO$_2$ (right).
Green, blue, red and black indicate Zr, N, O and \VO respectively.
The Figures have been generated using the VESTA software package.
}
\label{fgr:fe_os}
\end{figure}
Figure~\ref{fgr:fe_soh}
shows the free energy for CV = S$_\text{O-H}$ of PT involving on each O$_\text{s}$ .
Figure~\ref{fgr:fe_os}
shows the positions of the O$_\text{s}$s in the collective variable by number.
In Zr$_7$O$_8$N$_4$, since the O atoms on the surface are not equivalent, 
the free energy differs depending on the O$_\text{s}$.
For O$_\text{s}$=1,2,4, 
the S$_\text{O-H}$ does not reach a minimum close to 1 and there is no surface PT.
The fact that dissociation PT is unlikely to occur in these oxygens near \VO1 
is consistent with the fact that 
there is almost no dissociated adsorption density near \VO1 
(Figure~\ref{fgr:density_3d}).
On the other hand, in ZrO$_2$, 
all O$_\text{s}$s are equivalent, so the free energy is the same for all O$_\text{s}$s.
In any case, with this CV, it is unclear 
between O$_\text{s}$ and which oxygen (adsorbate or solvent) PT has occurred.

\begin{figure}
\begin{tabular}{cc}
\includegraphics[width=3.2in]{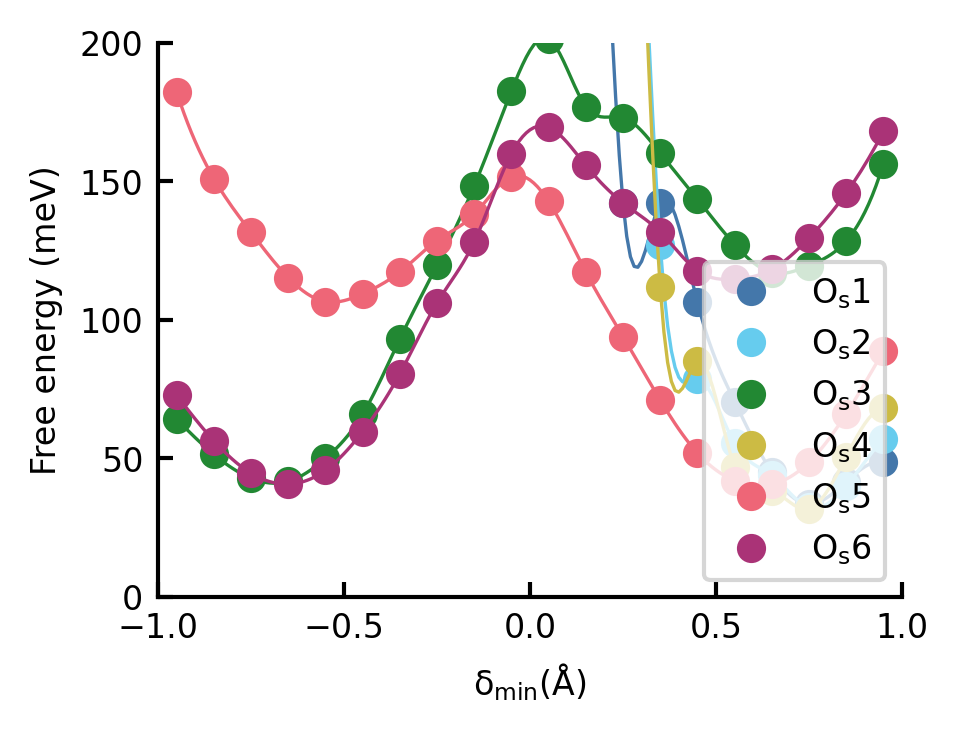} & 
\includegraphics[width=3.2in]{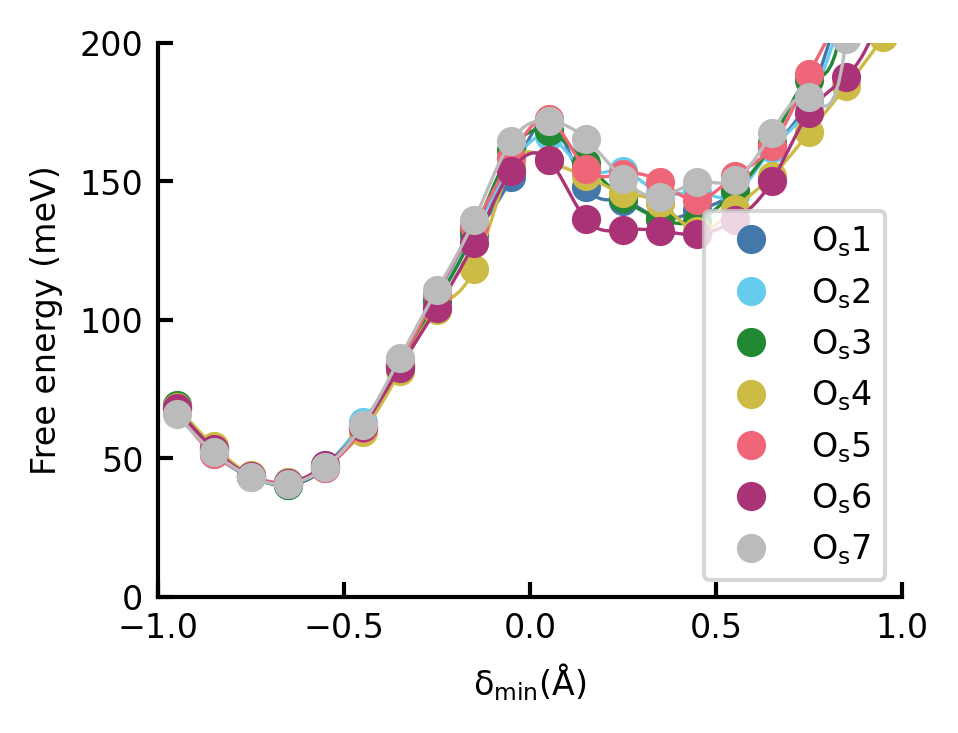} \\
\end{tabular}
\caption{
Free energy for CV = $\delta_\text{min}$ of surface-PT on each O$_\text{s}$
in Zr$_7$O$_8$N$_4$ (left) and ZrO$_2$ (right).
Positions of each O$_\text{s}$ sites is shown in Figure~\ref{fgr:fe_os}.
The PT reaction formulas are summarised in the Table~\ref{tbl:PT_list}.
}
\label{fgr:fe_suos}
\end{figure}
Figure~\ref{fgr:fe_suos} 
shows free energy for CV = $\delta_\text{min}$ of surface-PT on each O$_\text{s}$.
Figure~\ref{fgr:fe_os}
shows the positions of O$_\text{s}$s in the collective variable by number.
In Zr$_7$O$_8$N$_4$, since the O$_\text{s}$s are not equivalent, 
the free energy differs depending on the O$_\text{s}$.
When O$_\text{s}$=1,2,4, the free energy becomes infinite 
in the negative region of $\delta_\text{min}$, 
so it can be seen that dissociative adsorption does not occur.
On the other hand, in ZrO$_2$, 
all O$_\text{s}$s are equivalent, so the free energy is the same for all O$_\text{s}$s.
Contrary to Figure~\ref{fgr:fe_soh},
it is clear that this collective variable represents
a surface-PT between Os and adsorbed water.

\begin{figure}
\begin{tabular}{cc}
\includegraphics[width=3.2in]{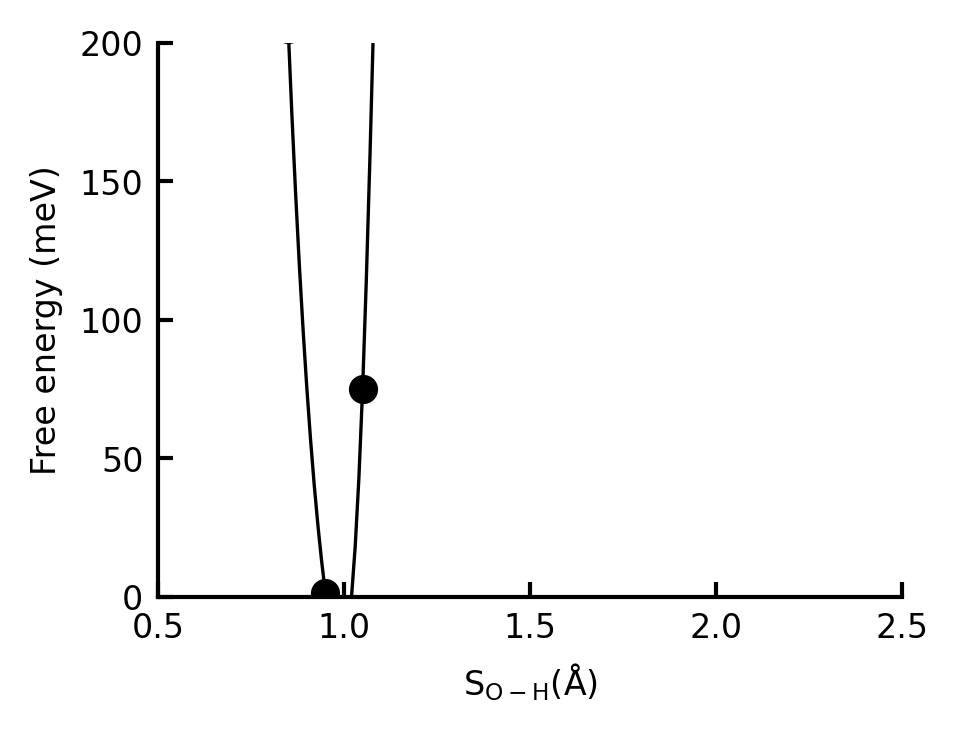} & 
\includegraphics[width=3.2in]{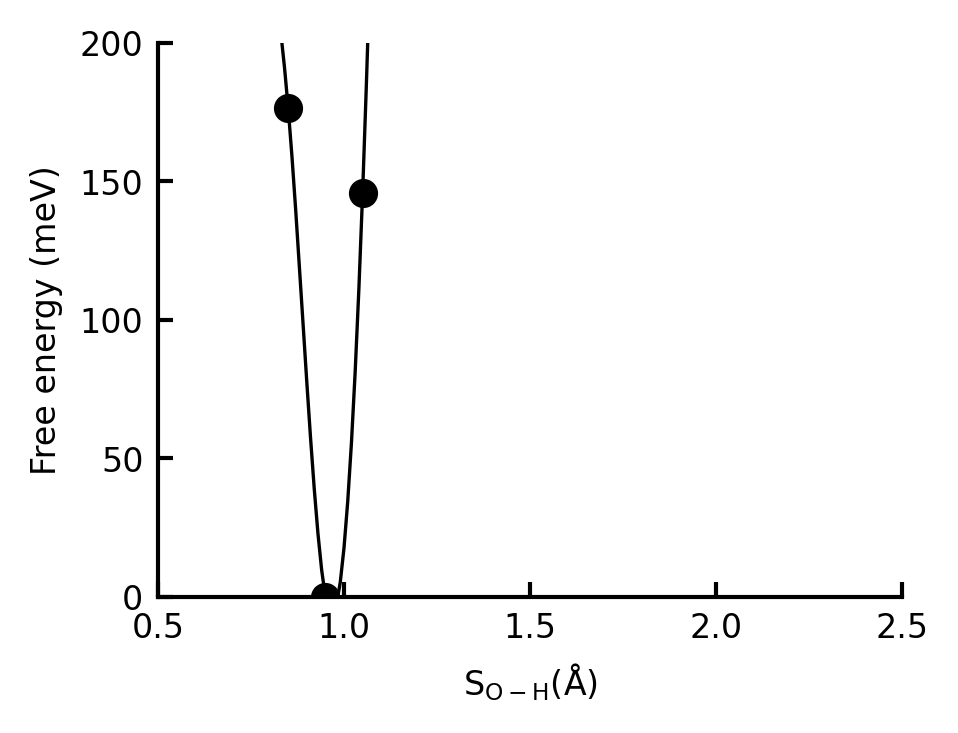} \\
\end{tabular}
\caption{
Free energy for CV = S$_\text{O-H}$ of PT involving all O$_\text{s}$ sites
in Zr$_7$O$_8$N$_4$ (left) and ZrO$_2$ (right).
}
\label{fgr:fe_sohall}
\end{figure}
Figure~\ref{fgr:fe_sohall}
represents free energy for CV = S$_\text{O-H}$ of PT involving all O$_\text{s}$ sites.
Figure~\ref{fgr:fe_suad}
While the S$_\text{O-H}$ in Figure~\ref{fgr:fe_soh} is 
the minimum distance between a "specific" O$_\text{s}$ and all Hs, 
this S$_\text{O-H}$ is the minimum distance between "all" O$_\text{s}$s and all Hs. 
In other words, if even one H is adsorbed among all O$_\text{s}$,
the minimum distance between that adsorbed O$_\text{s}$ and H is the S$_\text{O-H}$.
Even if H moves from another O$_\text{s}$, 
S$_\text{O-H}$ does not represent the distance between that the O and the H.
Therefore the free energy of the surface PT 
as shown in Figure~\ref{fgr:fe_sohall} 
cannot be determined from this CV.

The CVs such as those shown 
in Figures~\ref{fgr:fe_suos} and \ref{fgr:fe_sohall}
are not usually used.
We were interested in 
whether it would be possible to say which CV is most appropriate, 
since the free energy barrier seems to be calculated differently 
depending on how the CV is taken. 
We carried out calculations using such CVs.
This is a different matter from the exact calculation 
of the free energy surface, 
and the idea is to compare different CVs and the barriers.
We believe that it is not obvious which CVs are important, 
and that all of CVs seem to be important.
Therefore, the free energy for such CVs is also included.

\newpage

\begin{figure}
\begin{tabular}{cc}
\includegraphics[width=3.2in]{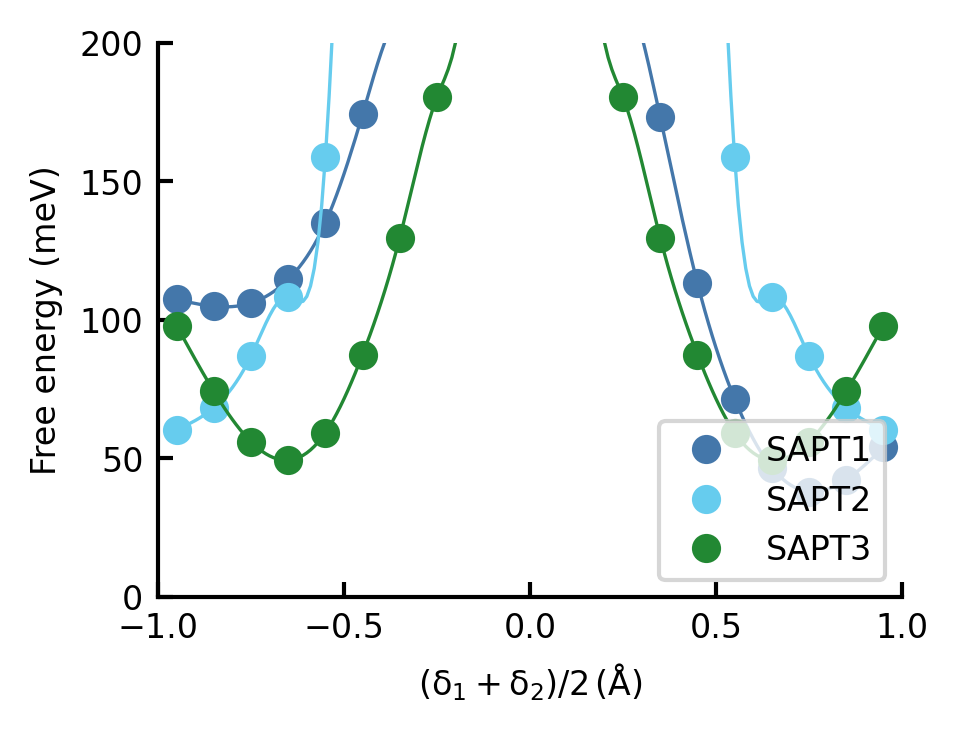} & 
\includegraphics[width=3.2in]{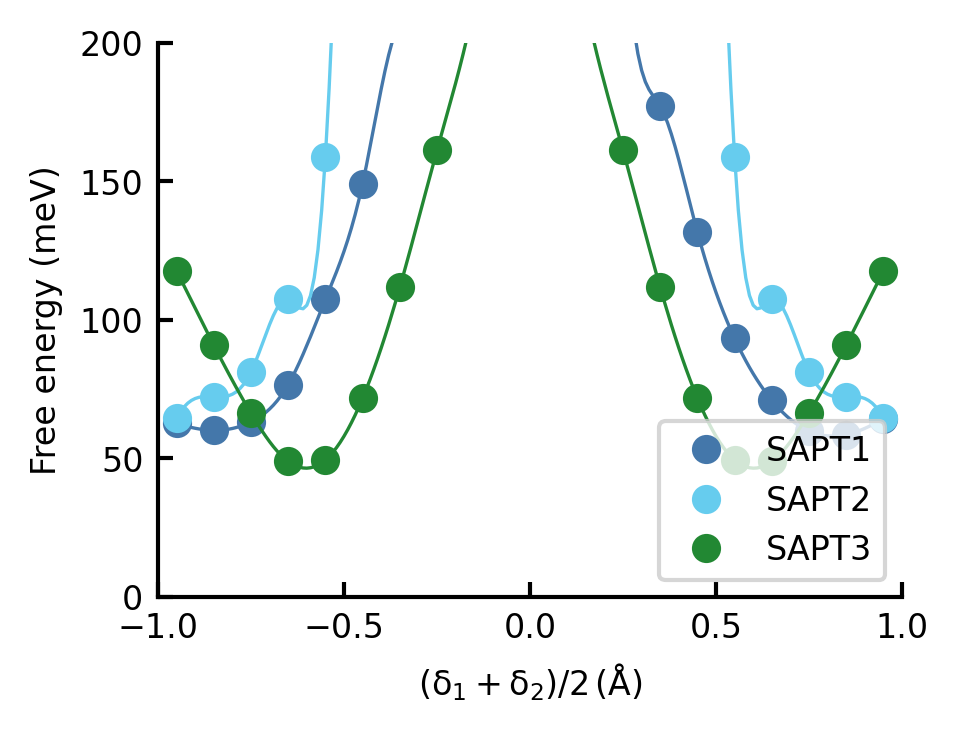} \\
\end{tabular}
\caption{
Free energy for CV = $(\delta_1 + \delta_2)/2$ 
of solvent-assisted PT (SAPT)
in Zr$_7$O$_8$N$_4$ (left) and ZrO$_2$ (right).
The PT reaction formulas are summarised in the Table~\ref{tbl:PT_list}.
}
\label{fgr:fe_sapt}
\end{figure}
Figure~\ref{fgr:fe_sapt} represents the free energy of solvent-assisted PT.
The PT reactions treated in this study are summarised in the Table~\ref{tbl:PT_list}.
All of these PTs have very high free energy barriers, so their probability is quite low.

\newpage

\begin{figure}
\begin{tabular}{cc}
\includegraphics[width=3.2in]{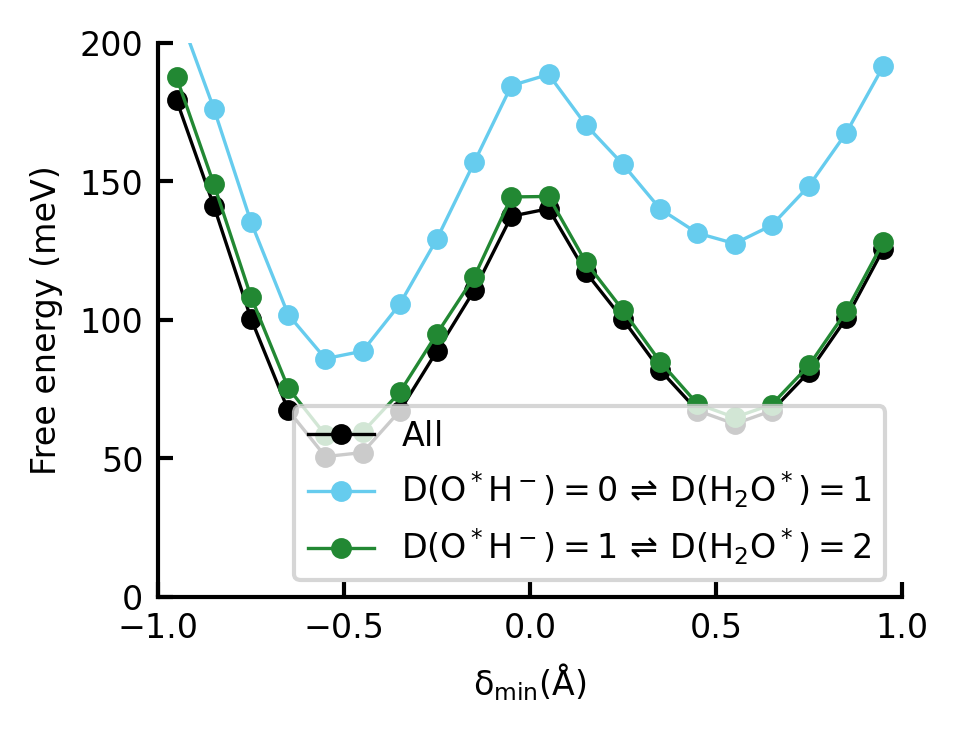} & 
\includegraphics[width=3.2in]{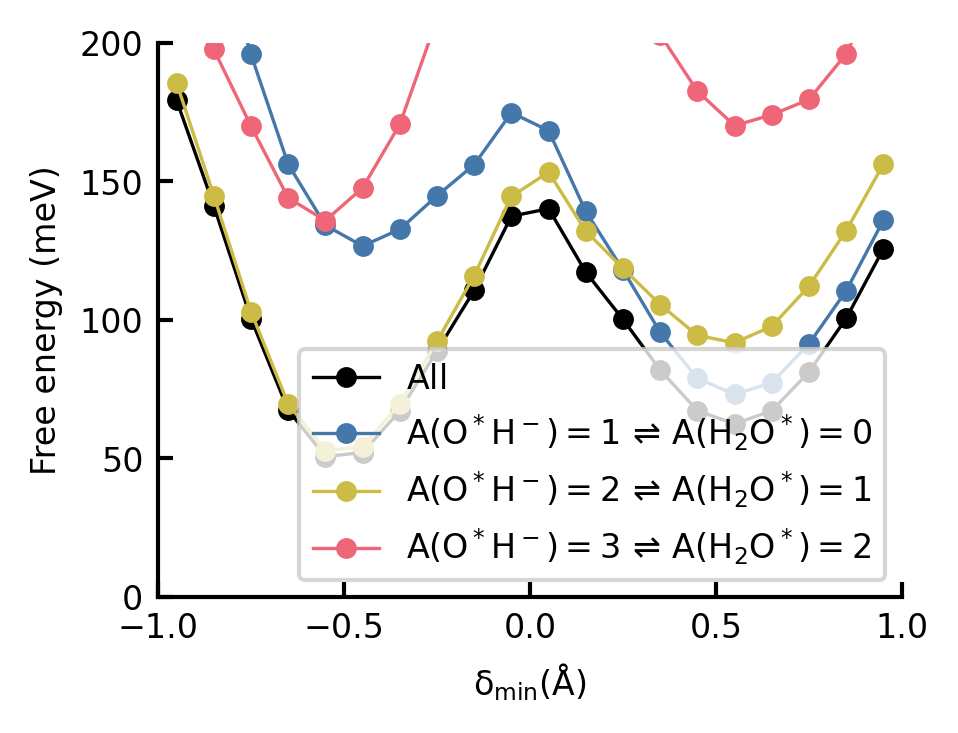} \\
\includegraphics[width=3.2in]{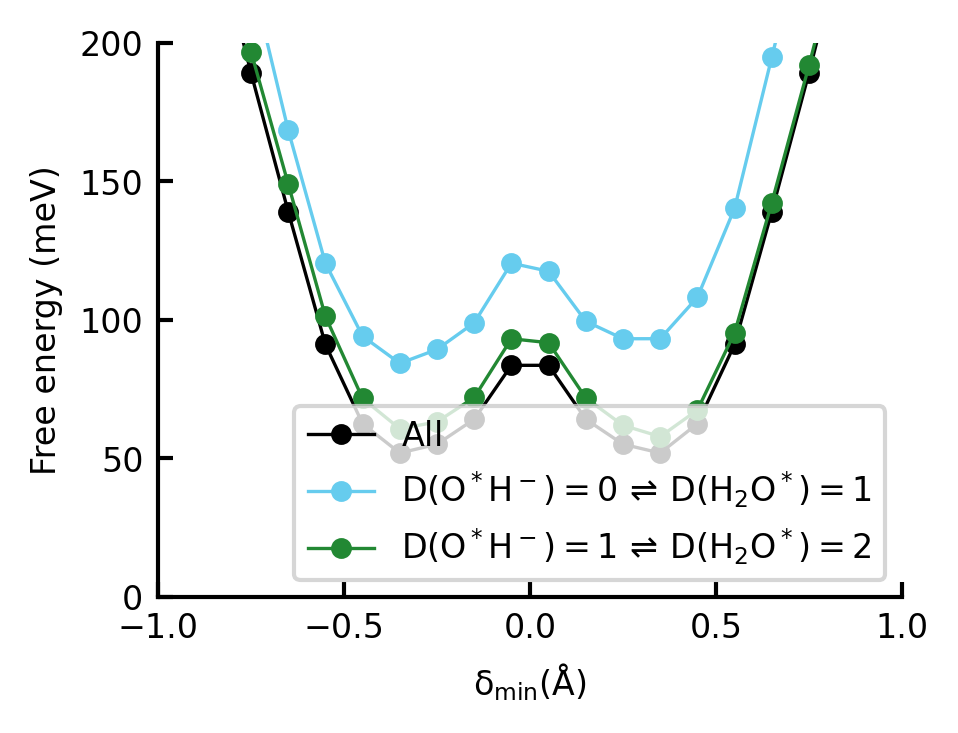} & 
\includegraphics[width=3.2in]{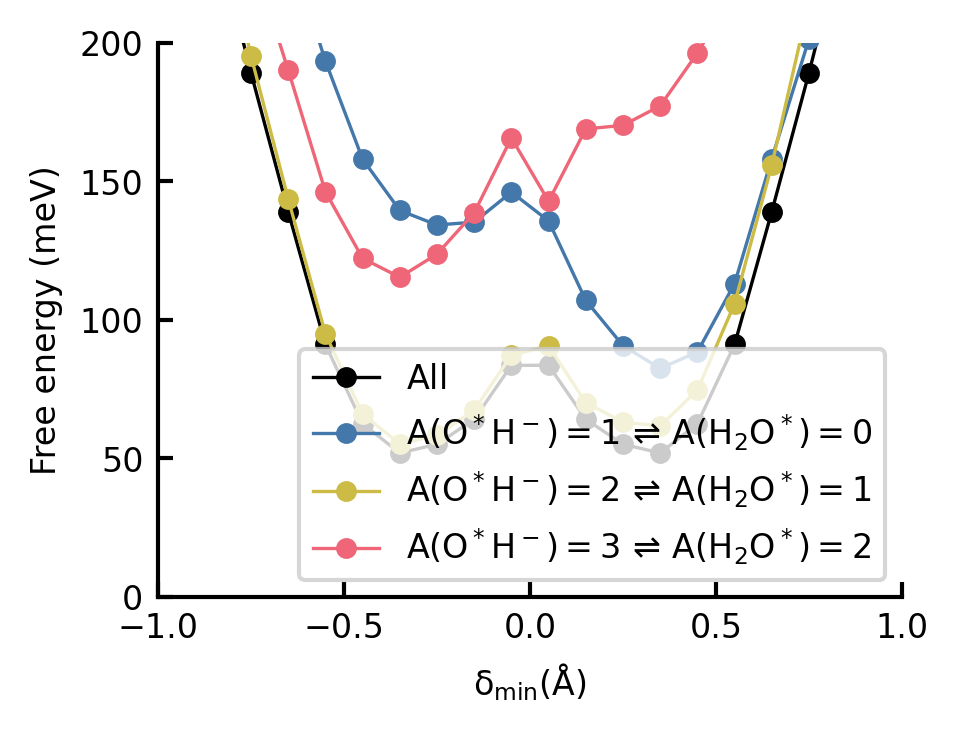} \\
\end{tabular}
\caption{
Free energy for CV = $\delta_\text{min}$ of surface-PT (top) and adlayer-PT (bottom) 
for the number of hydrogen bonds, 
where O$^*$H$^-$ in left-hand side (H$_2$O$^*$ in right-hand side) is
the donor (D) or acceptor (A) in Zr$_7$O$_8$N$_4$.
The PT reaction formulas are summarised in the Table~\ref{tbl:PT_list}.
}
\label{fgr:fe_HBD_ZON}
\end{figure}
Figure~\ref{fgr:fe_HBD_ZON}
shows free energy for CV = $\delta_\text{min}$ of surface-PT and adlayer-PT
for the number of hydrogen bonds, 
where O$^*$H$^-$ in left-hand side (H$_2$O$^*$ in right-hand side) is
the donor or acceptor in Zr$_7$O$_8$N$_4$.
For the surface PT, 
the free energy curves cross at A(H$_2$O$^*$)=0,1, as in previous studies on ZnO. 
\cite{Quaranta2017}
In other words, 
it can be said that there is a predominant pre-solvation mechanism. 
This means that the most stable value is A=0, 
the number at which H$_2$O$^*$ accepts hydrogen bonds, 
but if it increases to A=1 due to thermal fluctuations, 
the energy barrier decreases and proton transfer becomes more likely.
On the other hand, in the adlayer-PT, 
in contrast to the previous studies on ZnO, there is no such crossing.

\begin{figure}
\begin{tabular}{cc}
\includegraphics[width=3.2in]{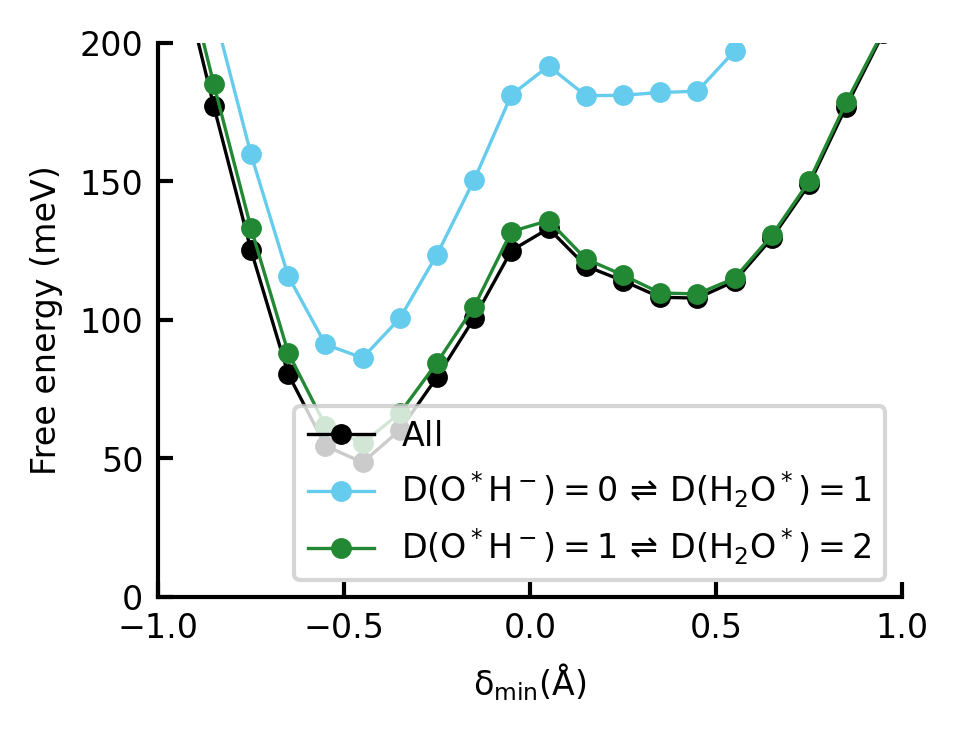} & 
\includegraphics[width=3.2in]{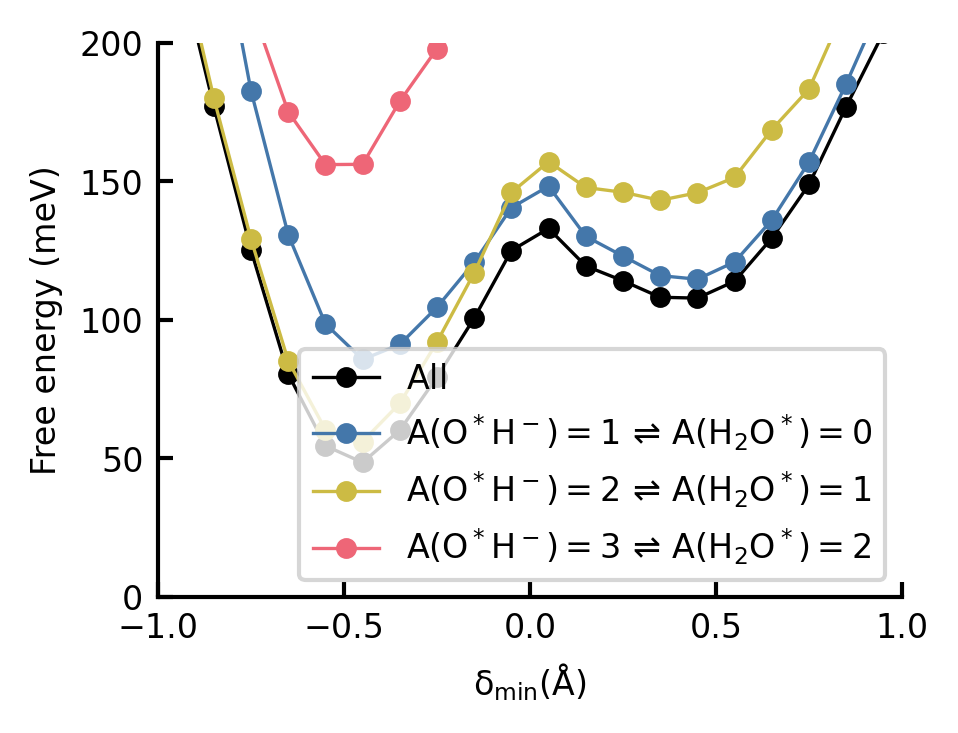} \\
\includegraphics[width=3.2in]{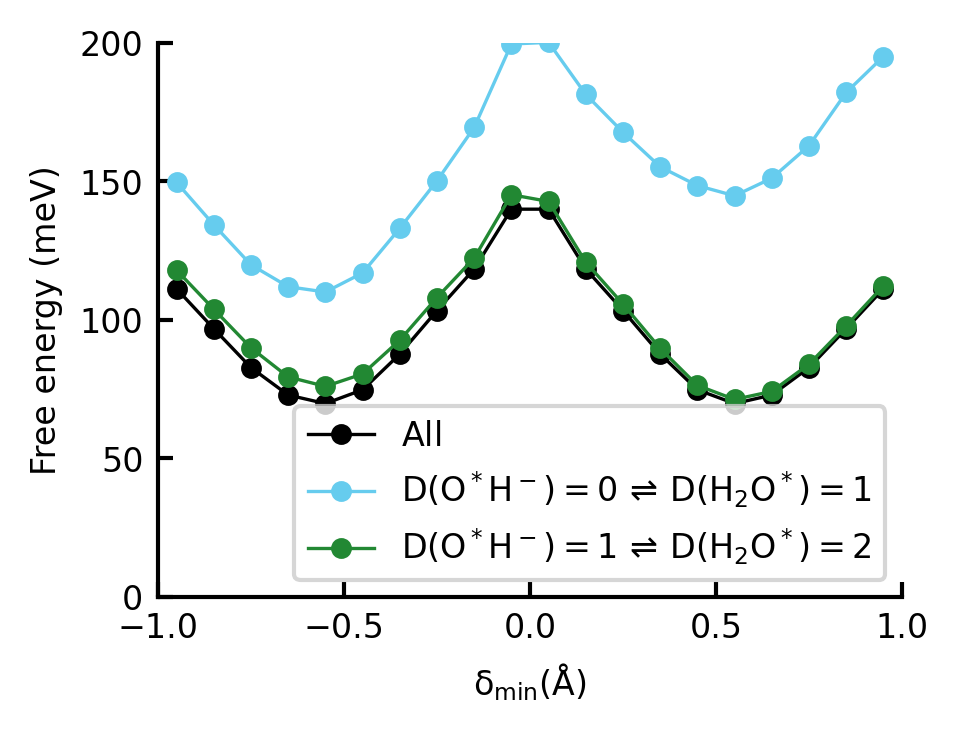} & 
\includegraphics[width=3.2in]{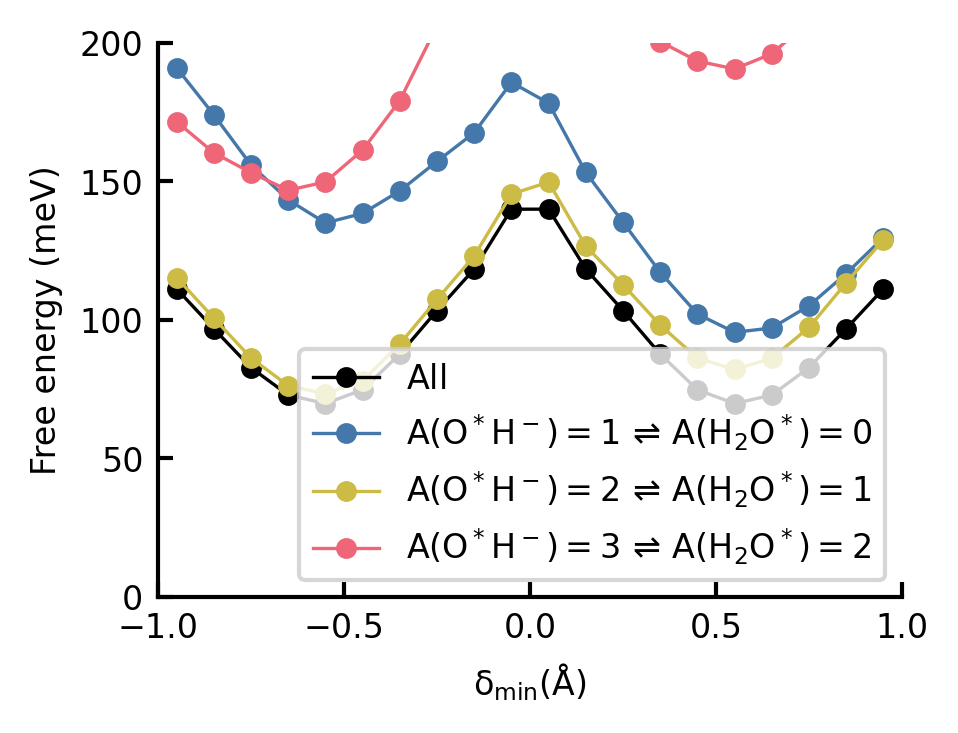} \\
\end{tabular}
\caption{
Free energy for CV = $\delta_\text{min}$ of surface-PT (top) and adlayer-PT (bottom) 
for the number of hydrogen bonds, 
where O$^*$H$^-$ in left-hand side (H$_2$O$^*$ in right-hand side) is
the donor (D) or acceptor (A) in ZrO$_2$.
The PT reaction formulas are summarised in the Table~\ref{tbl:PT_list}.
}
\label{fgr:fe_HBD_ZO}
\end{figure}
Figure~\ref{fgr:fe_HBD_ZO}
has been drawn in the same way as the Figure~\ref{fgr:fe_HBD_ZON} for ZrO$_2$.
Similar to Zr$_7$O$_8$N$_4$, there is an crossing at the surface-PT 
and no crossing at the adlayer-PT.
However, the difference is that 
the crossing is on the left-hand side instead of right-hand side.
It is unclear 
why such changes occur depending on the presence or absence of defects.

\begin{table}
\caption{
Percentage (top) and average (bottom) of the number of hydrogen bonds, 
where O$^*$H$^-$ in left-hand side (H$_2$O$^*$ in right-hand side) in PT formula is
the donor (D) or acceptor (A)
in Zr$_7$O$_8$N$_4$ and ZrO$_2$.
See also Figures ~\ref{fgr:fe_HBD_ZON} and ~\ref{fgr:fe_HBD_ZO}.
The PT reaction formulas are summarised in the Table~\ref{tbl:PT_list}.
}
  \label{tbl:fe_nhb_percentage}
  \begin{tabular}{rrrrr}
    \hline
    percentage& \multicolumn{2}{r}{Zr$_7$O$_8$N$_4$} & \multicolumn{2}{r}{ZrO$_2$} \\
              & surface & adlayer & surface & adlayer \\
    \hline
    D(O$^*$H$^-$)=0 &      28 &      23 &      27 &      10 \\
            1 &      72 &      77 &      73 &      90 \\
    A(O$^*$H$^-$)=1 &      16 &      38 &      33 &      44 \\
            2 &      81 &      61 &      66 &      55 \\
            3 &       3 &       1 &       1 &       1 \\
    D(H$_2$O$^*$)=1 &      12 &      23 &       9 &      10 \\
            2 &      88 &      77 &      91 &      90 \\
    A(H$_2$O$^*$)=0 &      66 &      38 &      75 &      44 \\
            1 &      33 &      61 &      24 &      55 \\
            2 &       1 &       1 &       1 &       1 \\
    \hline
    average   & \multicolumn{2}{r}{Zr$_7$O$_8$N$_4$} & \multicolumn{2}{r}{ZrO$_2$} \\
              & surface & adlayer & surface & adlayer \\
    \hline
    D(O$^*$H$^-$)   &    0.72 &    0.77 &    0.73 &    0.90  \\
    A(O$^*$H$^-$)   &    1.87 &    1.63 &    1.68 &    1.57  \\
    D(H$_2$O$^*$)   &    1.88 &    1.77 &    1.91 &    1.90  \\
    A(H$_2$O$^*$)   &    0.35 &    0.63 &    0.24 &    0.57  \\
    \hline
  \end{tabular}
\end{table}
Table~\ref{tbl:fe_nhb_percentage}
shows percentage and average of the number of hydrogen bonds, 
where O$^*$H$^-$ in left-hand side (H$_2$O$^*$ in right-hand side) in PT formula is
the donor (D) or acceptor (A).
In the case of ZnO in previous research, 
\cite{Quaranta2017}
the more (less) O$^*$H$^-$(H$_2$O$^*$) gives and accepts fewer (more) hydrogen bonds, 
the smaller the free energy barrier becomes.
Furthermore, the average size of the hydrogen bonds given (accepted) explains 
why the barrier is smaller in the adlayer-PT than in the surface-PT.
In the case of Zr$_7$O$_8$N$_4$ and ZrO$_2$ in this study, 
the average size relationship of the number of hydrogen bonds is 
the same as that of ZnO, but the shape of the free energy is different, 
so a similar argument does not hold.

\begin{figure}
\begin{tabular}{cc}
\multicolumn{2}{r}{
\includegraphics[width=6.4in]{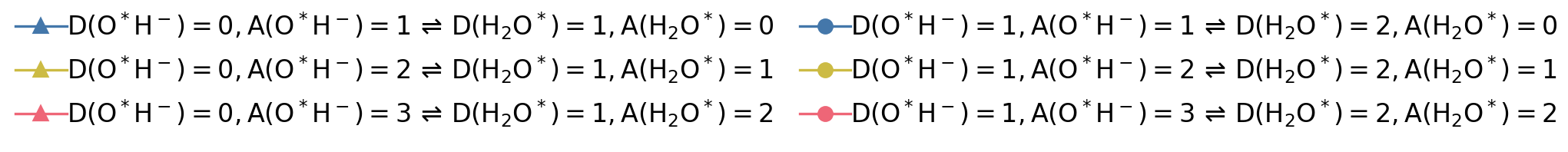}} \\
\includegraphics[width=3.2in]{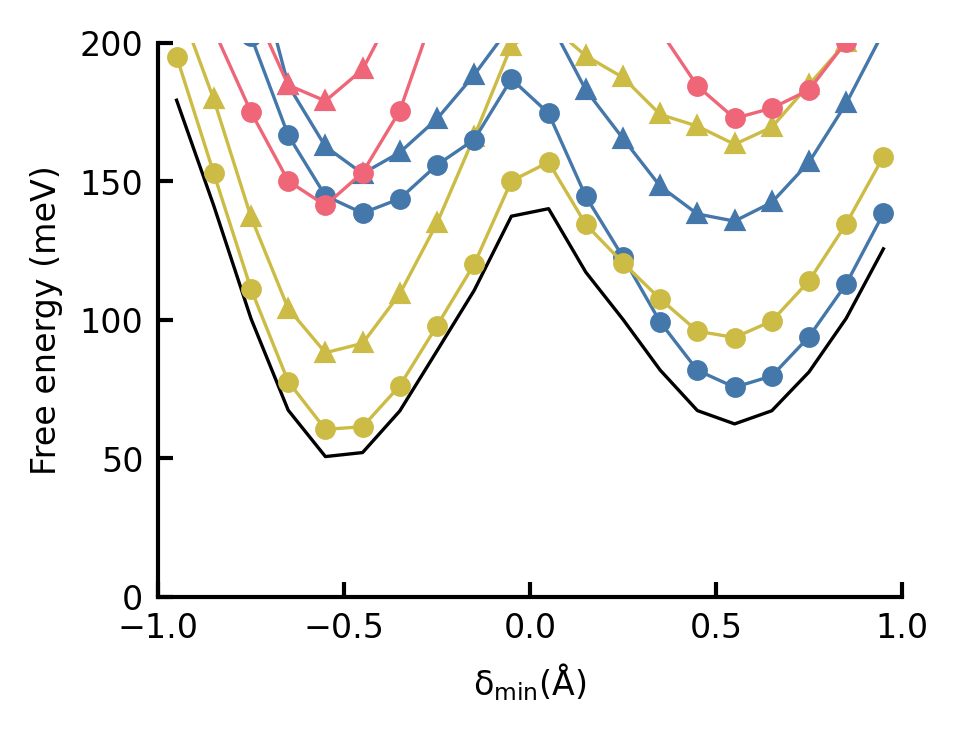} & 
\includegraphics[width=3.2in]{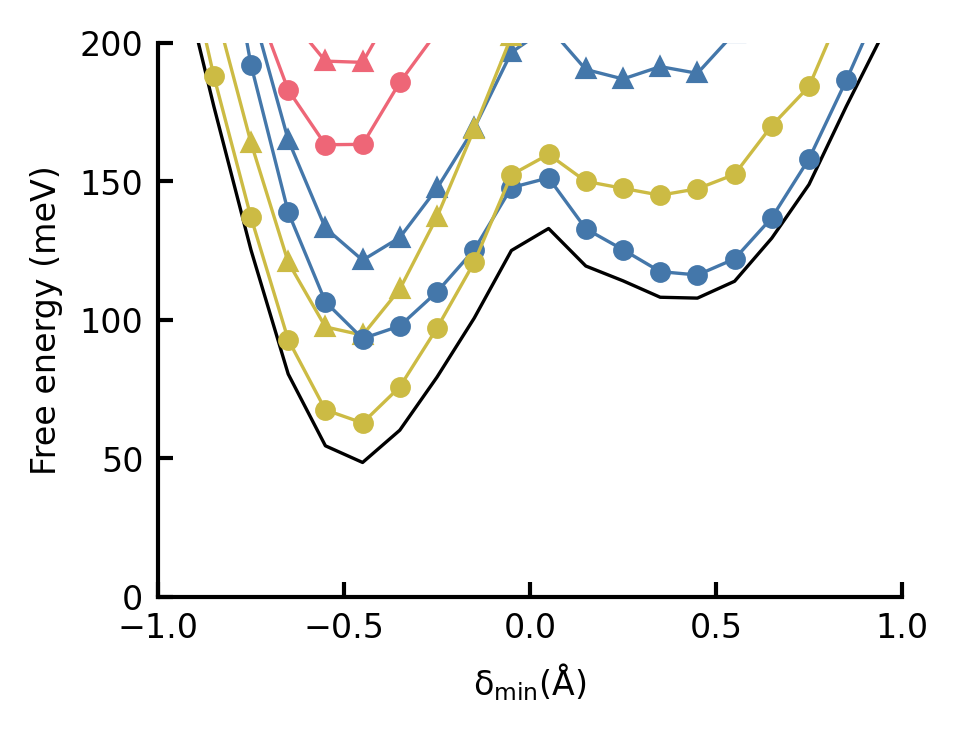} \\
\includegraphics[width=3.2in]{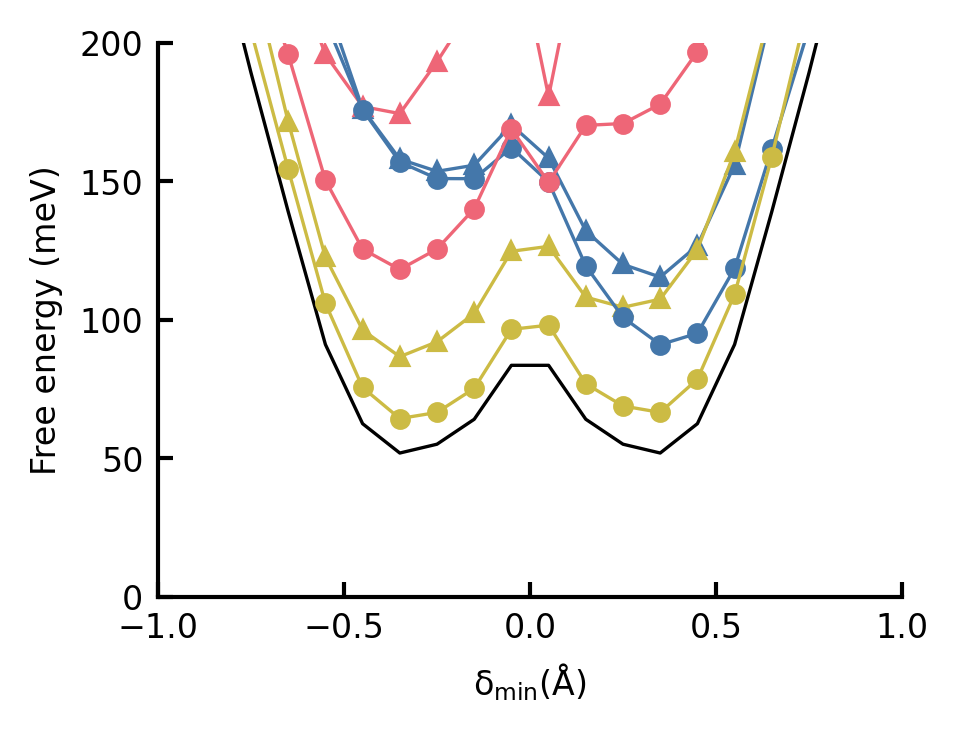} &
\includegraphics[width=3.2in]{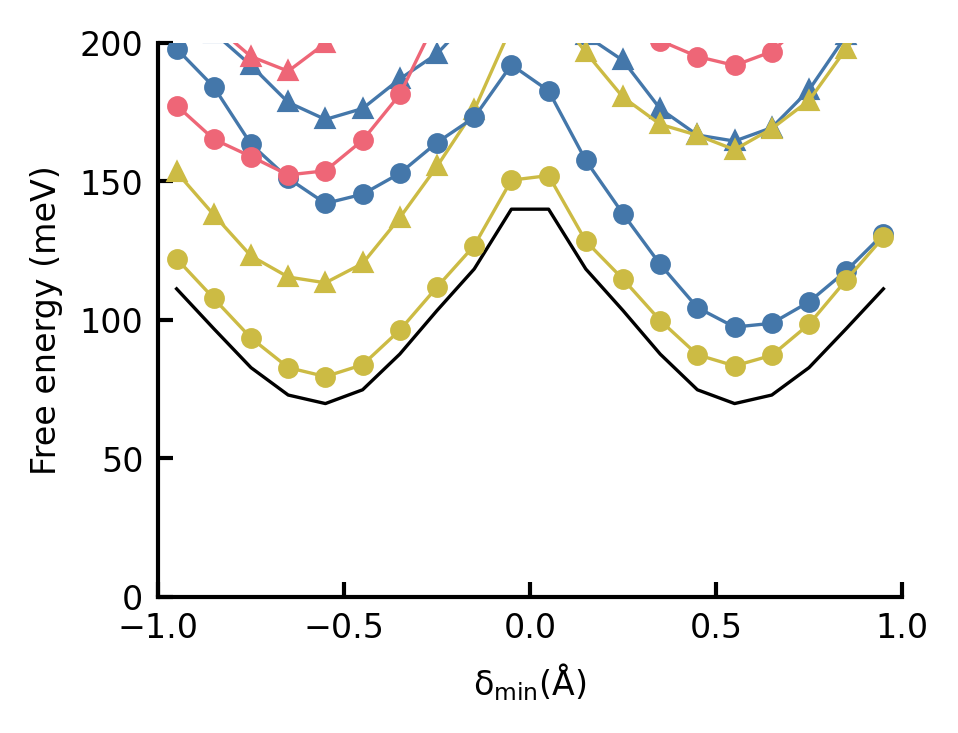} \\
\end{tabular}
\caption{
Free energy of surface-PT (top) and adlayer-PT (bottom)
for the number of hydrogen bonds, 
where O$^*$H$^-$ in left-hand side (H$_2$O$^*$ in right-hand side) is
the donor (D) and acceptor (A) in Zr$_7$O$_8$N$_4$ (left), ZrO$_2$ (right).
In Figure~\ref{fgr:fe_HBD_ZON} and Figure~\ref{fgr:fe_HBD_ZO},
the classification is based on D or A only.
Here, the classification is based on the combination of D and A.
The PT reaction formulas are summarised in the Table~\ref{tbl:PT_list}.
}
\label{fgr:fe_HBDA}
\end{figure}
Figure~\ref{fgr:fe_HBDA} 
shows free energy of surface-PT and adlayer-PT
for the number of hydrogen bonds, 
where O$^*$H$^-$ in left-hand side (H$_2$O$^*$ in right-hand side) is
the donor and acceptor.
In Figure~\ref{fgr:fe_HBD_ZON} and Figure~\ref{fgr:fe_HBD_ZO},
the classification is based on D or A only.
In this Figure, the classification is based on the combination of D and A.
The PT reaction formulas are summarised in the Table~\ref{tbl:PT_list}.
From Figure~\ref{fgr:fe_HBD_ZON} and Figure~\ref{fgr:fe_HBD_ZO},
it was said that they cross at A(H$_2$O$^*$)=0 and 1.
From this figure we can see in more detail that
D(H$_2$O$^*$),A(H$_2$O$^*$) = 2,0 and 2,1 cross.

\newpage

\begin{figure}
\begin{tabular}{cc}
\includegraphics[width=3.2in]{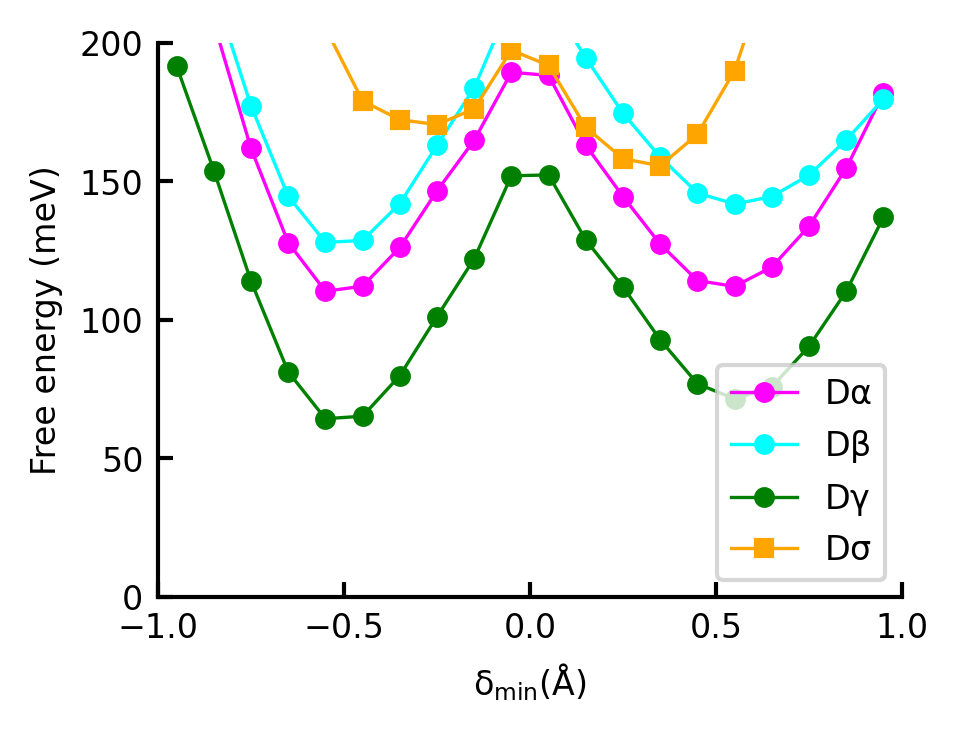} & 
\includegraphics[width=3.2in]{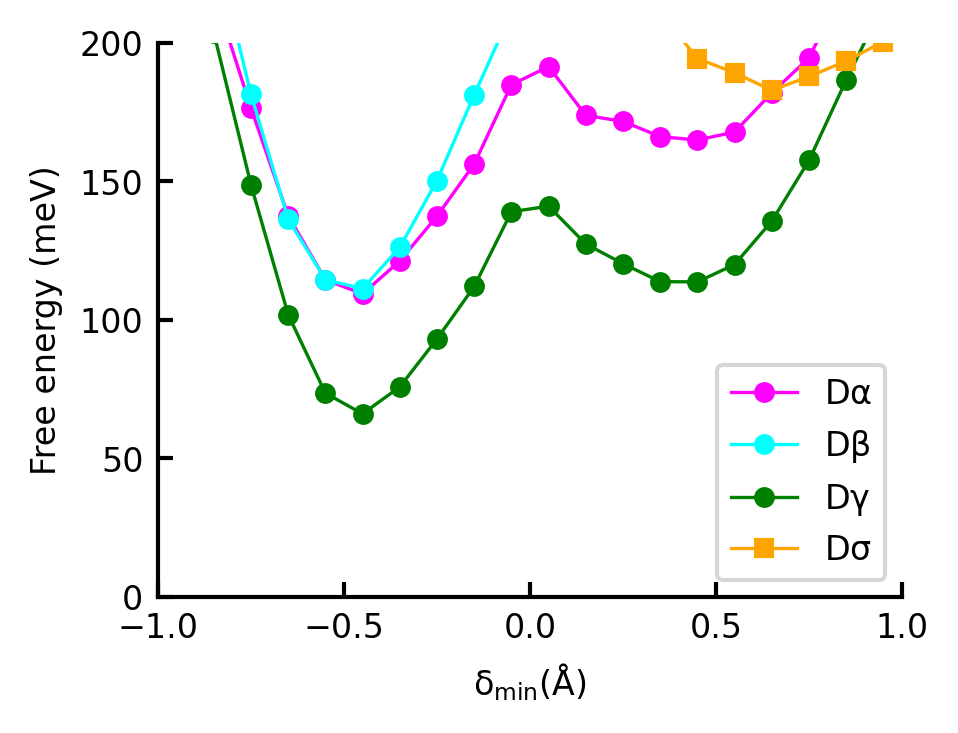} \\
\includegraphics[width=3.2in]{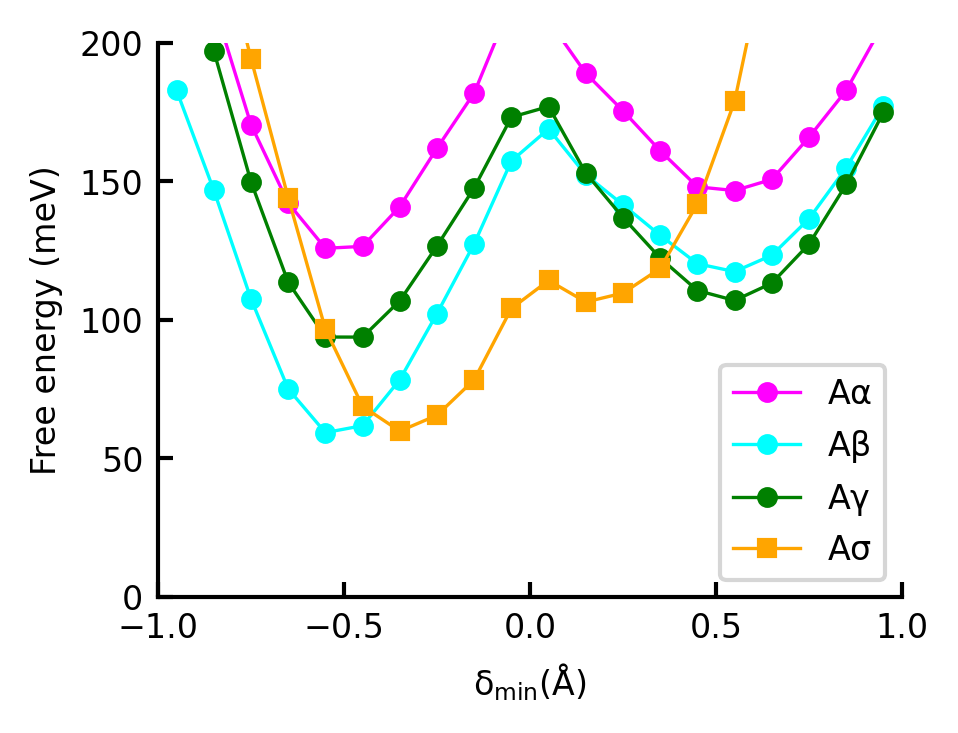} &
\includegraphics[width=3.2in]{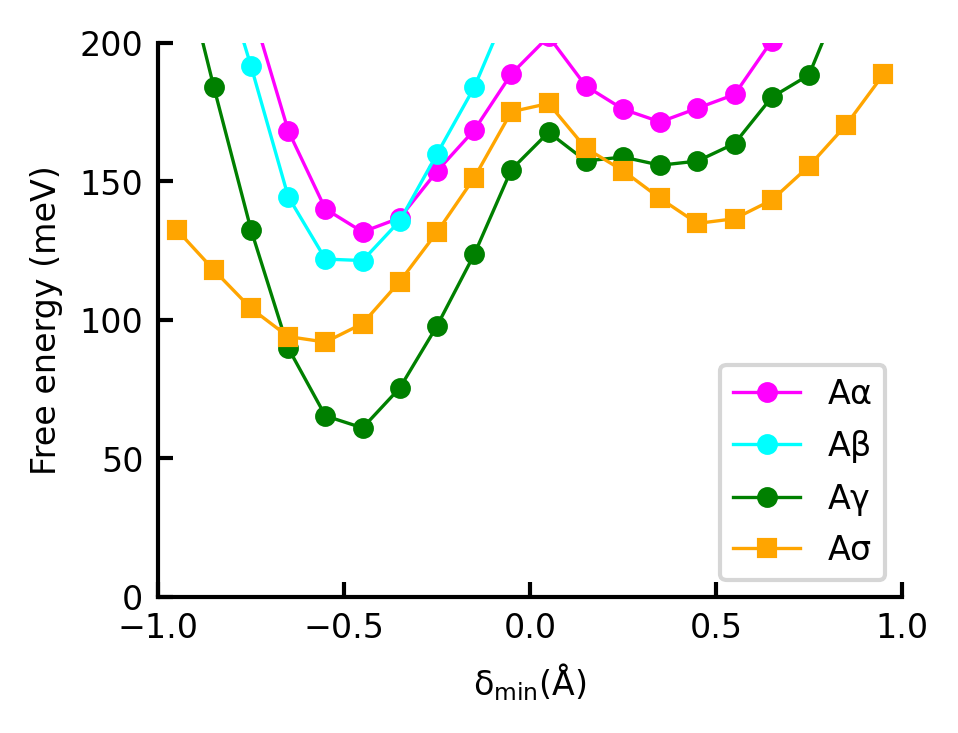} \\
\end{tabular}
\caption{
Free energy for CV = $\delta_\text{min}$ of surface-PT or adlayer-PT 
for the partner of hydrogen bonds, 
where O$^*$H$^-$ in left-hand side (H$_2$O$^*$ in right-hand side) is
the donor (top) or acceptor (bottom) in Zr$_7$O$_8$N$_4$ (left), ZrO$_2$ (right).
D$\alpha$,D$\beta$,D$\gamma$ mean
O$^*$H$^-$ in surface-PT is a hydrogen bond donor with O$^*$H$^-$, H$_2$O$^*$, H$_2$O$^\sim$, respectively.
D$\sigma$ means
O$^*$H$^-$ in adlayer-PT is a hydrogen bond donor with O$_\text{s}^{2-}$.
A$\alpha,\beta,\gamma,\sigma$ are similar symbols for acceptors.
Free energy properties (minimum, maximum and barriers) 
are shown in Table~\ref{tbl:fe_DAX}.
The PT reaction formulas are summarised in the Table~\ref{tbl:PT_list}.
}
\label{fgr:fe_DAX}
\end{figure}
Figure~\ref{fgr:fe_DAX}
represents free energy for CV = $\delta_\text{min}$ of surface-PT or adlayer-PT 
for the partner of hydrogen bonds, 
where O$^*$H$^-$ in left-hand side (H$_2$O$^*$ in right-hand side) is the donor or acceptor.
D$\alpha,\beta,\gamma$ mean
O$^*$H$^-$ in surface-PT is a hydrogen bond donor with O$^*$H$^-$, H$_2$O$^*$, H$_2$O$^\sim$, respectively.
D$\sigma$ means
O$^*$H$^-$ in adlayer-PT is a hydrogen bond donor with O$_\text{s}^{2-}$.
A$\alpha$,A$\beta$,A$\gamma$,A$\sigma$ are similar symbols for acceptors.
Free energy properties (minimum, maximum and barriers) 
are shown in Table~\ref{tbl:fe_DAX}.
The PT reaction formulas are summarised in the Table~\ref{tbl:PT_list}.
In previous research on ZnO, 
\cite{Quaranta2017}
the minimum values of the free energy 
in the right-hand side of D$\alpha$ and D$\sigma$ and 
in the left-hand side of A$\beta$ and A$\sigma$ are close to each other.
This suggests that these states are competitive.
In this study, the minimum values of A$\beta$ and A$\sigma$ are close, 
but the minimum values of D$\alpha$ and D$\sigma$ are far apart.

\begin{table}
\caption{
Free energy properties (meV) for CV = $\delta_\text{min}$ of surface-PT and adlayer-PT 
for the partner of hydrogen bonds, 
where O$^*$H$^-$ in left-hand side (H$_2$O$^*$ in right-hand side) is
the donor (top) or acceptor (bottom) 
in Zr$_7$O$_8$N$_4$ and ZrO$_2$ (See also Figure~\ref{fgr:fe_DAX}).
LHS, RHS: left- and right-hand sides minimum.
MAX: maximum.
FW, BW: forward ($\rightarrow$) and backward ($\leftarrow$) PT barriers.
The PT reaction formulas are summarised in the Table~\ref{tbl:PT_list}.
}
  \label{tbl:fe_DAX}
  \begin{tabular}{rrrrrr}
    \hline
    Zr$_7$O$_8$N$_4$              & LHS &  FW & MAX &  BW & RHS \\
    \hline
    surface-PT D$\alpha$ & 110 &  79 & 189 &  77 & 112 \\
    surface-PT D$\beta $ & 128 &  88 & 216 &  74 & 142 \\
    surface-PT D$\gamma$ &  64 &  88 & 152 &  81 &  71 \\
    adlayer-PT D$\sigma$ & 170 &  27 & 197 &  41 & 156 \\
    surface-PT A$\alpha$ & 126 &  88 & 214 &  67 & 147 \\
    surface-PT A$\beta $ &  59 & 110 & 169 &  52 & 117 \\
    surface-PT A$\gamma$ &  94 &  87 & 177 &  70 & 107 \\
    adlayer-PT A$\sigma$ &  60 &  54 & 114 &   8 & 106 \\
    \hline
    ZrO$_2$                 & LHS &  FW & MAX &  BW & RHS \\
    \hline
    surface-PT D$\alpha$ & 109 &  82 & 191 &  26 & 165 \\
    surface-PT D$\beta $ & 111 & 118 & 229 &  21 & 208 \\
    surface-PT D$\gamma$ &  66 &  75 & 141 &  27 & 114 \\
    adlayer-PT D$\sigma$ & 205 &  93 & 298 & 115 & 183 \\
    surface-PT A$\alpha$ & 132 &  70 & 202 &  31 & 171 \\
    surface-PT A$\beta $ & 121 & 112 & 233 &  15 & 218 \\
    surface-PT A$\gamma$ &  61 & 107 & 168 &  12 & 156 \\
    adlayer-PT A$\sigma$ &  92 &  86 & 178 &  43 & 135 \\
    \hline
  \end{tabular}
\end{table}

In surface PT, the least common hydrogen bond partner (high free energy) 
is often H$_2$O$^*$ ($\beta$).
However, as an exception, 
A$\alpha$ has a higher free energy than A$\beta$ in Zr$_7$O$_8$N$_4$.
Following previous research on ZnO, 
\cite{Quaranta2017}
we compare the maximum free energy values, 
assuming that the relative velocity between two PT mechanisms i and j 
(including both forward and backward reactions) is 
$ r_i/r_j = \exp(-MAX_i/k_BT) / \exp(-MAX_j/k_BT)$.
In Zr$_7$O$_8$N$_4$, the relative velocities are close 
in D$\alpha$ and D$\sigma$, A$\beta$ and A$\gamma$, 
and in ZrO$_2$, A$\gamma$ and A$\sigma$.
The adlayer PT (A$\sigma$) has a lower barrier 
in competitive PT (A$\beta$ and A$\sigma$), 
but the adlayer PT (D$\sigma$) has a higher barrier 
in non-competitive PT (D$\alpha$ and A$\sigma$).
From the perspective of H$_2$O$^*$, which provides protons, 
D$\alpha$ is lower in left-hand side, and D$\sigma$ is lower in right-hand side.
In ZrO$_2$, the free energies of D$\alpha$ and D$\beta$ are 
almost the same at left-hand side, 
but $\alpha$ is significantly more stable than $\beta$ at right-hand side, 
indicating that it is more preferred as an acceptor for hydrogen bond.

\newpage

\begin{figure}
\begin{tabular}{cc}
\includegraphics[width=3.2in]{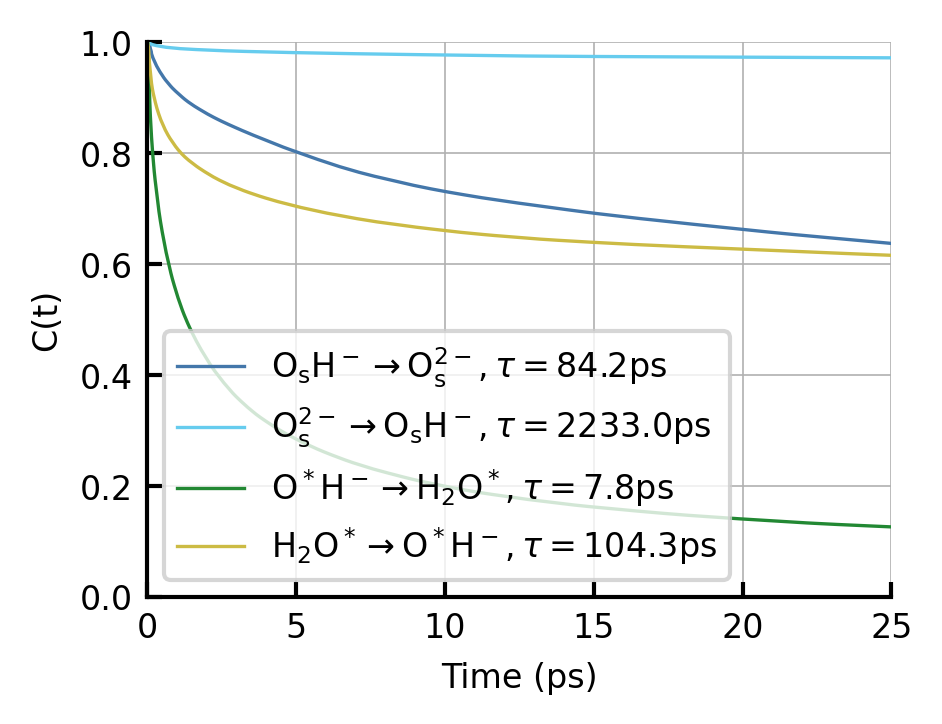} &
\includegraphics[width=3.2in]{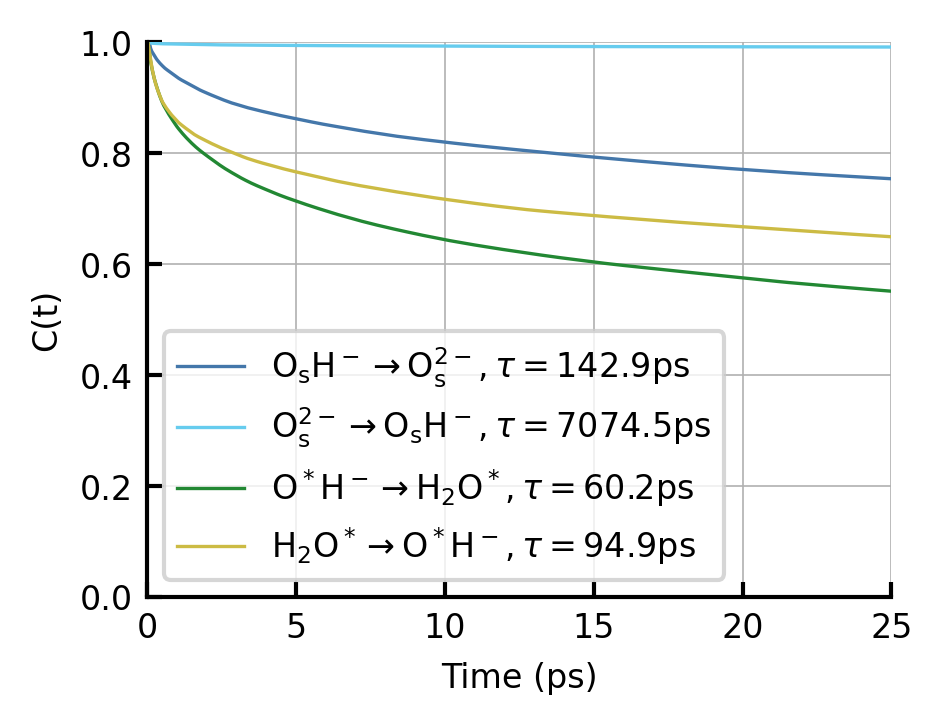} \\
\end{tabular}
\caption{
Time dependence of correlation function $C(t)$ of 
surface-PT and adlayer-PT in Zr$_7$O$_8$N$_4$ (left), ZrO$_2$ (right).
The PT reaction formulas are summarised in the Table~\ref{tbl:PT_list}.
}
\label{fgr:cf_pt}
\end{figure}
Figure~\ref{fgr:cf_pt}
shows time dependence of correlation function of surface-PT and adlayer-PT.
Zr$_7$O$_8$N$_4$ has a lower lifetime than ZrO$_2$ for most PTs, 
which means faster proton transfer.
The lifetime ratios,
$\tau$(O$_\text{s}^{2-}$ $\rightarrow$ O$_\text{s}$H$^-$)/$\tau$(O$_\text{s}$H$^-$ $\rightarrow$ O$_\text{s}^{2-}$),
are 26.5 and 49.5 for Zr$_7$O$_8$N$_4$ and ZrO$_2$ respectively.
If this ratio is considered as the ease of dissociation, 
ZrO$_2$ is easier to dissociate, 
which is consistent with the fact that defects reduce dissociative adsorption
(Table~\ref{tbl:nads}).

\begin{figure}
\begin{tabular}{cc}
\includegraphics[width=3.2in]{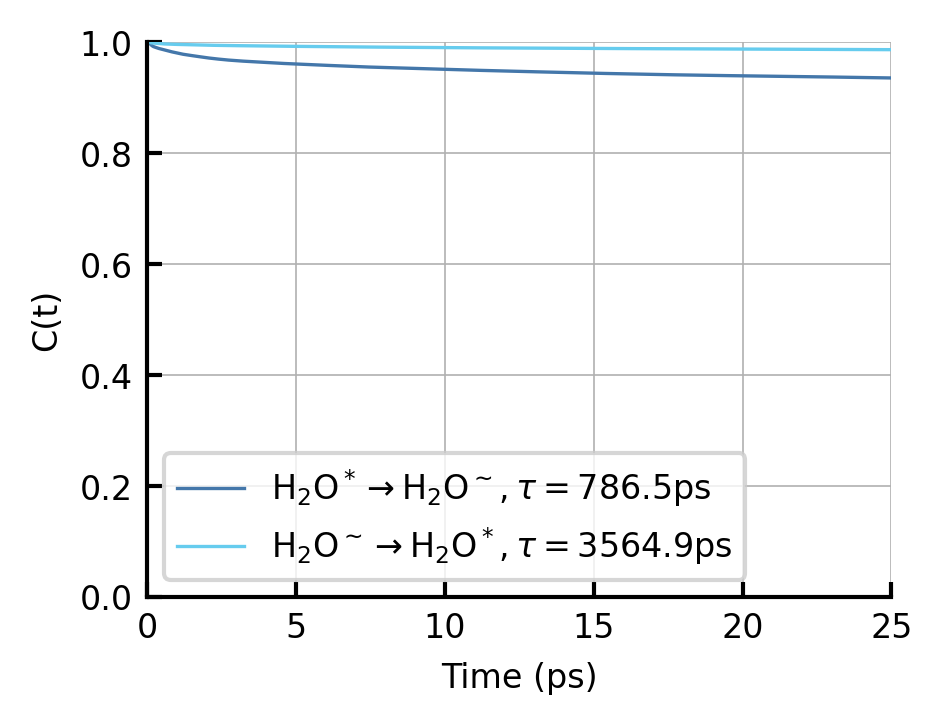} &
\includegraphics[width=3.2in]{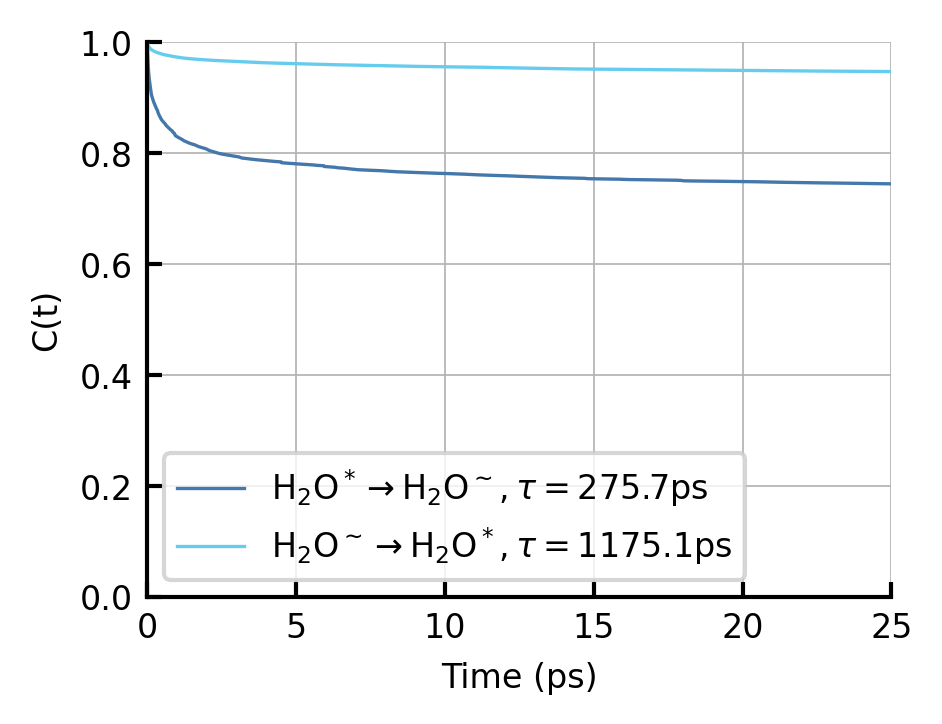} \\
\end{tabular}
\caption{
Time dependence of correlation function $C(t)$ of 
water adsorption/desorption in Zr$_7$O$_8$N$_4$ (left), ZrO$_2$ (right).
}
\label{fgr:cf_h2o}
\end{figure}
Figure~\ref{fgr:cf_h2o}
shows time dependence of correlation function of water adsorption/desorption.
ZrO$_2$ has a shorter lifetime and water is easily released.
In previous research, 
\cite{Quaranta2019}
based on their calculation results, 
ZnO($11\bar{2}0$) is easier to desorb adsorbed water 
than ZnO($10\bar{1}0$) (shorter lifetime), 
so the H$_2$O$^*$ $\rightarrow$ O$^*$H$^-$ reaction is 
less likely to occur (longer lifetime).
In the case of this study, 
analogy does not hold
because ZrO$_2$ has a shorter lifetime than Zr$_7$O$_8$N$_4$ for both reaction.

\newpage

\begin{figure}
\begin{tabular}{cc}
\includegraphics[width=3.2in]{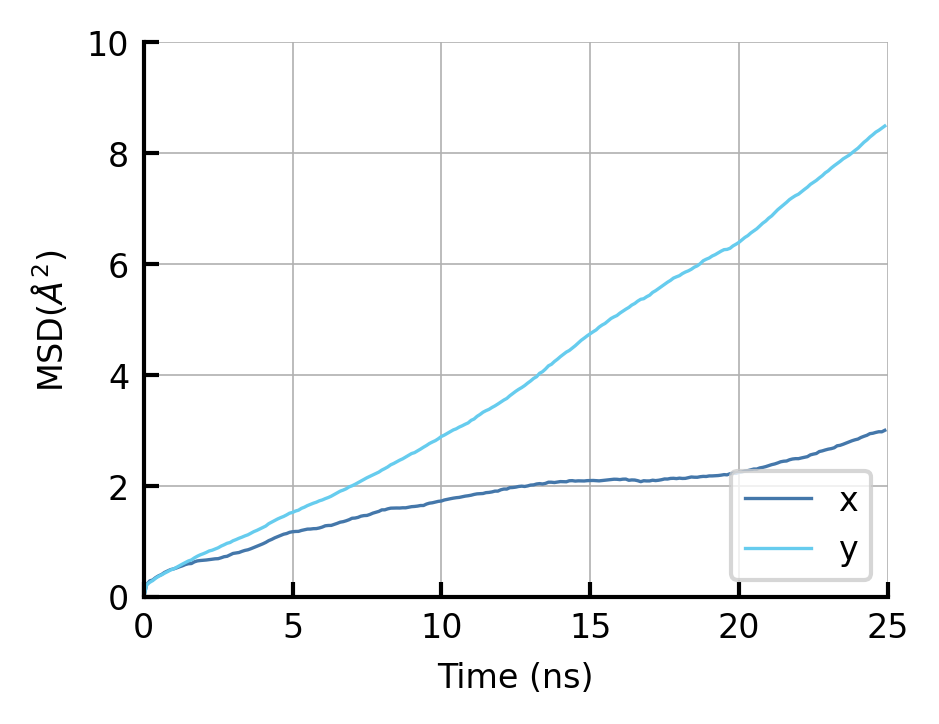}
\includegraphics[width=3.2in]{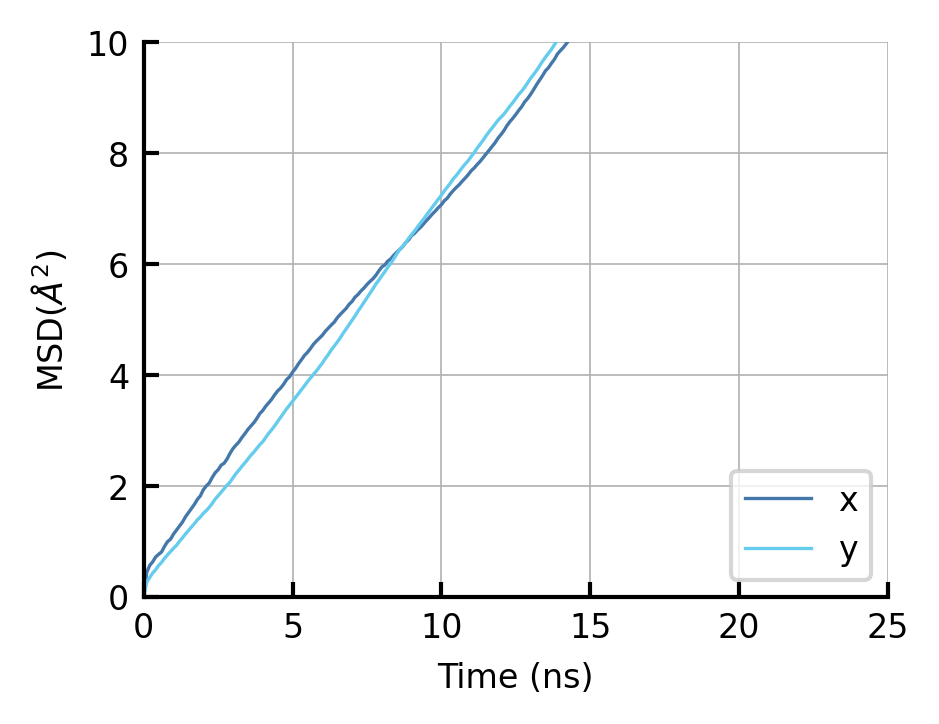}
\end{tabular}
\caption{
Mean squared displacement (MSD) of proton hole centers (PHCs)
in Zr$_7$O$_8$N$_4$ (left), ZrO$_2$ (right).
}
\label{fgr:phc_msd}
\end{figure}
Figure~\ref{fgr:phc_msd}
shows mean squared displacement (MSD) of proton hole centers (PHCs)
The diffusion coefficients $(10^{-12}\,m^2/s)$  in the x and y directions
calculated from this MSD are respectively
Zr$_7$O$_8$N$_4$ is (0.5, 1.7) and ZrO$_2$ is (3.0, 3.7).
While ZrO$_2$ diffuses isotropically, 
Zr$_7$O$_8$N$_4$ diffuses anisotropically and is small.
Because Zr$_7$O$_8$N$_4$ has defects, 
the adsorbate tends to concentrate (Figure~\ref{fgr:density_3d}).
This is probably because adsorbates tend to be densely packed in Zr$_7$O$_8$N$_4$ and 
even if an adlayer-PT often occurs between them, 
protons will circulate in the same place and 
will not contribute to effective diffusion.

\begin{figure}
\begin{tabular}{cc}
\includegraphics[width=3.2in]{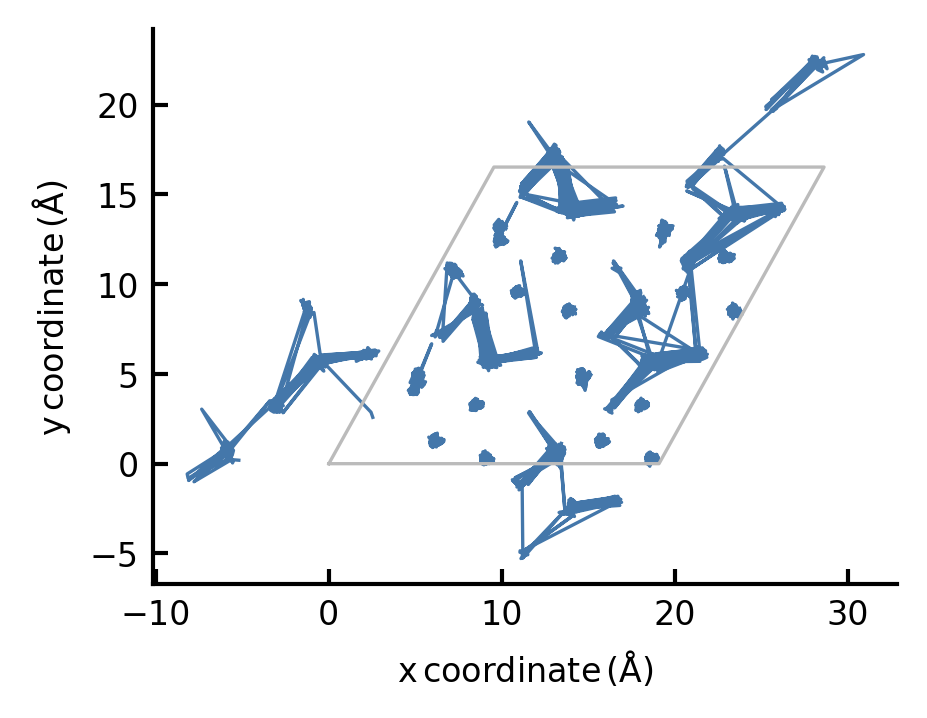}
\includegraphics[width=3.2in]{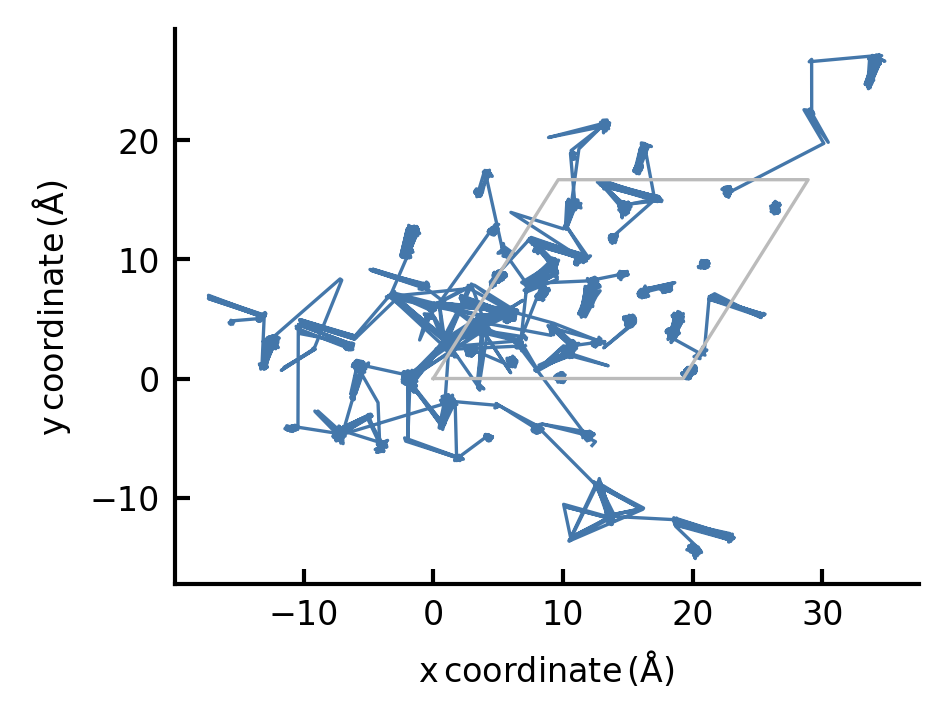}
\end{tabular}
\caption{
Trajectries of proton hole centers (PHCs)
in Zr$_7$O$_8$N$_4$ (left), ZrO$_2$ (right).
}
\label{fgr:phc_traj}
\end{figure}
Figure~\ref{fgr:phc_traj}
shows trajectries of proton hole centers (PHCs)
It can be seen that ZrO$_2$ moves more from its initial position than Zr$_7$O$_8$N$_4$.

\begin{figure}
\begin{tabular}{cc}
\includegraphics[width=3.2in]{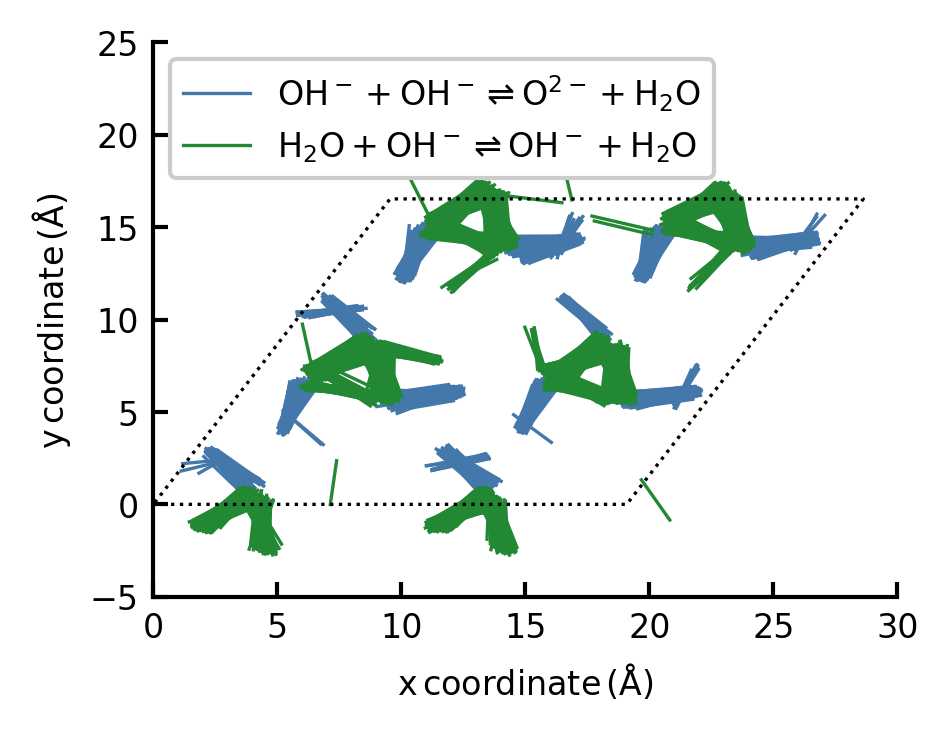}
\includegraphics[width=3.2in]{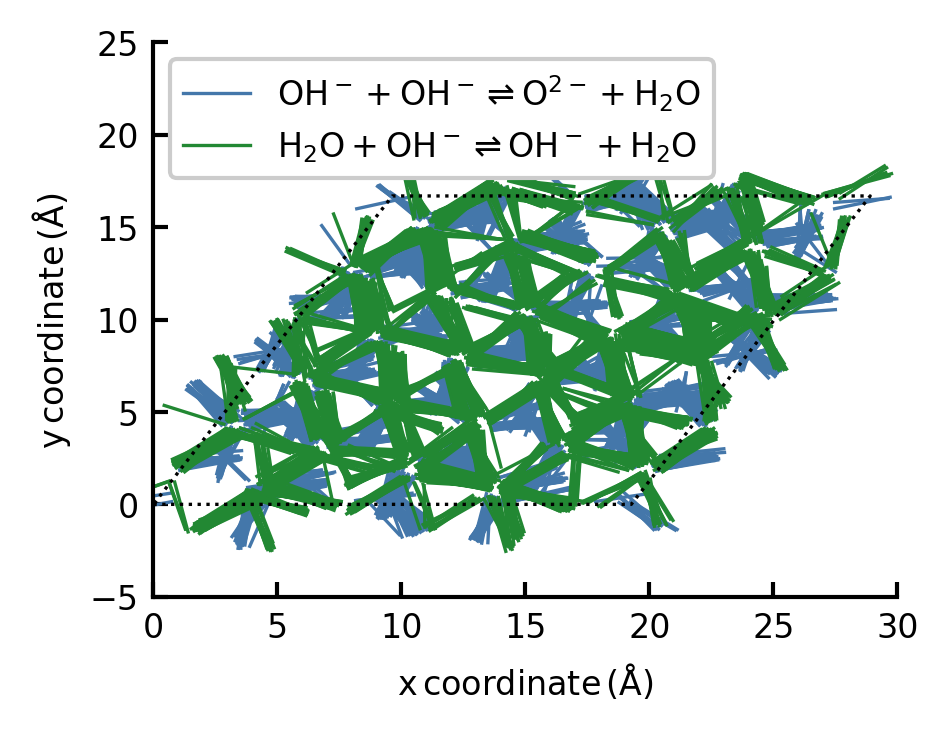}
\end{tabular}
\caption{
Pathways of PT
in Zr$_7$O$_8$N$_4$ (left), ZrO$_2$ (right).
}
\label{fgr:pt_path}
\end{figure}
Figure~\ref{fgr:pt_path}
shows pathways of PT.
It is thought that protons can only diffuse effectively over 
long distances through these PTs, 
depending on whether the paths in multiple PTs connect 
from a particular location within one cell 
to an equivalent location in a neighbouring cell.
In Zr$_7$O$_8$N$_4$, PT pathways are clustered around \VO and do not connect 
from a location within a cell 
to an equivalent location in an adjacent cell.
In ZrO$_2$, on the other hand, PT pathways exist throughout the cell, 
suggesting that they can connect from one cell to another.
However, Figure~\ref{fgr:pt_path} was calculated 
from 100 trajectories of 50\,ps MD, 
and it does not mean that 
protons move between different cells within 50\,ps in a single trajectory.
A longer-term MD is required for a more detailed discussion.
In the case of the 50 ns MD trajectory mentioned above, 
the total duration is sufficient.
However, due to file capacity constraints, 
the trajectories were recorded at 100 ps intervals, 
so PT judgments tend to be incorrect.
Future tasks will include performing long-term MD at smaller time intervals 
to accurately examine PT trajectories 
and determine which PTs contribute to long-range diffusion.

\newpage

\subsection{Anharmonic OH vibrational spectrum}
\begin{figure}
\begin{tabular}{cc}
\includegraphics[width=3.2in]{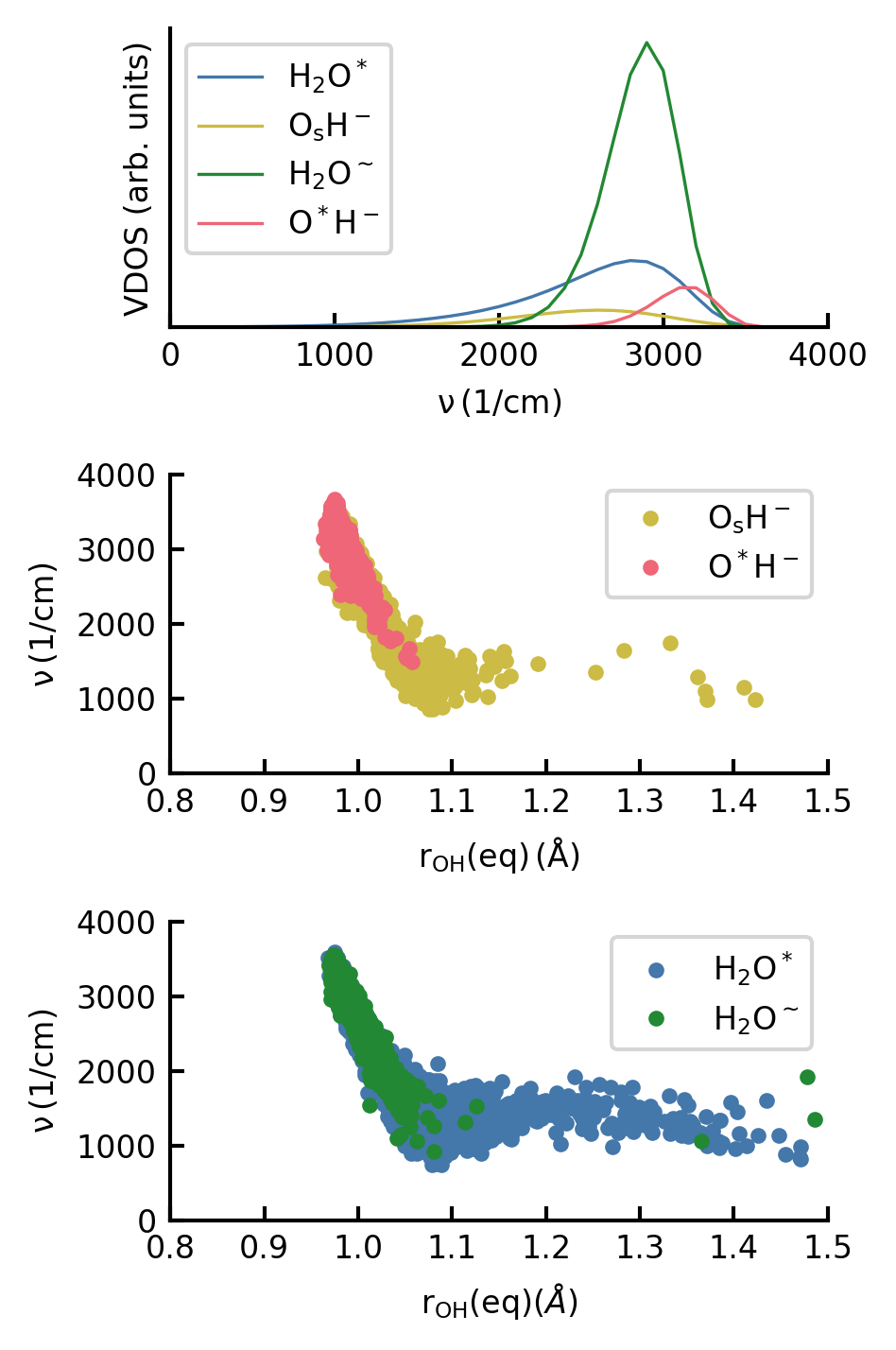}
\includegraphics[width=3.2in]{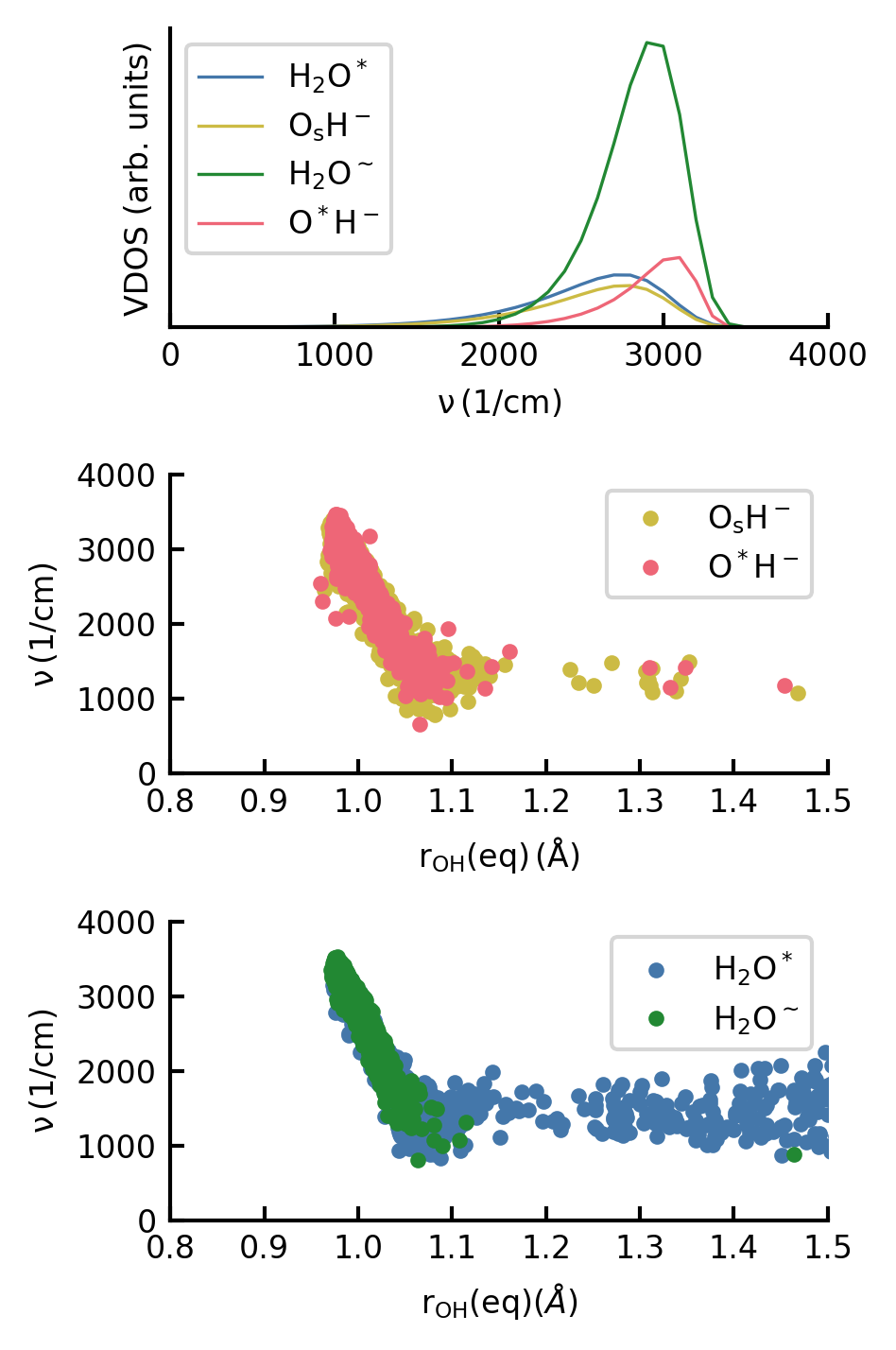}
\end{tabular}
\caption{
Vibrational density of states (VDOS) (top) and
scatter plot of $r_\text{OH}$(eq) and $\nu_\text{max}$ for OH (middle) and H$_2$O (bottom)
in Zr$_7$O$_8$N$_4$ (left), ZrO$_2$ (right).
Adsorbed oxygen, oxygen of the slab surface and oxygen in the solvent are expressed
as O$^*$, O$_\text{s}$ and O$^\sim$ respectively.
}
\label{fgr:vdos_tot}
\end{figure}

The upper part of Figure~\ref{fgr:vdos_tot} shows 
the OH vibrational density of states (VDOS) for each oxygen state.
Adsorbed oxygen, oxygen of the slab surface and oxygen in the solvent are expressed
as O$^*$, O$_\text{s}$ and O$^\sim$ respectively.
For O$_\text{s}$H$^-$, the frequency $\nu_\text{max}$ was lowered (red shift) due to the defect.
As mentioned above, the presence of the defect reduces the dissociative adsorption, 
so the surface PT reaction O$_\text{s}$H$^-$ $\rightarrow$ O$_\text{s}^{2-}$ 
is more likely to occur than the reverse reaction, 
which is likely due to the broadening of the bottom of the potential $V(r_\text{OH})$.
Conversely, for O$^*$H$^-$ the frequency $\nu_\text{max}$ increases due to defects (blue shift).
The reason for this is still not well understood.
The middle and lower parts of Figure~\ref{fgr:vdos_tot} 
are scatter plots of the frequency $\nu$ 
and the equilibrium distance $r_\text{OH}$(eq) for each oxygen state.
There is a negative correlation between $\nu$ and $r_\text{OH}$(eq).
This is because the weaker the OH bond, the lower the $\nu$ and the longer $r_\text{OH}$(eq).

\newpage

\begin{figure}
\begin{tabular}{cc}
\includegraphics[width=3.2in]{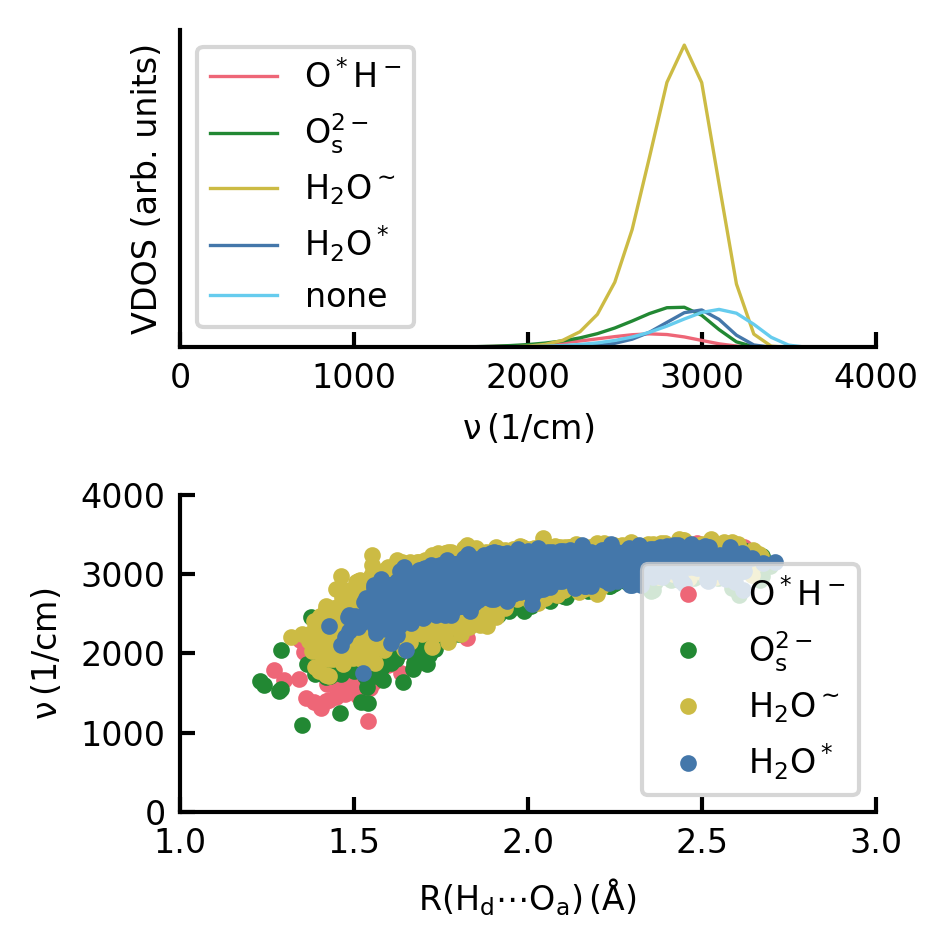}
\includegraphics[width=3.2in]{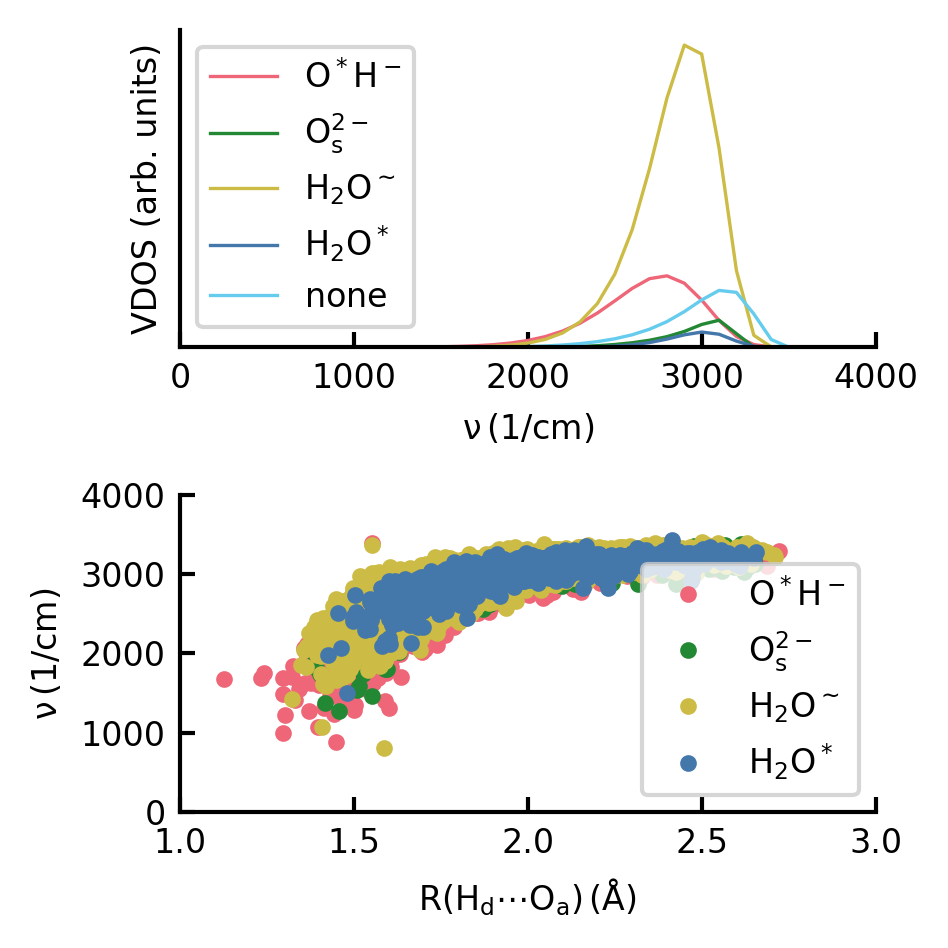}
\end{tabular}
\caption{
VDOS of H$_2$O$^\sim$ for each acceptor of Zr$_7$O$_8$N$_4$ (left), ZrO$_2$ (right).
}
\label{fgr:vdos_of2}
\end{figure}
The upper part of Figure~\ref{fgr:vdos_of2} 
represents the VDOS when O in H$_2$O$^\sim$ is the hydrogen bond donor 
and the acceptors are O$^*$H$^-$, O$_\text{s}^{2-}$, H$_2$O$^\sim$, H$_2$O$^*$, and none.
The lower part of Figure~\ref{fgr:vdos_of2} 
is a scatter plot of $\nu$ and the distance R(H$_\text{d}$$\cdots$O$_\text{a}$)
between the acceptors O and donated H.
In Zr$_7$O$_8$N$_4$, $\nu_\text{max}$ decreases when the acceptor is O$_\text{s}^{2-}$, but the reason is unknown.
For other acceptors,
$\nu_\text{max}$ is almost the same depending on whether there is a defect or not.

\newpage

\begin{figure}
\begin{tabular}{cc}
\includegraphics[width=3.2in]{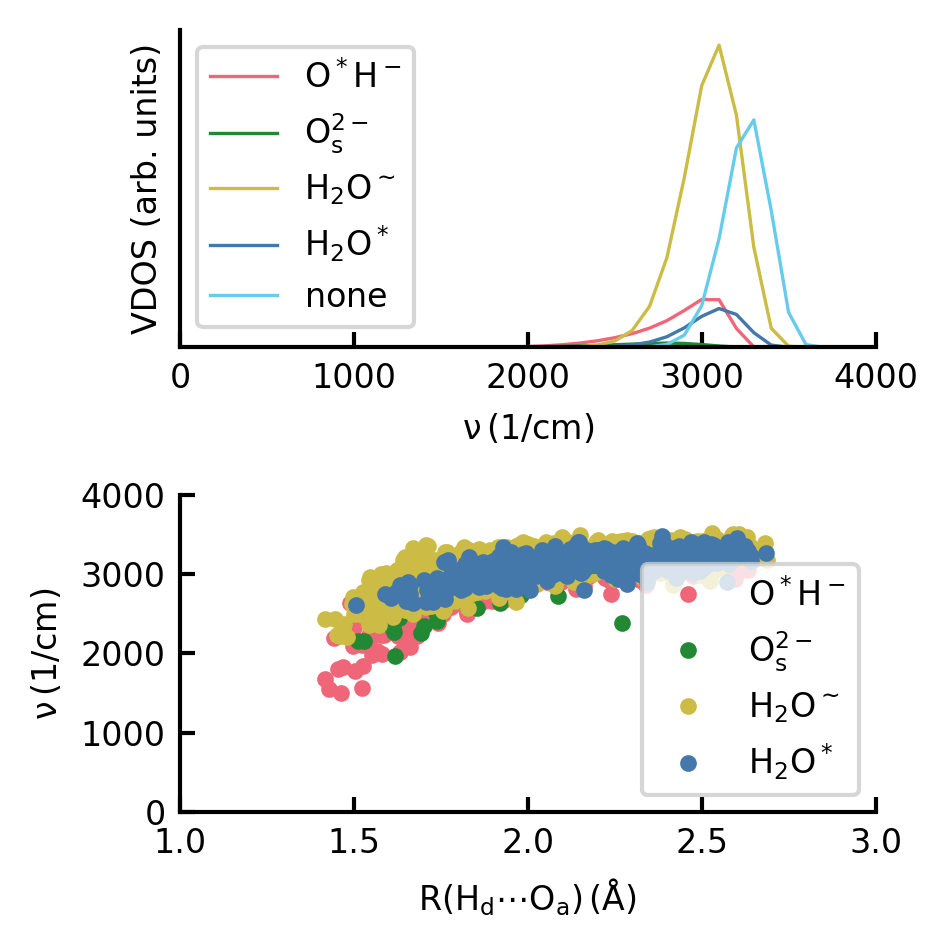}
\includegraphics[width=3.2in]{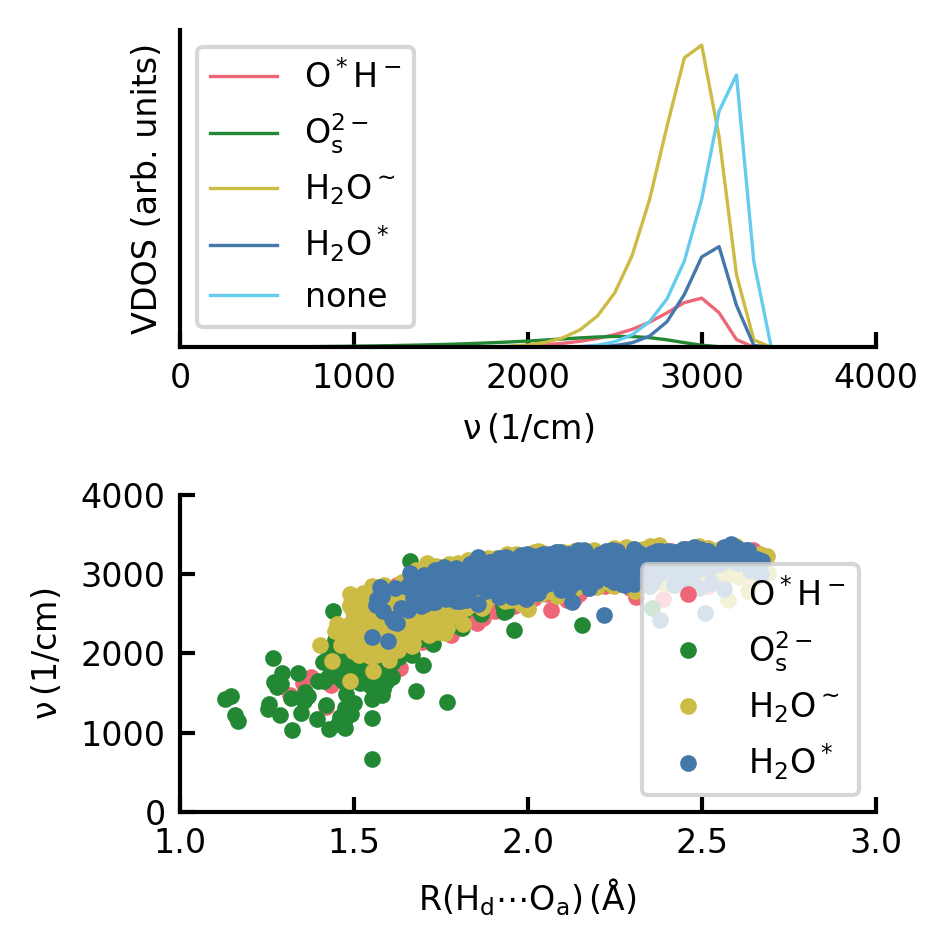}
\end{tabular}
\caption{
VDOS of O$^*$H$^-$ for each acceptor of Zr$_7$O$_8$N$_4$ (left), ZrO$_2$ (right).
}
\label{fgr:vdos_oz1}
\end{figure}
Figure~\ref{fgr:vdos_oz1}
has been drawn in the same way as the Figure~\ref{fgr:vdos_of2} 
for the case where O in O$^*$H$^-$ is the hydrogen bond donor and 
the acceptors are O$^*$H$^-$, O$_\text{s}^{2-}$, H$_2$O$^\sim$, H$_2$O$^*$, and none.
In the scatter plot below, the samples are distributed in an elongated shape. 
This is because the hydrogen bonding environment has a uniform effect 
on the correlation between hydrogen bonds and vibrational frequency.

\newpage

\begin{figure}
\begin{tabular}{cc}
\includegraphics[width=3.2in]{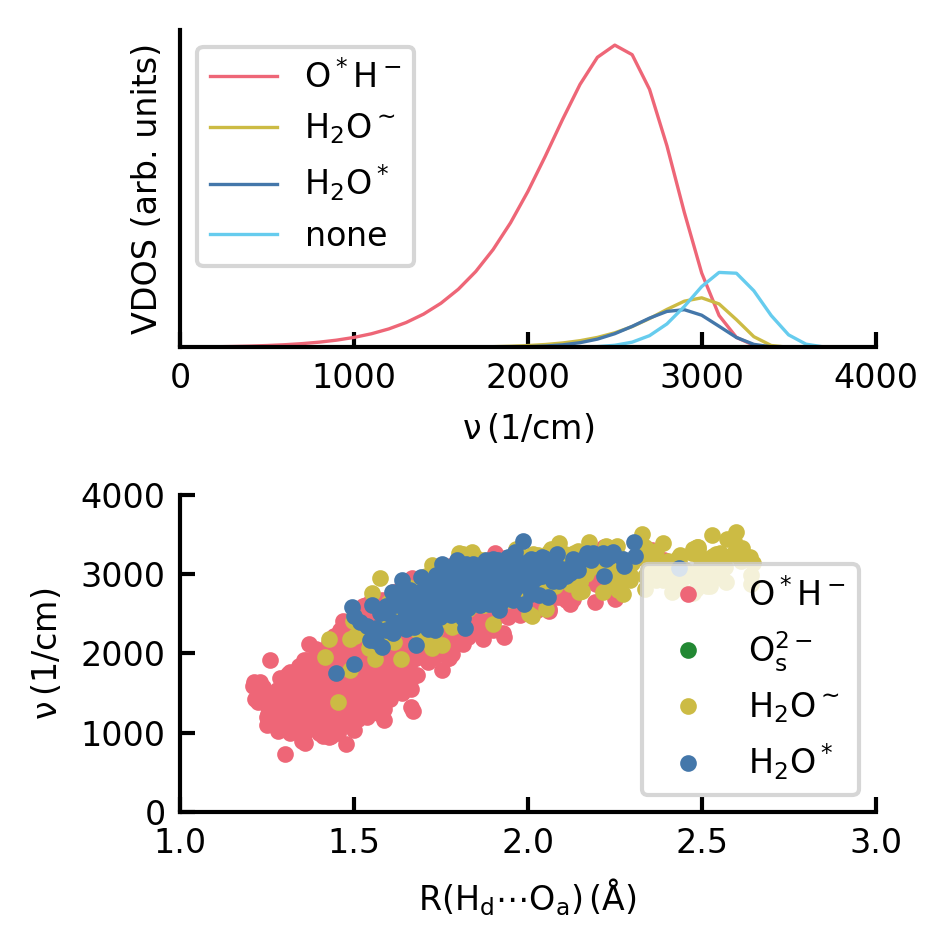}
\includegraphics[width=3.2in]{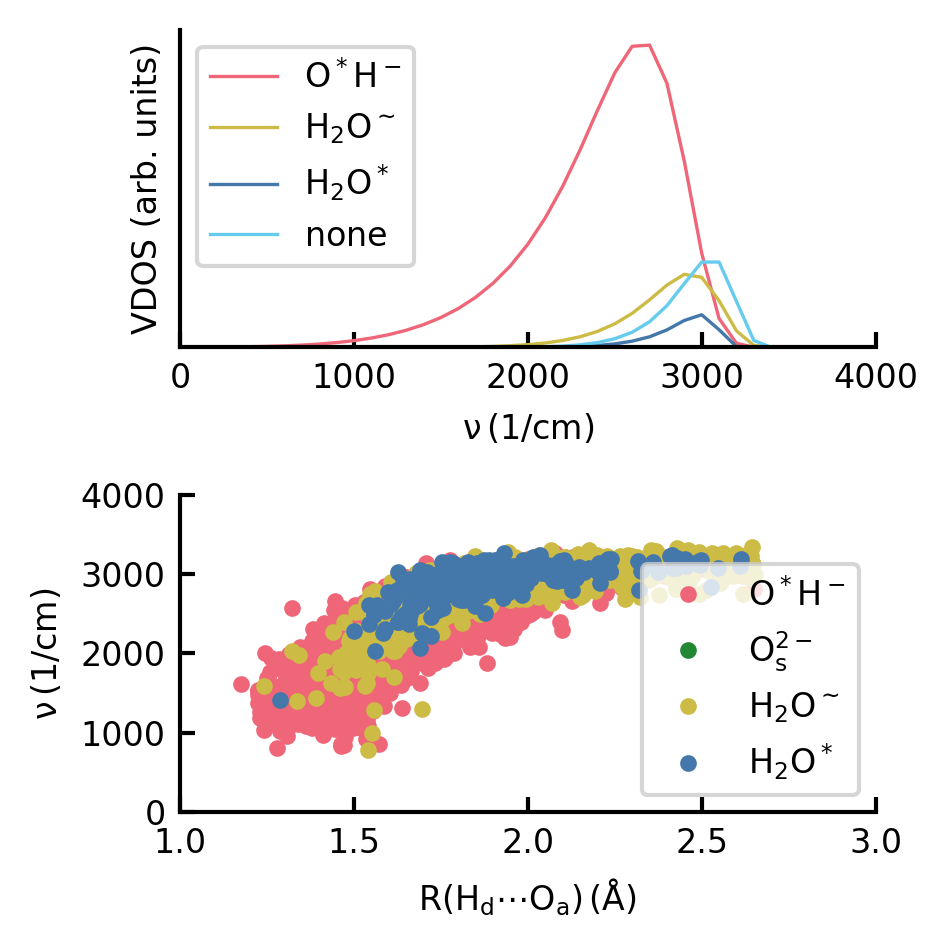}
\end{tabular}
\caption{
VDOS of O$_\text{s}^{2-}$ for each acceptor of Zr$_7$O$_8$N$_4$ (left), ZrO$_2$ (right).
}
\label{fgr:vdos_os1}
\end{figure}
Figure~\ref{fgr:vdos_os1}
has been drawn in the same way as the Figure~\ref{fgr:vdos_of2} 
for the case where O in O$_\text{s}$H$^-$ is the hydrogen bond donor and 
the acceptors are O$^*$H$^-$, H$_2$O$^\sim$, H$_2$O$^*$, and none.
The data for the acceptor O$_\text{s}^{2-}$ are too small (less than 10) and are therefore omitted.
As O$^*$H$^-$ is involved in the surface PT, its $\nu_\text{max}$ as an acceptor is the lowest.

\newpage

\begin{figure}
\begin{tabular}{cc}
\includegraphics[width=3.2in]{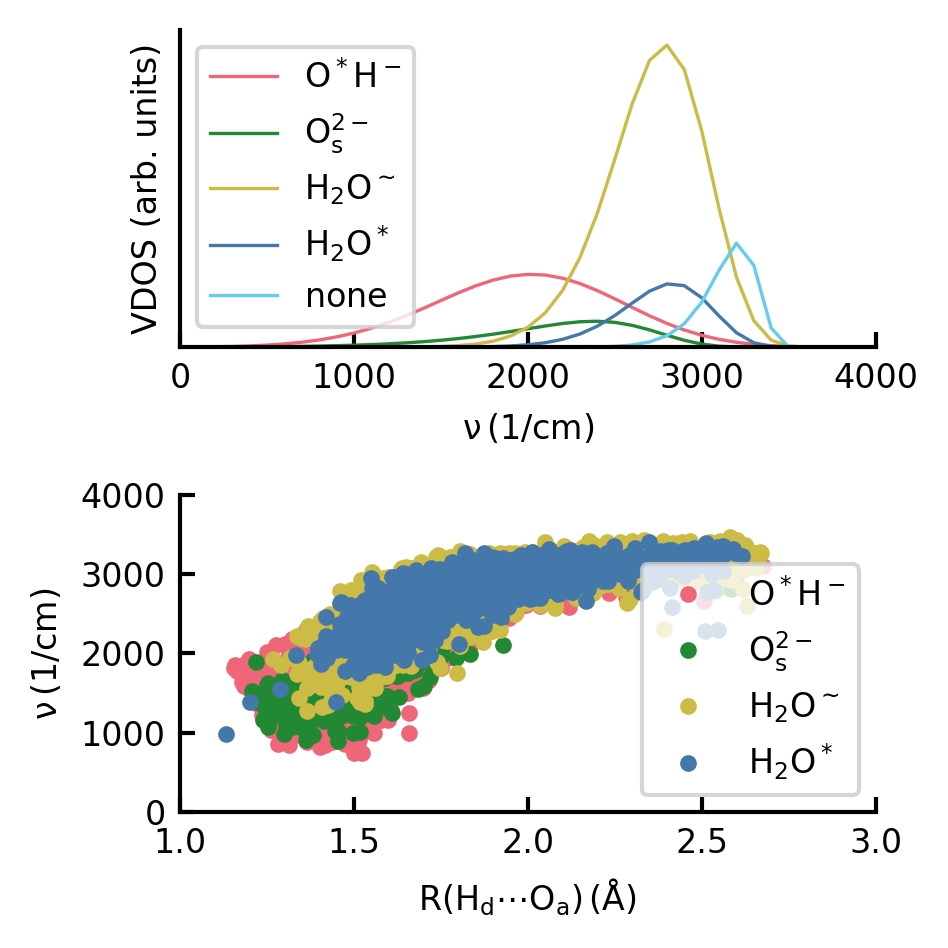}
\includegraphics[width=3.2in]{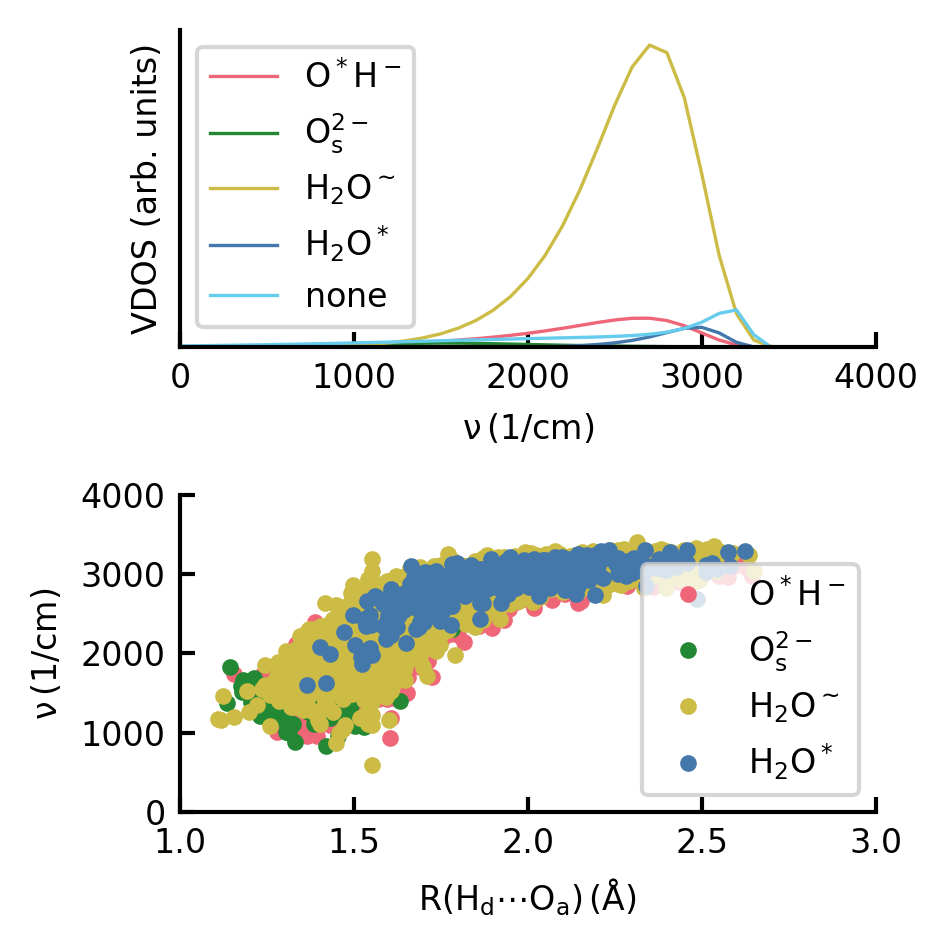}
\end{tabular}
\caption{
VDOS of H$_2$O$^*$ for each acceptor of Zr$_7$O$_8$N$_4$ (left), ZrO$_2$ (right).
}
\label{fgr:vdos_oz2}
\end{figure}
Figure~\ref{fgr:vdos_oz2}
has been drawn in the same way as the Figure~\ref{fgr:vdos_of2} 
for the case where O in H$_2$O$^*$ is the hydrogen bond donor and 
the acceptors are O$^*$H$^-$, O$_\text{s}^{2-}$, H$_2$O$^\sim$, H$_2$O$^*$, and none.
$\nu_\text{max}$ is low when the acceptor is O$^*$H$^-$ or O$_\text{s}^{2-}$.
This is because each is related to the adlayer PT and the surface PT.
Since ZrO$_2$ is easily dissociated, the $\nu_\text{max}$ of O$_\text{s}^{2-}$ is lower than that of Zr$_7$O$_8$N$_4$.

\newpage

\begin{table}
\caption{
$\nu_\text{max}$ at which the VDOS is maximum, 
the distance R$_\text{max}$ at which the histogram of the distance R(H$_\text{d}$$\cdots$O$_\text{a}$) is maximum and 
the number of data N$_\text{data}$ for each hydrogen bond donor (D) and acceptor (A). 
}
  \label{tbl:numax_rmax_ndata}
  \begin{tabular}{rrrrrrrr}
    \hline
    \multicolumn{2}{r}{} & \multicolumn{3}{r}{Zr$_7$O$_8$N$_4$} & \multicolumn{3}{r}{ZrO$_2$} \\
   D &     A &$\nu_\text{max}$ & R$_\text{max}$ &N$_\text{data}$ &$\nu_\text{max}$ & R$_\text{max}$ &N$_\text{data}$ \\
    \hline
H$_2$O$^\sim$ & O$^*$H$^-$  &  2675 & 1.625 &  1898 &  2747 & 1.675 &  7856 \\
     & O$_\text{s}^{2-}$  &  2814 & 1.775 &  4764 &  3041 & 1.825 &  1774 \\
     & H$_2$O$^\sim$  &  2886 & 1.775 & 27949 &  2910 & 1.775 & 25022 \\
     & H$_2$O$^*$  &  2956 & 1.825 &  3136 &  2991 & 1.825 &  1066 \\
     & none  &  3061 &       &  4280 &  3091 &       &  4932 \\
     & total &  2889 &       & 42027 &  2910 &       & 40650 \\
O$^*$H$^-$ & O$^*$H$^-$  &  2995 & 2.075 &   537 &  2929 & 1.925 &   722 \\
     & O$_\text{s}^{2-}$  &  2756 & 1.625 &    63 &  2420 & 1.675 &   334 \\
     & H$_2$O$^\sim$  &  3076 & 1.875 &  2715 &  2935 & 1.825 &  4095 \\
     & H$_2$O$^*$  &  3095 & 2.125 &   349 &  3049 & 2.025 &   932 \\
     & none  &  3266 &       &  1681 &  3142 &       &  2964 \\
     & total &  3132 &       &  5345 &  3015 &       &  9047 \\
O$_\text{s}$H$^-$ & O$^*$H$^-$  &  2451 & 1.675 &  4062 &  2579 & 1.625 &  6639 \\
     & O$_\text{s}^{2-}$  &       &       &     0 &       &       &     5 \\
     & H$_2$O$^\sim$  &  2949 & 1.825 &   435 &  2892 & 1.825 &  1083 \\
     & H$_2$O$^*$  &  2854 & 1.875 &   307 &  2956 & 1.875 &   314 \\
     & none  &  3135 &       &   541 &  3020 &       &  1006 \\
     & total &  2570 &       &  5345 &  2692 &       &  9047 \\
H$_2$O$^*$ & O$^*$H$^-$  &  2008 & 1.525 &  4403 &  2543 & 1.675 &  1149 \\
     & O$_\text{s}^{2-}$  &  2313 & 1.675 &  1294 &  1615 & 1.475 &   187 \\
     & H$_2$O$^\sim$  &  2767 & 1.775 &  9666 &  2674 & 1.725 &  8823 \\
     & H$_2$O$^*$  &  2809 & 1.725 &  1879 &  2932 & 1.775 &   376 \\
     & none  &  3194 &       &  1761 &  2948 &       &  1204 \\
     & total &  2743 &       & 19003 &  2675 &       & 11739 \\
    \hline
  \end{tabular}
\end{table}
Table~\ref{tbl:numax_rmax_ndata} 
shows the frequency $\nu_\text{max}$ at which the VDOS is maximum, 
the distance R$_\text{max}$ at which the histogram of the distance R(H$_\text{d}$$\cdots$O$_\text{a}$) is maximum and 
the number of data N$_\text{data}$ for each hydrogen bond donor (D) and acceptor (A). 
There is nothing special to discuss on the basis of this table alone.

\begin{figure}
\begin{tabular}{cc}
\multicolumn{2}{r}{
\includegraphics[width=6.4in]{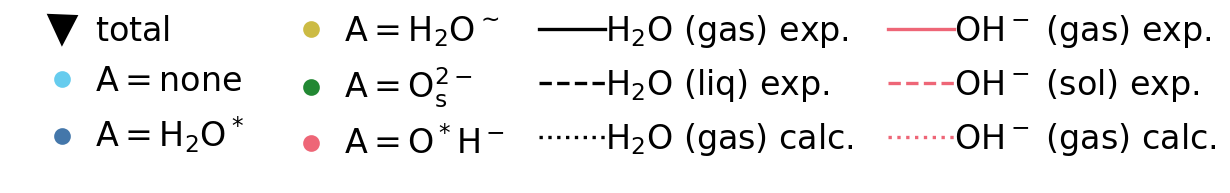}} \\
\includegraphics[width=3.2in]{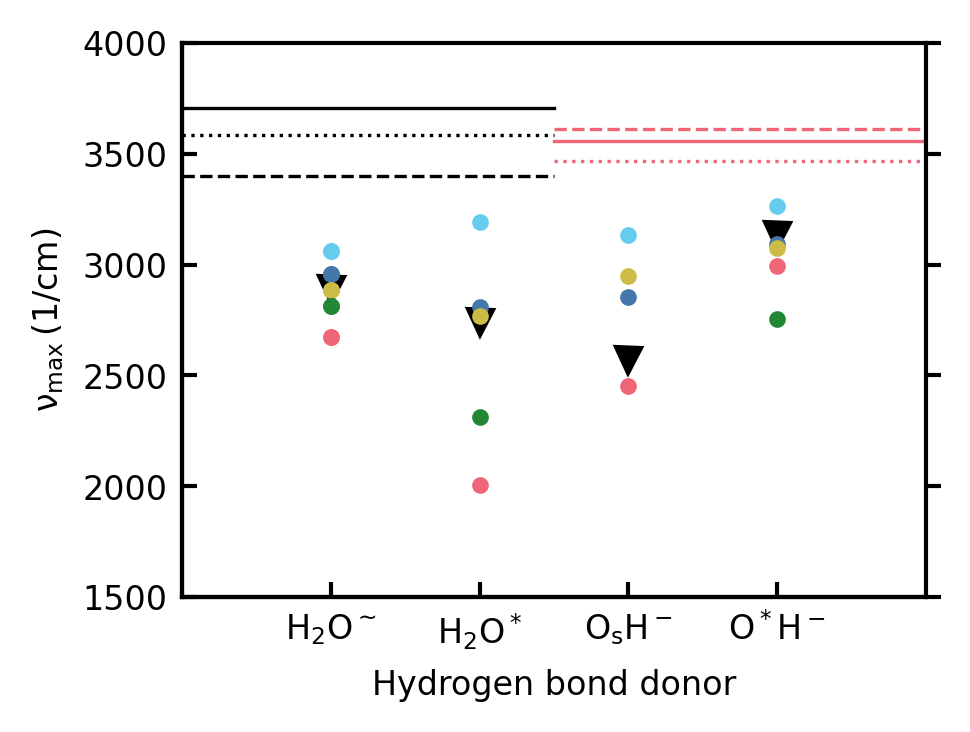}
\includegraphics[width=3.2in]{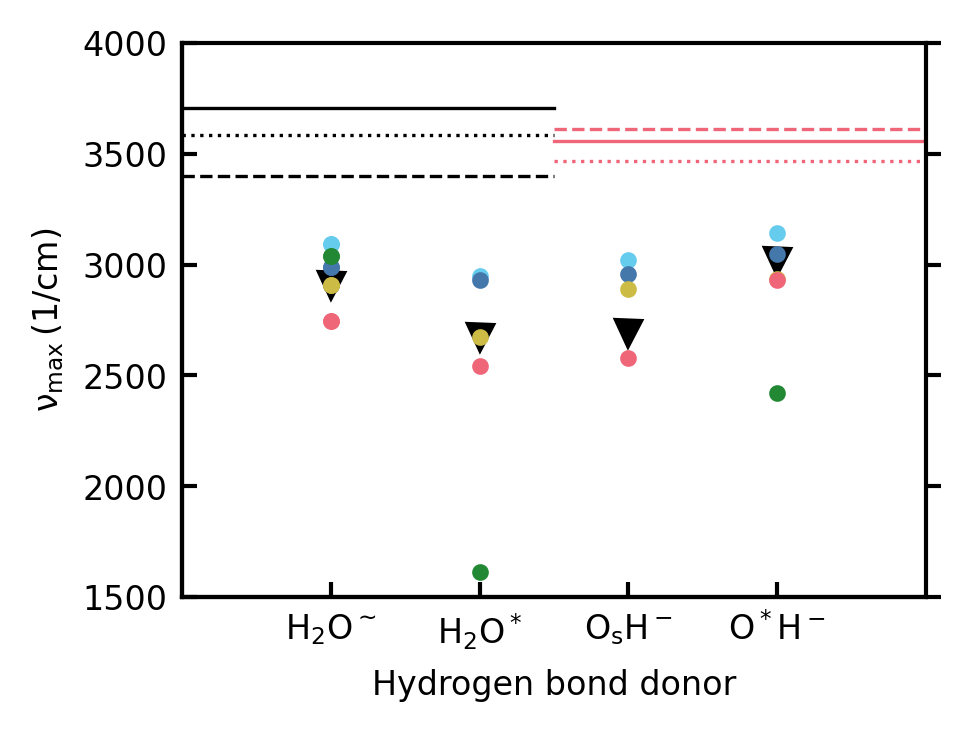}
\end{tabular}
\caption{
Summary of $\nu_\text{max}$ for H$_2$O and OH of Zr$_7$O$_8$N$_4$ (left), ZrO$_2$ (right).
For comparison, 
we also show 
first-principles calculation data of gas phase H$_2$O and OH$^-$,
\cite{Quaranta2018}
and experimental data of gas phase H$_2$O, 
\cite{Benedict1956}
liquid phase H$_2$O, 
\cite{Fecko2003}
gas phase OH$^-$, 
\cite{Owrutsky1985}
and solvated OH$^-$, 
\cite{Corridoni2007}
}
\label{fgr:vdos_numax}
\end{figure}
The Figure~\ref{fgr:vdos_numax} summarises 
the frequency $\nu_\text{max}$ at which the VDOS becomes maximum.
When the acceptor is H$_2$O$^*$, 
most of the $\nu_\text{max}$ are higher than the total (except when the donor is O$^*$H$^-$ in Zr$_7$O$_8$N$_4$).
When the acceptor is O$^*$H$^-$, all the $\nu_\text{max}$ are lower than the total.
This is because O$^*$H$^-$ is a good acceptor, 
accepting H in both the surface PT and the adlayer PT.
All $\nu_\text{max}$ in this study are lower than
experimental results and first-principles calculations.

\begin{figure}
\begin{tabular}{cc}
\includegraphics[width=3.2in]{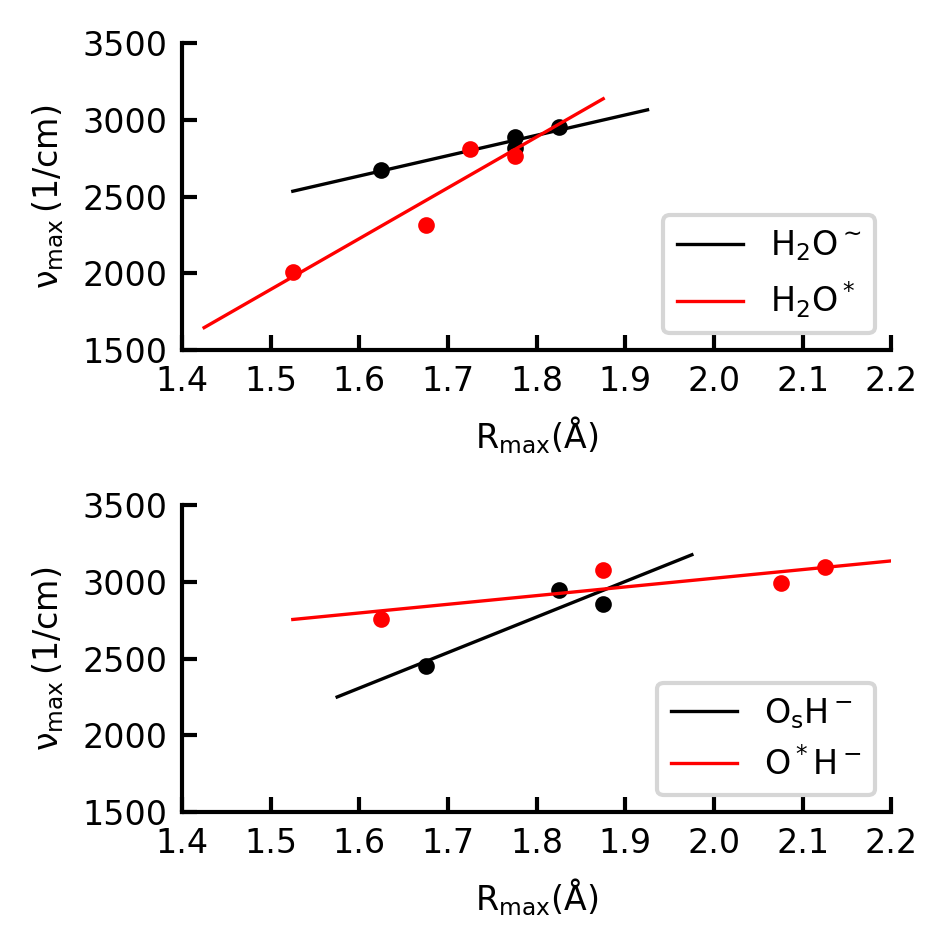}
\includegraphics[width=3.2in]{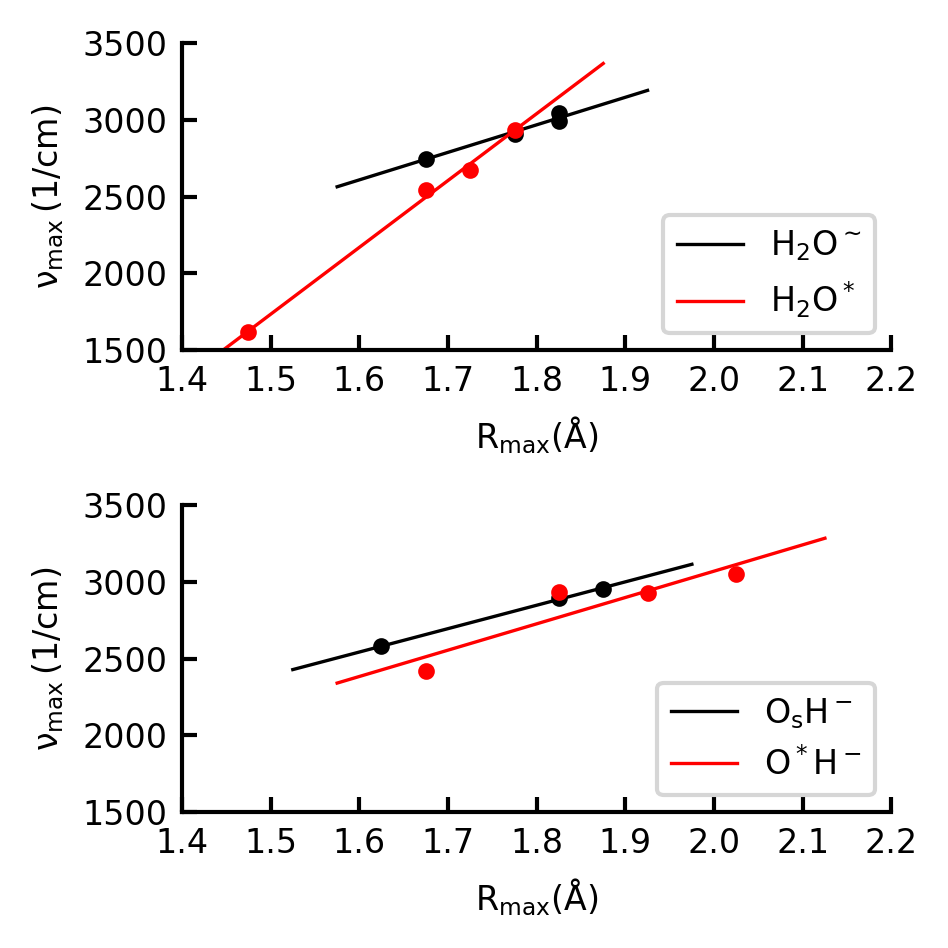}
\end{tabular}
\caption{
R$_\text{max}$ and $\nu_\text{max}$ of Zr$_7$O$_8$N$_4$ (left), ZrO$_2$ (right).
}
\label{fgr:vdos_rmax_numax}
\end{figure}
Figure~\ref{fgr:vdos_rmax_numax} is a scatter plot of 
the frequency $\nu_\text{max}$ where VDOS is maximum 
and the R$_\text{max}$ where the histogram of distance R(H$_\text{d}$$\cdots$O$_\text{a}$) is maximum.
In addition, the correlation for each donor 
has been approximated by a linear function.
There's nothing special to say about this.

\begin{table}
\caption{
The slope of the linear function for $\nu_\text{max}$-R$_\text{max}$, the range of R$_\text{max}$ 
and the full width at half maximum $\gamma_0$ of the VDOS for each donor.
}
  \label{tbl:slope_range_gamma0}
  \begin{tabular}{rrrrrrr}
    \hline
     & \multicolumn{3}{r}{Zr$_7$O$_8$N$_4$} & \multicolumn{3}{r}{ZrO$_2$} \\
   D & slope & range & $\gamma_0$ & slope & range & $\gamma_0$  \\
    \hline
H$_2$O$^*$ &  3313 & 0.25  &       982 &  4356 & 0.3   &       876  \\
O$_\text{s}$H$^-$ &  2316 & 0.2   &      1089 &  1524 & 0.25  &       857  \\
H$_2$O$^\sim$ &  1326 & 0.2   &       505 &  1793 & 0.15  &       548  \\
O$^*$H$^-$ &   565 & 0.5   &       446 &  1715 & 0.35  &       492  \\
    \hline
  \end{tabular}
\end{table}
Table~\ref{tbl:slope_range_gamma0} summarises 
the slope of the linear function in the Figure~\ref{fgr:vdos_rmax_numax},
the range of R$_\text{max}$ (the difference between the maximum and minimum values) 
and the full width at half maximum $\gamma_0$ of the VDOS for each donor.
In ZnO, these orders coincide and are correlated.
\cite{Quaranta2018}
In both Zr$_7$O$_8$N$_4$ and ZrO$_2$ the order does not match completely 
and there is no correlation as in previous studies.

\newpage

\begin{figure}
\begin{tabular}{cc}
\includegraphics[width=3.2in]{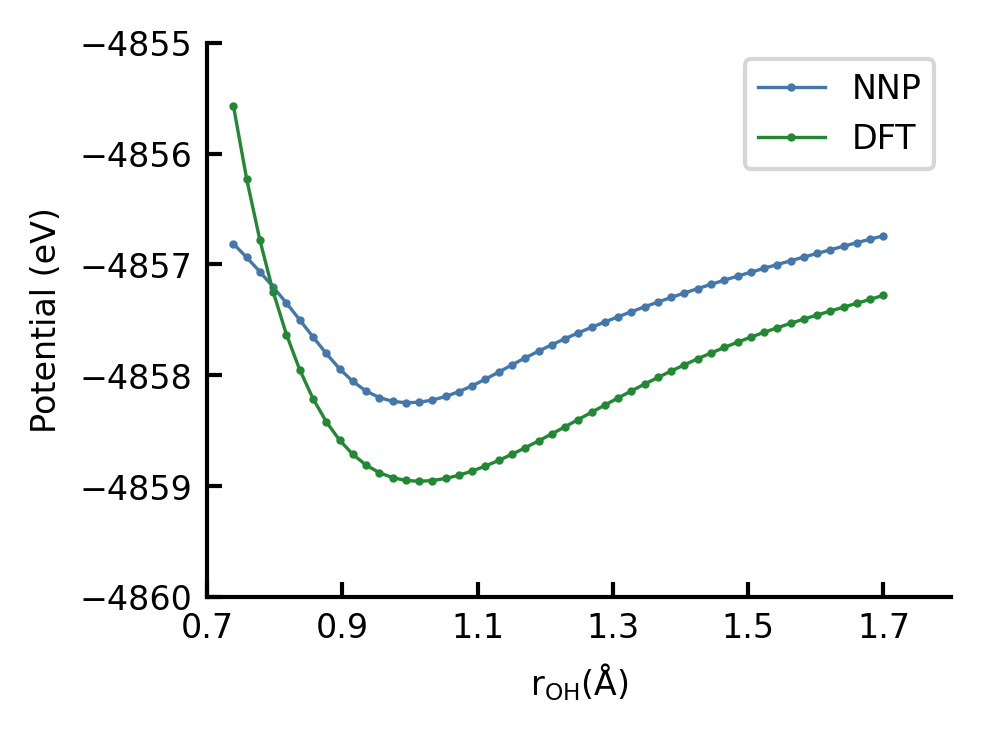}
\includegraphics[width=3.2in]{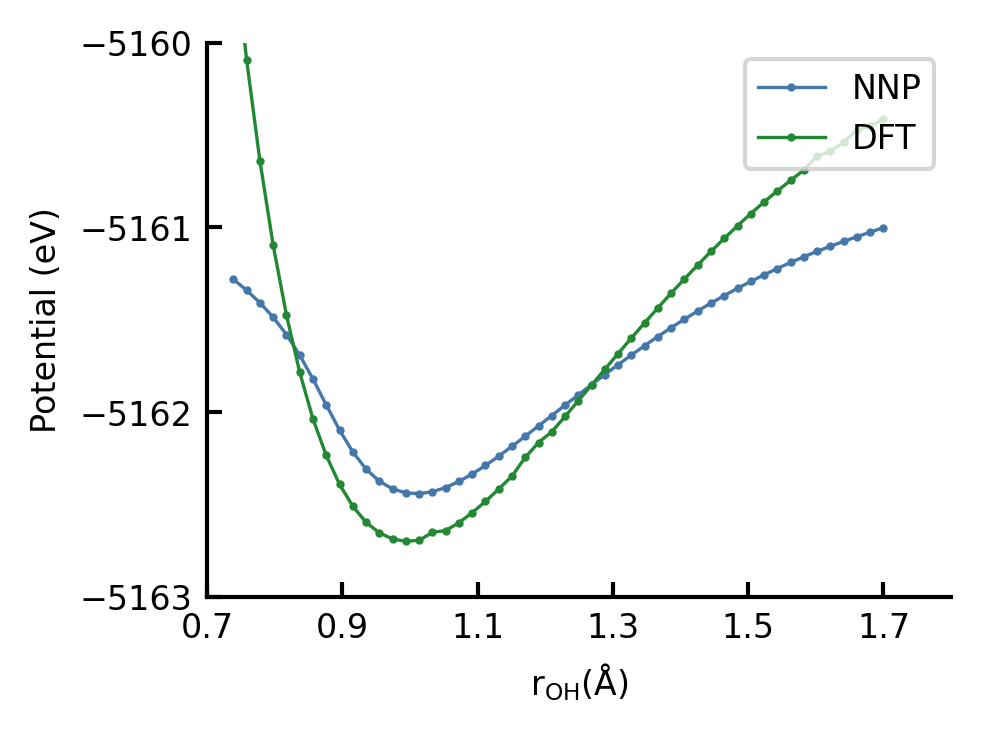}
\end{tabular}
\caption{
Potential energy for the OH bond distance $r_\text{OH}$ 
of solvent water molecules calculated by NNP (blue) and DFT (green).
}
\label{fgr:vdos_vroh}
\end{figure}
Figure~\ref{fgr:vdos_vroh} is an example of the potential energy 
for stretching and contracting the OH bond distance $r_\text{OH}$ 
of solvent water molecules, 
calculated using NNP and Density Functional Theory (DFT).
The RMSEs of energy in Zr$_7$O$_8$N$_4$,ZrO$_2$ were 1.064, 0.661 (meV/atom).
The frequencies of Zr$_7$O$_8$N$_4$ calculated from the NNP and DFT potentials 
were 3084,2969 (1/cm) and these of ZrO$_2$ were 2853, 3302 (1/cm).
ZrO$_2$ has a smaller error in $V(r_\text{OH})$,
but the difference in frequency is large, reaching more than 300 (1/cm).
This is thought to be due to the change in curvature of the potential.
This may be one of the causes of the problem mentioned above that 
the $\nu_\text{max}$ calculated by NNP is far from the experimental values of 
the vibrational frequencies in H$_2$O and OH$^-$.
A method to solve this problem has not yet been found.
In addition, as for ZrO$_2$, 
if the potential in the range of $r_\text{OH}$(eq)$\pm$0.2{\AA} 
is approximated by a quadratic polynomial (harmonic approximation),
the calculated frequencies are 3079, 3081 (1/cm) (NNP, DFT).
While the difference between NNP and DFT almost disappears, 
the difference to the 11th degree polynomial is large.
This means that the influence of the anharmonic effect cannot be ignored.

\newpage

\subsection{Future work}
In addition to structural properties such as those in this study, 
other studies of solid/water interface systems using NNPMD 
include 
that for water dissociation
\cite{Zeng2023,Wen2023a}
and solvation dynamics.
\cite{Schienbein2022}
Systems with intermediates in the ORR
could also be added to the data to learn NNPs, 
sampled by NNPMD and calculate the free energies on the reaction pathways, 
so that the influence of defects and 
the proportion of molecular and dissociative adsorption 
on catalytic activity could also be discussed. 
There is also much to be explored in studies 
based on conventional first-principles calculations.
In contrast to the present study, 
there are previous studies that have discussed the electric double layer (EDL)
by calculating the electrostatic potential and density 
when the system is about a system with ions in water.
\cite{Ando2013,Ikeda2014,Li2020,Huang2021,Shin2022}
For example, Ando et al. performed AIMD at the interface 
between Pt and water containing Na$^+$ ions and
calculated the electrostatic potential due to the EDL
based on the difference in the potentials 
between the case with and without Na$^+$ ions.
\cite{Ando2013}
For Zr$_7$O$_8$N$_4$ and ZrO$_2$, 
the electrostatic potential due to the EDL
could also be evaluated in NNPMD by training NNP 
with additional data for the case of the solvent containing ions. 
Alternatively, as in the work of Otani et al.,
\cite{Otani2008}
the electrostatic potential of H$_3$O$^+$ in oxygen can be discussed 
by creating a system with excess H$^+$ and electrons in water 
and performing an AIMD based on the effective screening medium method.
Other studies,
\cite{Hormann2019,Guo2019,Sun2021,Wen2023b}
have performed AIMD and structural optimisation to discuss band offsets, 
which are important values in interface systems, taking into account 
electron affinity, charge neutrality level and pinning factor,
and have discussed the band offset, 
which is an important value in interface systems.
This multifaceted study will allow a more detailed investigation 
of the effects of defects at the oxide/water interface.

\section{Conclusion}

In this study, NNPs were constructed by active learning 
for interfaces between the water and defective Zr$_7$O$_8$N$_4$ or pristine ZrO$_2$.
Based on 1000 NNPMD trajectories, 
the structures of water molecule adsorption on these interfaces 
were investigated and it was found that Zr$_7$O$_8$N$_4$ has a bilayer structure,
which is different from a monolayer structure of ZrO$_2$.
On the Zr$_7$O$_8$N$_4$ surface,
water molecules do not adsorb on the \VO\,
but only on some of the surrounding Zr atoms,
suggesting that the \VO\,is expected to 
act as an active site for the ORR.
While the proportion of dissociative adsorption in Zr$_7$O$_8$N$_4$ is less than in ZrO$_2$ 
due to the less oxygen on which the H$^+$ can adsorb, 
the total number of molecular and dissociative adsorption
is the same for both materials.
Therefore, Zr$_7$O$_8$N$_4$ may promote the ORR more than ZrO$_2$ 
because the \VO s act as additional active sites.
In the future, 
the influence of the solvent on the ORR activity of Zr$_7$O$_8$N$_4$ 
could be discussed by updating the NNP with new data 
for systems with intermediate oxygen and hydrogen molecules 
in addition to the solvent, water molecules.
We also calculated things that had been calculated in previous studies, 
such as the free energy of proton transfer 
and the vibrational density of states of OH.
We hope that the various findings from this study 
will be useful for research into the solid-water interface.

\begin{acknowledgement}
This research was supported by the New Energy and 
Industrial Technology Development Organization (NEDO) project, 
MEXT as ``Program for Promoting Researches on the Supercomputer Fugaku''
(Fugaku battery and Fuel Cell Project) 
(Grant No. JPMXP1020200301, Project No.: hp220177, hp210173, hp200131), 
and the Japan Society for the Promotion of Science (JSPS) 
as Grants in Aid for Scientific Research 
on Innovative Area ``Hydrogenomics'' (Grant No. 18H05519). 
The calculations were conducted at the ISSP Supercomputers Center, 
The University of Tokyo. 
\end{acknowledgement}

\bibliography{main}

\providecommand{\latin}[1]{#1}
\makeatletter
\providecommand{\doi}
  {\begingroup\let\do\@makeother\dospecials
  \catcode`\{=1 \catcode`\}=2 \doi@aux}
\providecommand{\doi@aux}[1]{\endgroup\texttt{#1}}
\makeatother
\providecommand*\mcitethebibliography{\thebibliography}
\csname @ifundefined\endcsname{endmcitethebibliography}
  {\let\endmcitethebibliography\endthebibliography}{}
\begin{mcitethebibliography}{75}
\providecommand*\natexlab[1]{#1}
\providecommand*\mciteSetBstSublistMode[1]{}
\providecommand*\mciteSetBstMaxWidthForm[2]{}
\providecommand*\mciteBstWouldAddEndPuncttrue
  {\def\EndOfBibitem{\unskip.}}
\providecommand*\mciteBstWouldAddEndPunctfalse
  {\let\EndOfBibitem\relax}
\providecommand*\mciteSetBstMidEndSepPunct[3]{}
\providecommand*\mciteSetBstSublistLabelBeginEnd[3]{}
\providecommand*\EndOfBibitem{}
\mciteSetBstSublistMode{f}
\mciteSetBstMaxWidthForm{subitem}{(\alph{mcitesubitemcount})}
\mciteSetBstSublistLabelBeginEnd
  {\mcitemaxwidthsubitemform\space}
  {\relax}
  {\relax}

\bibitem[Debe(2012)]{Debe2012}
Debe,~M.~K. Electrocatalyst approaches and challenges for automotive fuel
  cells. \emph{Nature} \textbf{2012}, \emph{486}, 43--51\relax
\mciteBstWouldAddEndPuncttrue
\mciteSetBstMidEndSepPunct{\mcitedefaultmidpunct}
{\mcitedefaultendpunct}{\mcitedefaultseppunct}\relax
\EndOfBibitem
\bibitem[Doi \latin{et~al.}(2007)Doi, Ishihara, Mitsushima, Kamiya, and
  Ota]{Doi2007}
Doi,~S.; Ishihara,~A.; Mitsushima,~S.; Kamiya,~N.; Ota,~K. Zirconium-Based
  Compounds for Cathode of Polymer Electrolyte Fuel Cell. \emph{J. Electrochem.
  Soc.} \textbf{2007}, \emph{154}, B362\relax
\mciteBstWouldAddEndPuncttrue
\mciteSetBstMidEndSepPunct{\mcitedefaultmidpunct}
{\mcitedefaultendpunct}{\mcitedefaultseppunct}\relax
\EndOfBibitem
\bibitem[Chisaka \latin{et~al.}(2017)Chisaka, Ishihara, Morioka, Nagai, Yin,
  Ohgi, Matsuzawa, Mitsushima, and Ota]{Chisaka2017}
Chisaka,~M.; Ishihara,~A.; Morioka,~H.; Nagai,~T.; Yin,~S.; Ohgi,~Y.;
  Matsuzawa,~K.; Mitsushima,~S.; Ota,~K. Zirconium Oxynitride-Catalyzed Oxygen
  Reduction Reaction at Polymer Electrolyte Fuel Cell Cathodes. \emph{ACS
  Omega} \textbf{2017}, \emph{2}, 678–684\relax
\mciteBstWouldAddEndPuncttrue
\mciteSetBstMidEndSepPunct{\mcitedefaultmidpunct}
{\mcitedefaultendpunct}{\mcitedefaultseppunct}\relax
\EndOfBibitem
\bibitem[Ishihara \latin{et~al.}(2019)Ishihara, Nagai, Ukita, Arao, Matsumoto,
  Yu, Nakamura, Sekizawa, Takagi, Matsuzawa, Napporn, Mitsushima, Uruga,
  Yokoyama, Iwasawa, Imai, and Ota]{Ishihara2019}
Ishihara,~A. \latin{et~al.}  Emergence of Oxygen Reduction Activity in
  Zirconium Oxide-Based Compounds in Acidic Media: Creation of Active Sites for
  the Oxygen Reduction Reaction. \emph{J. Phys. Chem. C} \textbf{2019},
  \emph{123}, 18150–18159\relax
\mciteBstWouldAddEndPuncttrue
\mciteSetBstMidEndSepPunct{\mcitedefaultmidpunct}
{\mcitedefaultendpunct}{\mcitedefaultseppunct}\relax
\EndOfBibitem
\bibitem[Arashi \latin{et~al.}(2014)Arashi, Seo, Takanabe, Kubota, and
  Domen]{Arashi2014}
Arashi,~T.; Seo,~J.; Takanabe,~K.; Kubota,~J.; Domen,~K. Nb-doped TiO$_2$
  cathode catalysts for oxygen reduction reaction of polymer electrolyte fuel
  cells. \emph{Catal. Today} \textbf{2014}, \emph{233}, 181--186\relax
\mciteBstWouldAddEndPuncttrue
\mciteSetBstMidEndSepPunct{\mcitedefaultmidpunct}
{\mcitedefaultendpunct}{\mcitedefaultseppunct}\relax
\EndOfBibitem
\bibitem[Ishihara \latin{et~al.}(2020)Ishihara, Arao, Matsumoto, Tokai, Nagai,
  Kuroda, Matsuzawa, Imai, Mitsushima, and Ota]{Ishihara2020a}
Ishihara,~A.; Arao,~M.; Matsumoto,~M.; Tokai,~T.; Nagai,~T.; Kuroda,~Y.;
  Matsuzawa,~K.; Imai,~H.; Mitsushima,~S.; Ota,~K. Niobium-added titanium
  oxides powders as non-noble metal cathodes for polymer electrolyte fuel cells
  – Electrochemical evaluation and effect of added amount of niobium.
  \emph{Int. J. Hydrog. Energy} \textbf{2020}, \emph{45}, 5438--5448\relax
\mciteBstWouldAddEndPuncttrue
\mciteSetBstMidEndSepPunct{\mcitedefaultmidpunct}
{\mcitedefaultendpunct}{\mcitedefaultseppunct}\relax
\EndOfBibitem
\bibitem[Ishihara \latin{et~al.}(2016)Ishihara, Hamazaki, Arao, Matsumoto,
  Imai, Kohno, Matsuzawa, Mitsushima, and Ota]{Ishihara2016}
Ishihara,~A.; Hamazaki,~M.; Arao,~M.; Matsumoto,~M.; Imai,~H.; Kohno,~Y.;
  Matsuzawa,~K.; Mitsushima,~S.; Ota,~K. Titanium-Niobium Oxides Mixed with
  Ti$_4$O$_7$ as Precious-Metal- and Carbon-Free Cathodes for Polymer
  Electrolyte Fuel Cells. \emph{J. Electrochem. Soc.} \textbf{2016},
  \emph{163}, F603\relax
\mciteBstWouldAddEndPuncttrue
\mciteSetBstMidEndSepPunct{\mcitedefaultmidpunct}
{\mcitedefaultendpunct}{\mcitedefaultseppunct}\relax
\EndOfBibitem
\bibitem[Ishihara \latin{et~al.}(2018)Ishihara, Wu, Nagai, Ohara, Nakada,
  Matsuzawa, Napporn, Arao, Kuroda, Tominaka, Mitsushima, Imai, and
  Ota]{Ishihara2018}
Ishihara,~A.; Wu,~C.; Nagai,~T.; Ohara,~K.; Nakada,~K.; Matsuzawa,~K.;
  Napporn,~T.; Arao,~M.; Kuroda,~Y.; Tominaka,~S.; Mitsushima,~S.; Imai,~H.;
  Ota,~K. Factors affecting oxygen reduction activity of Nb$_2$O$_5$-doped
  TiO$_2$ using carbon nanotubes as support in acidic solution.
  \emph{Electrochim. Acta} \textbf{2018}, \emph{283}, 1779--1788\relax
\mciteBstWouldAddEndPuncttrue
\mciteSetBstMidEndSepPunct{\mcitedefaultmidpunct}
{\mcitedefaultendpunct}{\mcitedefaultseppunct}\relax
\EndOfBibitem
\bibitem[Ishihara \latin{et~al.}(2020)Ishihara, Tominaka, Mitsushima, Imai,
  Sugino, and Ota]{Ishihara2020b}
Ishihara,~A.; Tominaka,~S.; Mitsushima,~S.; Imai,~H.; Sugino,~O.; Ota,~K.
  Challenge of advanced low temperature fuel cells based on high degree of
  freedom of group 4 and 5 metal oxides. \emph{Curr. Opin. Electrochem.}
  \textbf{2020}, \emph{21}, 234--241\relax
\mciteBstWouldAddEndPuncttrue
\mciteSetBstMidEndSepPunct{\mcitedefaultmidpunct}
{\mcitedefaultendpunct}{\mcitedefaultseppunct}\relax
\EndOfBibitem
\bibitem[Ota \latin{et~al.}(2012)Ota, Ohgi, Matsuzawa, Mitsushima, and
  Ishihara]{Ota2012}
Ota,~K.; Ohgi,~Y.; Matsuzawa,~K.; Mitsushima,~S.; Ishihara,~A. Transition Metal
  Oxide Based Materials for Cathode of Polymer Electrolyte Fuel Cells.
  \emph{ECS Trans.} \textbf{2012}, \emph{45}, 27\relax
\mciteBstWouldAddEndPuncttrue
\mciteSetBstMidEndSepPunct{\mcitedefaultmidpunct}
{\mcitedefaultendpunct}{\mcitedefaultseppunct}\relax
\EndOfBibitem
\bibitem[Ukita \latin{et~al.}(2011)Ukita, Ishihara, Ohgi, Matsuzawa,
  Mitsushima, and Ota]{Ukita2011}
Ukita,~K.; Ishihara,~A.; Ohgi,~Y.; Matsuzawa,~K.; Mitsushima,~S.; Ota,~K.
  Zirconium Oxide-Based Compounds as Non-Pt Cathode for Polymer Electrolyte
  Fuel Cell. \emph{Electrochemistry} \textbf{2011}, \emph{79}, 340--342\relax
\mciteBstWouldAddEndPuncttrue
\mciteSetBstMidEndSepPunct{\mcitedefaultmidpunct}
{\mcitedefaultendpunct}{\mcitedefaultseppunct}\relax
\EndOfBibitem
\bibitem[Yin \latin{et~al.}(2013)Yin, Ishihara, Kohno, Matsuzawa, Mitsushima,
  Ota, Matsumoto, and Imai]{Yin2013}
Yin,~S.; Ishihara,~A.; Kohno,~Y.; Matsuzawa,~K.; Mitsushima,~S.; Ota,~K.;
  Matsumoto,~M.; Imai,~H. Enhancement of Oxygen Reduction Activity of Zirconium
  Oxide-Based Cathode for PEFC. \emph{ECS Trans.} \textbf{2013}, \emph{58},
  1489\relax
\mciteBstWouldAddEndPuncttrue
\mciteSetBstMidEndSepPunct{\mcitedefaultmidpunct}
{\mcitedefaultendpunct}{\mcitedefaultseppunct}\relax
\EndOfBibitem
\bibitem[Bredow and Lerch(2004)Bredow, and Lerch]{Bredow2004}
Bredow,~T.; Lerch,~M. Anion {Distribution} in {Zr2ON2}. \emph{Z. Anorg. Allg.
  Chem.} \textbf{2004}, \emph{630}, 2262--2266\relax
\mciteBstWouldAddEndPuncttrue
\mciteSetBstMidEndSepPunct{\mcitedefaultmidpunct}
{\mcitedefaultendpunct}{\mcitedefaultseppunct}\relax
\EndOfBibitem
\bibitem[Bredow and Lerch(2007)Bredow, and Lerch]{Bredow2007}
Bredow,~T.; Lerch,~M. On the Anion Distribution in Zr$_7$O$_8$N$_4$. \emph{Z.
  Anorg. Allg. Chem.} \textbf{2007}, \emph{633}, 2598--2602\relax
\mciteBstWouldAddEndPuncttrue
\mciteSetBstMidEndSepPunct{\mcitedefaultmidpunct}
{\mcitedefaultendpunct}{\mcitedefaultseppunct}\relax
\EndOfBibitem
\bibitem[Maekawa \latin{et~al.}(2008)Maekawa, Ishihara, Kim, Mitsushima, and
  Ota]{Maekawa2008}
Maekawa,~Y.; Ishihara,~A.; Kim,~J.-H.; Mitsushima,~S.; Ota,~K.-i. Catalytic
  {Activity} of {Zirconium} {Oxynitride} {Prepared} by {Reactive} {Sputtering}
  for {ORR} in {Sulfuric} {Acid}. \emph{Electrochem. Solid-State Lett.}
  \textbf{2008}, \emph{11}, B109\relax
\mciteBstWouldAddEndPuncttrue
\mciteSetBstMidEndSepPunct{\mcitedefaultmidpunct}
{\mcitedefaultendpunct}{\mcitedefaultseppunct}\relax
\EndOfBibitem
\bibitem[Yamamoto \latin{et~al.}(2019)Yamamoto, Kasamatsu, and
  Sugino]{Yamamoto2019}
Yamamoto,~Y.; Kasamatsu,~S.; Sugino,~O. Scaling Relation of Oxygen Reduction
  Reaction Intermediates at Defective TiO$_2$ Surfaces. \emph{J. Phys. Chem. C}
  \textbf{2019}, \emph{123}, 19486–19492\relax
\mciteBstWouldAddEndPuncttrue
\mciteSetBstMidEndSepPunct{\mcitedefaultmidpunct}
{\mcitedefaultendpunct}{\mcitedefaultseppunct}\relax
\EndOfBibitem
\bibitem[Muhammady \latin{et~al.}(2022)Muhammady, Haruyama, Kasamatsu, and
  Sugino]{Muhammady2022a}
Muhammady,~S.; Haruyama,~J.; Kasamatsu,~S.; Sugino,~O. Tuning Oxygen Reduction
  on Monoclinic and Tetragonal Zirconia Surfaces Using Oxygen Vacancy and
  Nitrogen Doping: A Density-Functional Study. \emph{Meet. Abstr.}
  \textbf{2022}, \emph{MA2022-01}, 1517\relax
\mciteBstWouldAddEndPuncttrue
\mciteSetBstMidEndSepPunct{\mcitedefaultmidpunct}
{\mcitedefaultendpunct}{\mcitedefaultseppunct}\relax
\EndOfBibitem
\bibitem[Muhammady \latin{et~al.}(2022)Muhammady, Haruyama, Kasamatsu, and
  Sugino]{Muhammady2022b}
Muhammady,~S.; Haruyama,~J.; Kasamatsu,~S.; Sugino,~O. Effect of Nitrogen
  Doping and Oxygen Vacancy on the Oxygen Reduction Reaction on the Tetragonal
  Zirconia (101) Surface. \emph{J. Phys. Chem. C} \textbf{2022}, \emph{126},
  15662–15670\relax
\mciteBstWouldAddEndPuncttrue
\mciteSetBstMidEndSepPunct{\mcitedefaultmidpunct}
{\mcitedefaultendpunct}{\mcitedefaultseppunct}\relax
\EndOfBibitem
\bibitem[N\"orskov \latin{et~al.}(2004)N\"orskov, Rossmeisl, Logadottir,
  Lindqvist, Kitchin, Bligaard, and J\'onsson]{norskov_origin_2004}
N\"orskov,~J.~K.; Rossmeisl,~J.; Logadottir,~A.; Lindqvist,~L.; Kitchin,~J.~R.;
  Bligaard,~T.; J\'onsson,~H. Origin of the {Overpotential} for {Oxygen}
  {Reduction} at a {Fuel}-{Cell} {Cathode}. \emph{J. Phys. Chem. B}
  \textbf{2004}, \emph{108}, 17886--17892\relax
\mciteBstWouldAddEndPuncttrue
\mciteSetBstMidEndSepPunct{\mcitedefaultmidpunct}
{\mcitedefaultendpunct}{\mcitedefaultseppunct}\relax
\EndOfBibitem
\bibitem[Behler and Cs\'anyi(2021)Behler, and Cs\'anyi]{Behler2021}
Behler,~J.; Cs\'anyi,~G. Machine learning potentials for extended systems: a
  perspective. \emph{Eur. Phys. J.l B} \textbf{2021}, \emph{94}, 142\relax
\mciteBstWouldAddEndPuncttrue
\mciteSetBstMidEndSepPunct{\mcitedefaultmidpunct}
{\mcitedefaultendpunct}{\mcitedefaultseppunct}\relax
\EndOfBibitem
\bibitem[Natarajan and Behler(2016)Natarajan, and Behler]{Natarajan2016}
Natarajan,~S.~K.; Behler,~J. Neural network molecular dynamics simulations of
  solid–liquid interfaces: water at low-index copper surfaces. \emph{Phys.
  Chem. Chem. Phys.} \textbf{2016}, \emph{18}, 28704--28725\relax
\mciteBstWouldAddEndPuncttrue
\mciteSetBstMidEndSepPunct{\mcitedefaultmidpunct}
{\mcitedefaultendpunct}{\mcitedefaultseppunct}\relax
\EndOfBibitem
\bibitem[Natarajan and Behler(2017)Natarajan, and Behler]{Kondati2017}
Natarajan,~S.~K.; Behler,~J. Self-Diffusion of Surface Defects at
  Copper–Water Interfaces. \emph{J. Phys. Chem. C} \textbf{2017}, \emph{121},
  4368–4383\relax
\mciteBstWouldAddEndPuncttrue
\mciteSetBstMidEndSepPunct{\mcitedefaultmidpunct}
{\mcitedefaultendpunct}{\mcitedefaultseppunct}\relax
\EndOfBibitem
\bibitem[Quaranta \latin{et~al.}(2017)Quaranta, Hellstr\"{o}m, and
  Behler]{Quaranta2017}
Quaranta,~V.; Hellstr\"{o}m,~M.; Behler,~J. Proton-Transfer Mechanisms at the
  Water–ZnO Interface: The Role of Presolvation. \emph{J. Phys. Chem. Lett.}
  \textbf{2017}, \emph{8}, 1476–1483\relax
\mciteBstWouldAddEndPuncttrue
\mciteSetBstMidEndSepPunct{\mcitedefaultmidpunct}
{\mcitedefaultendpunct}{\mcitedefaultseppunct}\relax
\EndOfBibitem
\bibitem[Quaranta \latin{et~al.}(2018)Quaranta, Hellstr\"{o}m, Behler,
  Kullgren, Mitev, and Hermansson]{Quaranta2018}
Quaranta,~V.; Hellstr\"{o}m,~M.; Behler,~J.; Kullgren,~J.; Mitev,~P.~D.;
  Hermansson,~K. Maximally resolved anharmonic OH vibrational spectrum of the
  water/ZnO$(10\overline{1}0)$ interface from a high-dimensional neural network
  potential. \emph{J. Chem. Phys.} \textbf{2018}, \emph{148}, 241720\relax
\mciteBstWouldAddEndPuncttrue
\mciteSetBstMidEndSepPunct{\mcitedefaultmidpunct}
{\mcitedefaultendpunct}{\mcitedefaultseppunct}\relax
\EndOfBibitem
\bibitem[Quaranta \latin{et~al.}(2019)Quaranta, Behler, and
  Hellstr\"{o}m]{Quaranta2019}
Quaranta,~V.; Behler,~J.; Hellstr\"{o}m,~M. Structure and Dynamics of the
  Liquid–Water/Zinc-Oxide Interface from Machine Learning Potential
  Simulations. \emph{J. Phys. Chem. C} \textbf{2019}, \emph{123},
  1293–1304\relax
\mciteBstWouldAddEndPuncttrue
\mciteSetBstMidEndSepPunct{\mcitedefaultmidpunct}
{\mcitedefaultendpunct}{\mcitedefaultseppunct}\relax
\EndOfBibitem
\bibitem[Artrith(2019)]{Artrith2019}
Artrith,~N. Machine learning for the modeling of interfaces in energy storage
  and conversion materials. \emph{J. Phys. Energy} \textbf{2019}, \emph{1},
  032002\relax
\mciteBstWouldAddEndPuncttrue
\mciteSetBstMidEndSepPunct{\mcitedefaultmidpunct}
{\mcitedefaultendpunct}{\mcitedefaultseppunct}\relax
\EndOfBibitem
\bibitem[Andrade \latin{et~al.}(2020)Andrade, Ko, Zhang, Car, and
  Selloni]{Andrade2020}
Andrade,~M. F.~C.; Ko,~H.; Zhang,~L.; Car,~R.; Selloni,~A. Free energy of
  proton transfer at the water–TiO$_2$ interface from ab initio deep
  potential molecular dynamics. \emph{Chem. Sci.} \textbf{2020}, \emph{11},
  2335--2341\relax
\mciteBstWouldAddEndPuncttrue
\mciteSetBstMidEndSepPunct{\mcitedefaultmidpunct}
{\mcitedefaultendpunct}{\mcitedefaultseppunct}\relax
\EndOfBibitem
\bibitem[Ghorbanfekr \latin{et~al.}(2020)Ghorbanfekr, Behler, and
  Peeters]{Ghorbanfekr2020}
Ghorbanfekr,~H.; Behler,~J.; Peeters,~F.~M. Insights into Water Permeation
  through hBN Nanocapillaries by Ab Initio Machine Learning Molecular Dynamics
  Simulations. \emph{J. Phys. Chem. Lett.} \textbf{2020}, \emph{11},
  7363–7370\relax
\mciteBstWouldAddEndPuncttrue
\mciteSetBstMidEndSepPunct{\mcitedefaultmidpunct}
{\mcitedefaultendpunct}{\mcitedefaultseppunct}\relax
\EndOfBibitem
\bibitem[Eckhoff and Behler(2021)Eckhoff, and Behler]{Eckhoff2021}
Eckhoff,~M.; Behler,~J. Insights into lithium manganese oxide–water
  interfaces using machine learning potentials. \emph{J. Chem. Phys.}
  \textbf{2021}, \emph{155}, 244703\relax
\mciteBstWouldAddEndPuncttrue
\mciteSetBstMidEndSepPunct{\mcitedefaultmidpunct}
{\mcitedefaultendpunct}{\mcitedefaultseppunct}\relax
\EndOfBibitem
\bibitem[Schran \latin{et~al.}(2021)Schran, Thiemann, Rowe, M\"{u}ller,
  Marsalek, and Michaelides]{Schran2021}
Schran,~C.; Thiemann,~F.~L.; Rowe,~P.; M\"{u}ller,~E.~A.; Marsalek,~O.;
  Michaelides,~A. Machine learning potentials for complex aqueous systems made
  simple. \emph{Proc. Natl. Acad. Sci. U.S.A.} \textbf{2021}, \emph{118},
  e2110077118\relax
\mciteBstWouldAddEndPuncttrue
\mciteSetBstMidEndSepPunct{\mcitedefaultmidpunct}
{\mcitedefaultendpunct}{\mcitedefaultseppunct}\relax
\EndOfBibitem
\bibitem[Mikkelsen \latin{et~al.}(2021)Mikkelsen, Schi{\o}tz, Vegge, and
  Jacobsen]{Mikkelsen2021}
Mikkelsen,~A. E.~G.; Schi{\o}tz,~J.; Vegge,~T.; Jacobsen,~K.~W. Is the
  water/Pt(111) interface ordered at room temperature? \emph{J. Chem. Phys.}
  \textbf{2021}, \emph{155}, 224701\relax
\mciteBstWouldAddEndPuncttrue
\mciteSetBstMidEndSepPunct{\mcitedefaultmidpunct}
{\mcitedefaultendpunct}{\mcitedefaultseppunct}\relax
\EndOfBibitem
\bibitem[Schienbein and Blumberger(2022)Schienbein, and
  Blumberger]{Schienbein2022}
Schienbein,~P.; Blumberger,~J. Nanosecond solvation dynamics of the
  hematite/liquid water interface at hybrid DFT accuracy using committee neural
  network potentials. \emph{Phys. Chem. Chem. Phys.} \textbf{2022}, \emph{24},
  15365--15375\relax
\mciteBstWouldAddEndPuncttrue
\mciteSetBstMidEndSepPunct{\mcitedefaultmidpunct}
{\mcitedefaultendpunct}{\mcitedefaultseppunct}\relax
\EndOfBibitem
\bibitem[Mikkelsen \latin{et~al.}(2022)Mikkelsen, Kristoffersen, Schiøtz,
  Vegge, Hansen, and Jacobsen]{Mikkelsen2022}
Mikkelsen,~A. E.~G.; Kristoffersen,~H.~H.; Schiøtz,~J.; Vegge,~T.;
  Hansen,~H.~A.; Jacobsen,~K.~W. Structure and energetics of liquid
  water–hydroxyl layers on Pt(111). \emph{Phys. Chem. Chem. Phys.}
  \textbf{2022}, \emph{24}, 9885--9890\relax
\mciteBstWouldAddEndPuncttrue
\mciteSetBstMidEndSepPunct{\mcitedefaultmidpunct}
{\mcitedefaultendpunct}{\mcitedefaultseppunct}\relax
\EndOfBibitem
\bibitem[Fan \latin{et~al.}(2023)Fan, Wen, Zhuang, and Cheng]{Fan2023}
Fan,~X.; Wen,~X.; Zhuang,~Y.; Cheng,~J. Molecular insight into the
  GaP(110)–water interface using machine learning accelerated molecular
  dynamics. \emph{J. Energy Chem.} \textbf{2023}, \relax
\mciteBstWouldAddEndPunctfalse
\mciteSetBstMidEndSepPunct{\mcitedefaultmidpunct}
{}{\mcitedefaultseppunct}\relax
\EndOfBibitem
\bibitem[Wen \latin{et~al.}(2023)Wen, Andrade, Liu, and Selloni]{Wen2023a}
Wen,~B.; Andrade,~M. F.~C.; Liu,~L.; Selloni,~A. Water dissociation at the
  water–rutile TiO$_2$(110) interface from ab initio-based deep neural
  network simulations. \emph{Proc. Natl. Acad. Sci. U.S.A.} \textbf{2023},
  \emph{120}, e2212250120\relax
\mciteBstWouldAddEndPuncttrue
\mciteSetBstMidEndSepPunct{\mcitedefaultmidpunct}
{\mcitedefaultendpunct}{\mcitedefaultseppunct}\relax
\EndOfBibitem
\bibitem[Zeng \latin{et~al.}()Zeng, Wodaczek, Liu, Stein, Hutter, Chen, and
  Cheng]{Zeng2023}
Zeng,~Z.; Wodaczek,~F.; Liu,~K.; Stein,~F.; Hutter,~J.; Chen,~J.; Cheng,~B.
  Water dissociation on pristine low-index TiO$_2$ surfaces.
  \emph{arXiv:2303.07433 [cond-mat.mtrl-sci]} \relax
\mciteBstWouldAddEndPunctfalse
\mciteSetBstMidEndSepPunct{\mcitedefaultmidpunct}
{}{\mcitedefaultseppunct}\relax
\EndOfBibitem
\bibitem[Artrith \latin{et~al.}(2017)Artrith, Urban, and Ceder]{Artrith2017}
Artrith,~N.; Urban,~A.; Ceder,~G. Efficient and accurate machine-learning
  interpolation of atomic energies in compositions with many species.
  \emph{Phys. Rev. B} \textbf{2017}, \emph{96}, 014112\relax
\mciteBstWouldAddEndPuncttrue
\mciteSetBstMidEndSepPunct{\mcitedefaultmidpunct}
{\mcitedefaultendpunct}{\mcitedefaultseppunct}\relax
\EndOfBibitem
\bibitem[Kasamatsu \latin{et~al.}(2022)Kasamatsu, Motoyama, Yoshimi, Matsumoto,
  Kuwabara, and Ogawa]{Kasamatsu2022}
Kasamatsu,~S.; Motoyama,~Y.; Yoshimi,~K.; Matsumoto,~U.; Kuwabara,~A.;
  Ogawa,~T. Facilitating ab initio configurational sampling of multicomponent
  solids using an on-lattice neural network model and active learning. \emph{J.
  Chem. Phys.} \textbf{2022}, \emph{157}, 104114, Publisher: American Institute
  of Physics\relax
\mciteBstWouldAddEndPuncttrue
\mciteSetBstMidEndSepPunct{\mcitedefaultmidpunct}
{\mcitedefaultendpunct}{\mcitedefaultseppunct}\relax
\EndOfBibitem
\bibitem[Hoshino \latin{et~al.}(2023)Hoshino, Kasamatsu, Hyodo, Yamamoto,
  Setoyama, Okajima, and Yamazaki]{Hoshino2023}
Hoshino,~K.; Kasamatsu,~S.; Hyodo,~J.; Yamamoto,~K.; Setoyama,~H.; Okajima,~T.;
  Yamazaki,~Y. Probing {Local} {Environments} of {Oxygen} {Vacancies}
  {Responsible} for {Hydration} in {Sc}-{Doped} {Barium} {Zirconates} at
  {Elevated} {Temperatures}: {In} {Situ} {X}-ray {Absorption} {Spectroscopy},
  {Thermogravimetry}, and {Active} {Learning} {Ab} {Initio} {Replica}
  {Exchange} {Monte} {Carlo} {Simulations}. \emph{Chem. Mater.} \textbf{2023},
  \emph{35}, 2289--2301, Publisher: American Chemical Society\relax
\mciteBstWouldAddEndPuncttrue
\mciteSetBstMidEndSepPunct{\mcitedefaultmidpunct}
{\mcitedefaultendpunct}{\mcitedefaultseppunct}\relax
\EndOfBibitem
\bibitem[Behler and Parrinello(2007)Behler, and Parrinello]{Behler2007}
Behler,~J.; Parrinello,~M. Generalized Neural-Network Representation of
  High-Dimensional Potential-Energy Surfaces. \emph{Phys. Rev. Lett.}
  \textbf{2007}, \emph{98}, 14601\relax
\mciteBstWouldAddEndPuncttrue
\mciteSetBstMidEndSepPunct{\mcitedefaultmidpunct}
{\mcitedefaultendpunct}{\mcitedefaultseppunct}\relax
\EndOfBibitem
\bibitem[Behler(2015)]{Behler2015}
Behler,~J. Constructing high-dimensional neural network potentials: {A}
  tutorial review. \emph{Int. J. Quant. Chem.} \textbf{2015}, \emph{115},
  1032--1050\relax
\mciteBstWouldAddEndPuncttrue
\mciteSetBstMidEndSepPunct{\mcitedefaultmidpunct}
{\mcitedefaultendpunct}{\mcitedefaultseppunct}\relax
\EndOfBibitem
\bibitem[Momma and Izumi(2011)Momma, and Izumi]{Momma2011}
Momma,~K.; Izumi,~F. VESTA 3 for three-dimensional visualization of crystal,
  volumetric and morphology data. \emph{J. Appl. Cryst.} \textbf{2011},
  \emph{44}, 1272--1276\relax
\mciteBstWouldAddEndPuncttrue
\mciteSetBstMidEndSepPunct{\mcitedefaultmidpunct}
{\mcitedefaultendpunct}{\mcitedefaultseppunct}\relax
\EndOfBibitem
\bibitem[Kresse and Furthm\"{u}ller(1996)Kresse, and
  Furthm\"{u}ller]{Kresse1996}
Kresse,~G.; Furthm\"{u}ller,~J. Efficient iterative schemes for \textit{ab
  initio} total-energy calculations using a plane-wave basis set. \emph{Phys.
  Rev. B} \textbf{1996}, \emph{54}, 11169\relax
\mciteBstWouldAddEndPuncttrue
\mciteSetBstMidEndSepPunct{\mcitedefaultmidpunct}
{\mcitedefaultendpunct}{\mcitedefaultseppunct}\relax
\EndOfBibitem
\bibitem[VAS()]{VASP}
\url{https://www.vasp.at/}\relax
\mciteBstWouldAddEndPuncttrue
\mciteSetBstMidEndSepPunct{\mcitedefaultmidpunct}
{\mcitedefaultendpunct}{\mcitedefaultseppunct}\relax
\EndOfBibitem
\bibitem[Bl\"{o}chl(1994)]{Blochl1994}
Bl\"{o}chl,~P.~E. Projector augmented-wave method. \emph{Phys. Rev. B}
  \textbf{1994}, \emph{50}, 17953\relax
\mciteBstWouldAddEndPuncttrue
\mciteSetBstMidEndSepPunct{\mcitedefaultmidpunct}
{\mcitedefaultendpunct}{\mcitedefaultseppunct}\relax
\EndOfBibitem
\bibitem[Perdew \latin{et~al.}(1996)Perdew, Burke, and Ernzerhof]{Perdew1996}
Perdew,~J.~P.; Burke,~K.; Ernzerhof,~M. Generalized Gradient Approximation Made
  Simple. \emph{Phys. Rev. Lett.} \textbf{1996}, \emph{77}, 3865\relax
\mciteBstWouldAddEndPuncttrue
\mciteSetBstMidEndSepPunct{\mcitedefaultmidpunct}
{\mcitedefaultendpunct}{\mcitedefaultseppunct}\relax
\EndOfBibitem
\bibitem[Lee \latin{et~al.}(2019)Lee, Yoo, Jeong, and Han]{Lee2019}
Lee,~K.; Yoo,~D.; Jeong,~W.; Han,~S. SIMPLE-NN: An efficient package for
  training and executing neural-network interatomic potentials. \emph{Comput.
  Phys. Commun.} \textbf{2019}, \emph{242}, 95--103\relax
\mciteBstWouldAddEndPuncttrue
\mciteSetBstMidEndSepPunct{\mcitedefaultmidpunct}
{\mcitedefaultendpunct}{\mcitedefaultseppunct}\relax
\EndOfBibitem
\bibitem[SIM()]{SIMPLENNv2}
\url{https://simple-nn-v2.readthedocs.io/en/latest/}\relax
\mciteBstWouldAddEndPuncttrue
\mciteSetBstMidEndSepPunct{\mcitedefaultmidpunct}
{\mcitedefaultendpunct}{\mcitedefaultseppunct}\relax
\EndOfBibitem
\bibitem[Plimpton(1995)]{Plimpton1995}
Plimpton,~S. Fast Parallel Algorithms for Short-Range Molecular Dynamics.
  \emph{J. Comput. Phys.} \textbf{1995}, \emph{117}, 1--19\relax
\mciteBstWouldAddEndPuncttrue
\mciteSetBstMidEndSepPunct{\mcitedefaultmidpunct}
{\mcitedefaultendpunct}{\mcitedefaultseppunct}\relax
\EndOfBibitem
\bibitem[LAM()]{LAMMPS}
\url{https://www.lammps.org/}\relax
\mciteBstWouldAddEndPuncttrue
\mciteSetBstMidEndSepPunct{\mcitedefaultmidpunct}
{\mcitedefaultendpunct}{\mcitedefaultseppunct}\relax
\EndOfBibitem
\bibitem[Tuckerman \latin{et~al.}(2002)Tuckerman, Marx, and
  Parrinello]{Tuckerman2002}
Tuckerman,~M.~E.; Marx,~D.; Parrinello,~M. The nature and transport mechanism
  of hydrated hydroxide ions in aqueous solution. \emph{Nature} \textbf{2002},
  \emph{417}, 925–929\relax
\mciteBstWouldAddEndPuncttrue
\mciteSetBstMidEndSepPunct{\mcitedefaultmidpunct}
{\mcitedefaultendpunct}{\mcitedefaultseppunct}\relax
\EndOfBibitem
\bibitem[Laage and Hynes(2008)Laage, and Hynes]{Laage2008}
Laage,~D.; Hynes,~J.~T. On the Residence Time for Water in a Solute Hydration
  Shell: Application to Aqueous Halide Solutions. \emph{J. Phys. Chem. B}
  \textbf{2008}, \emph{112}, 7697–7701\relax
\mciteBstWouldAddEndPuncttrue
\mciteSetBstMidEndSepPunct{\mcitedefaultmidpunct}
{\mcitedefaultendpunct}{\mcitedefaultseppunct}\relax
\EndOfBibitem
\bibitem[Hellstr\"{o}m \latin{et~al.}(2019)Hellstr\"{o}m, Quaranta, and
  Behler]{Hellstrom2019}
Hellstr\"{o}m,~M.; Quaranta,~V.; Behler,~J. One-dimensional vs. two-dimensional
  proton transport processes at solid–liquid zinc-oxide–water interfaces.
  \emph{Chem. Sci.} \textbf{2019}, \emph{10}, 1232--1243\relax
\mciteBstWouldAddEndPuncttrue
\mciteSetBstMidEndSepPunct{\mcitedefaultmidpunct}
{\mcitedefaultendpunct}{\mcitedefaultseppunct}\relax
\EndOfBibitem
\bibitem[Mitev \latin{et~al.}(2015)Mitev, Eriksson, Boily, and
  Hermansson]{Mitev2015}
Mitev,~P.~D.; Eriksson,~A.; Boily,~J.-F.; Hermansson,~K. Vibrational models for
  a crystal with 36 water molecules in the unit cell: IR spectra from
  experiment and calculation. \emph{Phys. Chem. Chem. Phys.} \textbf{2015},
  \emph{17}, 10520--10531\relax
\mciteBstWouldAddEndPuncttrue
\mciteSetBstMidEndSepPunct{\mcitedefaultmidpunct}
{\mcitedefaultendpunct}{\mcitedefaultseppunct}\relax
\EndOfBibitem
\bibitem[Pejov \latin{et~al.}(2010)Pejov, ngberg, and Hermansson]{Pejov2010}
Pejov,~L.; ngberg,~D.~S.; Hermansson,~K. $Al^{3+}$, $Ca^{2+}$, $Mg^{2+}$, and
  $Li^{+}$ in aqueous solution: Calculated first-shell anharmonic OH vibrations
  at 300 K. \emph{J. Chem. Phys.} \textbf{2010}, \emph{133}, 174513\relax
\mciteBstWouldAddEndPuncttrue
\mciteSetBstMidEndSepPunct{\mcitedefaultmidpunct}
{\mcitedefaultendpunct}{\mcitedefaultseppunct}\relax
\EndOfBibitem
\bibitem[Lill \latin{et~al.}(1982)Lill, Parker, and Light]{Lill1982}
Lill,~J.~V.; Parker,~G.~A.; Light,~J.~C. Discrete variable representations and
  sudden models in quantum scattering theory. \emph{Chem. Phys. Lett.}
  \textbf{1982}, \emph{89}, 483--489\relax
\mciteBstWouldAddEndPuncttrue
\mciteSetBstMidEndSepPunct{\mcitedefaultmidpunct}
{\mcitedefaultendpunct}{\mcitedefaultseppunct}\relax
\EndOfBibitem
\bibitem[Light \latin{et~al.}(1985)Light, Hamilton, and Lill]{Light1985}
Light,~J.~C.; Hamilton,~I.~P.; Lill,~J.~V. Generalized discrete variable
  approximation in quantum mechanics. \emph{J. Chem. Phys.} \textbf{1985},
  \emph{82}, 14001409\relax
\mciteBstWouldAddEndPuncttrue
\mciteSetBstMidEndSepPunct{\mcitedefaultmidpunct}
{\mcitedefaultendpunct}{\mcitedefaultseppunct}\relax
\EndOfBibitem
\bibitem[Ba\u{c}i\'{c} and Light(1989)Ba\u{c}i\'{c}, and Light]{Bacic1989}
Ba\u{c}i\'{c},~Z.; Light,~J.~C. Theoretical Methods for Rovibrational States of
  Floppy Molecules. \emph{Annu. Rev. Phys. Chem.} \textbf{1989}, \emph{40},
  469--498\relax
\mciteBstWouldAddEndPuncttrue
\mciteSetBstMidEndSepPunct{\mcitedefaultmidpunct}
{\mcitedefaultendpunct}{\mcitedefaultseppunct}\relax
\EndOfBibitem
\bibitem[Stancik and Brauns(2008)Stancik, and Brauns]{Stancik2008}
Stancik,~A.~L.; Brauns,~E.~B. A simple asymmetric lineshape for fitting
  infrared absorption spectra. \emph{Vib. Spectrosc.} \textbf{2008}, \emph{47},
  66--69\relax
\mciteBstWouldAddEndPuncttrue
\mciteSetBstMidEndSepPunct{\mcitedefaultmidpunct}
{\mcitedefaultendpunct}{\mcitedefaultseppunct}\relax
\EndOfBibitem
\bibitem[Luzar and Chandler(1996)Luzar, and Chandler]{Luzar1996}
Luzar,~A.; Chandler,~D. Effect of Environment on Hydrogen Bond Dynamics in
  Liquid Water. \emph{Phys. Rev. Lett.} \textbf{1996}, \emph{76},
  928--931\relax
\mciteBstWouldAddEndPuncttrue
\mciteSetBstMidEndSepPunct{\mcitedefaultmidpunct}
{\mcitedefaultendpunct}{\mcitedefaultseppunct}\relax
\EndOfBibitem
\bibitem[Benedict \latin{et~al.}(1956)Benedict, Gailar, and
  Plyler]{Benedict1956}
Benedict,~W.~S.; Gailar,~N.; Plyler,~E.~K. Rotation‐Vibration Spectra of
  Deuterated Water Vapor. \emph{J. Chem. Phys.} \textbf{1956}, \emph{24},
  1139--1165\relax
\mciteBstWouldAddEndPuncttrue
\mciteSetBstMidEndSepPunct{\mcitedefaultmidpunct}
{\mcitedefaultendpunct}{\mcitedefaultseppunct}\relax
\EndOfBibitem
\bibitem[Fecko \latin{et~al.}(2003)Fecko, Eaves, Loparo, Tokmakoff, and
  Geissler]{Fecko2003}
Fecko,~C.~J.; Eaves,~J.~D.; Loparo,~J.~J.; Tokmakoff,~A.; Geissler,~P.~L.
  Ultrafast Hydrogen-Bond Dynamics in the Infrared Spectroscopy of Water.
  \emph{Science} \textbf{2003}, \emph{301}, 1698--1702\relax
\mciteBstWouldAddEndPuncttrue
\mciteSetBstMidEndSepPunct{\mcitedefaultmidpunct}
{\mcitedefaultendpunct}{\mcitedefaultseppunct}\relax
\EndOfBibitem
\bibitem[Owrutsky \latin{et~al.}(1985)Owrutsky, Rosenbaum, Tack, and
  Saykally]{Owrutsky1985}
Owrutsky,~J.~C.; Rosenbaum,~N.~H.; Tack,~L.~M.; Saykally,~R.~J. The
  vibration‐rotation spectrum of the hydroxide anion (OH$^{-}$). \emph{J.
  Chem. Phys.} \textbf{1985}, \emph{83}, 5338--5339\relax
\mciteBstWouldAddEndPuncttrue
\mciteSetBstMidEndSepPunct{\mcitedefaultmidpunct}
{\mcitedefaultendpunct}{\mcitedefaultseppunct}\relax
\EndOfBibitem
\bibitem[Corridoni \latin{et~al.}(2007)Corridoni, Sodo, Bruni, Ricci, and
  Nardone]{Corridoni2007}
Corridoni,~T.; Sodo,~A.; Bruni,~F.; Ricci,~M.~A.; Nardone,~M. Probing water
  dynamics with OH$^{-}$. \emph{Chem. Phys.} \textbf{2007}, \emph{336},
  183--187\relax
\mciteBstWouldAddEndPuncttrue
\mciteSetBstMidEndSepPunct{\mcitedefaultmidpunct}
{\mcitedefaultendpunct}{\mcitedefaultseppunct}\relax
\EndOfBibitem
\bibitem[Ando \latin{et~al.}(2013)Ando, Gohda, and Tsuneyuki]{Ando2013}
Ando,~Y.; Gohda,~Y.; Tsuneyuki,~S. \textit{Ab initio} molecular dynamics study
  of the Helmholtz layer formed on solid–liquid interfaces and its
  capacitance. \emph{Chem. Phys. Lett.} \textbf{2013}, \emph{556}, 9--12\relax
\mciteBstWouldAddEndPuncttrue
\mciteSetBstMidEndSepPunct{\mcitedefaultmidpunct}
{\mcitedefaultendpunct}{\mcitedefaultseppunct}\relax
\EndOfBibitem
\bibitem[Ikeda(2014)]{Ikeda2014}
Ikeda,~T. First principles molecular dynamics study of interlayer water and
  cations in vermiculite. \emph{Clay Sci.} \textbf{2014}, \emph{18},
  23--31\relax
\mciteBstWouldAddEndPuncttrue
\mciteSetBstMidEndSepPunct{\mcitedefaultmidpunct}
{\mcitedefaultendpunct}{\mcitedefaultseppunct}\relax
\EndOfBibitem
\bibitem[Li \latin{et~al.}(2020)Li, Yang, Wang, and Feng]{Li2020}
Li,~D.; Yang,~Y.; Wang,~X.; Feng,~G. Electrical Double Layer of Linear
  Tricationic Ionic Liquids at Graphite Electrode. \emph{J. Phys. Chem. C}
  \textbf{2020}, \emph{124}, 15723–15729\relax
\mciteBstWouldAddEndPuncttrue
\mciteSetBstMidEndSepPunct{\mcitedefaultmidpunct}
{\mcitedefaultendpunct}{\mcitedefaultseppunct}\relax
\EndOfBibitem
\bibitem[Huang \latin{et~al.}(2021)Huang, Zhao, Liu, Hao, Tang, Hu, Liu, and
  Chen]{Huang2021}
Huang,~C.; Zhao,~X.; Liu,~S.; Hao,~Y.; Tang,~Q.; Hu,~A.; Liu,~Z.; Chen,~X.
  Stabilizing Zinc Anodes by Regulating the Electrical Double Layer with
  Saccharin Anions. \emph{Adv. Mater.} \textbf{2021}, \emph{33}, 2100445\relax
\mciteBstWouldAddEndPuncttrue
\mciteSetBstMidEndSepPunct{\mcitedefaultmidpunct}
{\mcitedefaultendpunct}{\mcitedefaultseppunct}\relax
\EndOfBibitem
\bibitem[Shin \latin{et~al.}(2022)Shin, Kim, Bae, Ringe, Choi, Lim, Choi, and
  Kim]{Shin2022}
Shin,~S.; Kim,~D.~H.; Bae,~G.; Ringe,~S.; Choi,~H.; Lim,~H.; Choi,~C.~H.;
  Kim,~H. On the importance of the electric double layer structure in aqueous
  electrocatalysis. \emph{Nat. Commun.} \textbf{2022}, \emph{13}, 174\relax
\mciteBstWouldAddEndPuncttrue
\mciteSetBstMidEndSepPunct{\mcitedefaultmidpunct}
{\mcitedefaultendpunct}{\mcitedefaultseppunct}\relax
\EndOfBibitem
\bibitem[Otani \latin{et~al.}(2008)Otani, Hamada, Sugino, Morikawa, Okamoto,
  and Ikeshoji]{Otani2008}
Otani,~M.; Hamada,~I.; Sugino,~O.; Morikawa,~Y.; Okamoto,~Y.; Ikeshoji,~T.
  Electrode Dynamics from First Principles. \emph{J. Phys. Soc. Jpn.}
  \textbf{2008}, \emph{77}, 024802\relax
\mciteBstWouldAddEndPuncttrue
\mciteSetBstMidEndSepPunct{\mcitedefaultmidpunct}
{\mcitedefaultendpunct}{\mcitedefaultseppunct}\relax
\EndOfBibitem
\bibitem[H\"{o}rmann \latin{et~al.}(2019)H\"{o}rmann, Guo, Ambrosio, Andreussi,
  Pasquarello, and Marzari]{Hormann2019}
H\"{o}rmann,~N.~G.; Guo,~Z.; Ambrosio,~F.; Andreussi,~O.; Pasquarello,~A.;
  Marzari,~N. Absolute band alignment at semiconductor-water interfaces using
  explicit and implicit descriptions for liquid water. \emph{npj Comput.
  Mater.} \textbf{2019}, \emph{5}, 100\relax
\mciteBstWouldAddEndPuncttrue
\mciteSetBstMidEndSepPunct{\mcitedefaultmidpunct}
{\mcitedefaultendpunct}{\mcitedefaultseppunct}\relax
\EndOfBibitem
\bibitem[Guo \latin{et~al.}(2019)Guo, Li, Clark, and Robertson]{Guo2019}
Guo,~Y.; Li,~H.; Clark,~S.~J.; Robertson,~J. Band Offset Models of
  Three-Dimensionally Bonded Semiconductors and Insulators. \emph{J. Phys.
  Chem. C} \textbf{2019}, \emph{123}, 5562–5570\relax
\mciteBstWouldAddEndPuncttrue
\mciteSetBstMidEndSepPunct{\mcitedefaultmidpunct}
{\mcitedefaultendpunct}{\mcitedefaultseppunct}\relax
\EndOfBibitem
\bibitem[Sun \latin{et~al.}(2021)Sun, Lu, Wang, and Huang]{Sun2021}
Sun,~M.; Lu,~Q.; Wang,~Z.~L.; Huang,~B. Understanding contact electrification
  at liquid–solid interfaces from surface electronic structure. \emph{Nat.
  Commun.} \textbf{2021}, \emph{12}, 1752\relax
\mciteBstWouldAddEndPuncttrue
\mciteSetBstMidEndSepPunct{\mcitedefaultmidpunct}
{\mcitedefaultendpunct}{\mcitedefaultseppunct}\relax
\EndOfBibitem
\bibitem[Wen \latin{et~al.}(2023)Wen, Fan, Jin, and Cheng]{Wen2023b}
Wen,~X.; Fan,~X.; Jin,~X.; Cheng,~J. Band Alignment of 2D Material–Water
  Interfaces. \emph{J. Phys. Chem. C} \textbf{2023}, \emph{127},
  4132–4143\relax
\mciteBstWouldAddEndPuncttrue
\mciteSetBstMidEndSepPunct{\mcitedefaultmidpunct}
{\mcitedefaultendpunct}{\mcitedefaultseppunct}\relax
\EndOfBibitem
\end{mcitethebibliography}

\begin{suppinfo}

\begin{table}
  \caption{
The configurations of atoms (Zr, N, O) and vacancies (V) 
in the unit cell of Zr$_7$O$_8$N$_4$ and ZrO$_2$
are shown in coordinates relative to the lattice vector.
}
  \label{tbl:configuration1}
  \begin{tabular}{rrrr}
    \hline
    n$_1$    & n$_2$    & n$_3$    & atom \\
    \hline
    0.333333 & 0.333333 & 0.118933 & Zr   \\
    0.476190 & 0.619048 & 0.118933 & Zr   \\
    0.619048 & 0.904762 & 0.118933 & Zr   \\
    0.904762 & 0.476190 & 0.118933 & Zr   \\
    0.047619 & 0.761905 & 0.118933 & Zr   \\
    0.761905 & 0.190476 & 0.118933 & Zr   \\
    0.190476 & 0.047619 & 0.118933 & Zr   \\
    0.809524 & 0.952381 & 0.237867 & Zr   \\
    0.238095 & 0.809524 & 0.237867 & Zr   \\
    0.523809 & 0.380952 & 0.237867 & Zr   \\
    0.666667 & 0.666667 & 0.237867 & Zr   \\
    0.952381 & 0.238095 & 0.237867 & Zr   \\
    0.095238 & 0.523809 & 0.237867 & Zr   \\
    0.380952 & 0.095238 & 0.237867 & Zr   \\
    0.571429 & 0.142857 & 0.356800 & Zr   \\
    0.857143 & 0.714286 & 0.356800 & Zr   \\
    0.714286 & 0.428571 & 0.356800 & Zr   \\
    0.000000 & 0.000000 & 0.356800 & Zr   \\
    0.428571 & 0.857143 & 0.356800 & Zr   \\
    0.285714 & 0.571429 & 0.356800 & Zr   \\
    0.142857 & 0.285714 & 0.356800 & Zr   \\
    0.904762 & 0.476190 & 0.475733 & Zr   \\
    0.761905 & 0.190476 & 0.475733 & Zr   \\
    0.190476 & 0.047619 & 0.475733 & Zr   \\
    0.619048 & 0.904762 & 0.475733 & Zr   \\
    0.476190 & 0.619048 & 0.475733 & Zr   \\
    0.047619 & 0.761905 & 0.475733 & Zr   \\
    0.333333 & 0.333333 & 0.475733 & Zr   \\
    0.952381 & 0.238095 & 0.594667 & Zr   \\
    0.238095 & 0.809524 & 0.594667 & Zr   \\
    0.095238 & 0.523809 & 0.594667 & Zr   \\
    0.523809 & 0.380952 & 0.594667 & Zr   \\
    0.666667 & 0.666667 & 0.594667 & Zr   \\
    0.809524 & 0.952381 & 0.594667 & Zr   \\
    0.380952 & 0.095238 & 0.594667 & Zr   \\
    \hline
  \end{tabular}
\end{table}

\begin{table}
  \caption{
Continued from Table~\ref{tbl:configuration1}
}
  \label{tbl:configuration2}
  \begin{tabular}{rrrr}
    \hline
    n$_1$    & n$_2$    & n$_3$    & atom \\
    \hline
    0.000000 & 0.000000 & 0.089200 & O    \\
    0.571429 & 0.142857 & 0.089200 & O    \\
    0.857143 & 0.714286 & 0.089200 & O    \\
    0.714286 & 0.428571 & 0.089200 & O    \\
    0.428571 & 0.857143 & 0.089200 & V(O) \\
    0.285714 & 0.571429 & 0.089200 & O    \\
    0.142857 & 0.285714 & 0.089200 & O    \\
    0.380952 & 0.095238 & 0.148667 & O    \\
    0.952381 & 0.238095 & 0.148667 & N(O) \\
    0.238095 & 0.809524 & 0.148667 & O    \\
    0.095238 & 0.523809 & 0.148667 & V(O) \\
    0.809524 & 0.952381 & 0.148667 & N(O) \\
    0.666667 & 0.666667 & 0.148667 & O    \\
    0.523809 & 0.380952 & 0.148667 & O    \\
    0.047619 & 0.761905 & 0.208133 & O    \\
    0.333333 & 0.333333 & 0.208133 & N(O) \\
    0.476190 & 0.619048 & 0.208133 & N(O) \\
    0.904762 & 0.476190 & 0.208133 & O    \\
    0.761905 & 0.190476 & 0.208133 & V(O) \\
    0.190476 & 0.047619 & 0.208133 & N(O) \\
    0.619048 & 0.904762 & 0.208133 & N(O) \\
    0.000000 & 0.000000 & 0.267600 & O    \\
    0.571429 & 0.142857 & 0.267600 & N(O) \\
    0.714286 & 0.428571 & 0.267600 & N(O) \\
    0.857143 & 0.714286 & 0.267600 & O    \\
    0.142857 & 0.285714 & 0.267600 & O    \\
    0.285714 & 0.571429 & 0.267600 & O    \\
    0.428571 & 0.857143 & 0.267600 & V(O) \\
    0.952381 & 0.238095 & 0.327067 & O    \\
    0.380952 & 0.095238 & 0.327067 & O    \\
    0.523809 & 0.380952 & 0.327067 & O    \\
    0.238095 & 0.809524 & 0.327067 & O    \\
    0.095238 & 0.523809 & 0.327067 & V(O) \\
    0.666667 & 0.666667 & 0.327067 & N(O) \\
    0.809524 & 0.952381 & 0.327067 & O    \\
    \hline
  \end{tabular}
\end{table}

\begin{table}
  \caption{
Continued from Table~\ref{tbl:configuration2}
}
  \label{tbl:configuration3}
  \begin{tabular}{rrrr}
    \hline
    n$_1$    & n$_2$    & n$_3$    & atom \\
    \hline
    0.190476 & 0.047619 & 0.386533 & O    \\
    0.761905 & 0.190476 & 0.386533 & V(O) \\
    0.904762 & 0.476190 & 0.386533 & N(O) \\
    0.047619 & 0.761905 & 0.386533 & O    \\
    0.333333 & 0.333333 & 0.386533 & O    \\
    0.619048 & 0.904762 & 0.386533 & N(O) \\
    0.476190 & 0.619048 & 0.386533 & N(O) \\
    0.714286 & 0.428571 & 0.446000 & O    \\
    0.571429 & 0.142857 & 0.446000 & O    \\
    0.000000 & 0.000000 & 0.446000 & O    \\
    0.142857 & 0.285714 & 0.446000 & O    \\
    0.857143 & 0.714286 & 0.446000 & N(O) \\
    0.428571 & 0.857143 & 0.446000 & V(O) \\
    0.285714 & 0.571429 & 0.446000 & N(O) \\
    0.095238 & 0.523809 & 0.505467 & V(O) \\
    0.952381 & 0.238095 & 0.505467 & N(O) \\
    0.380952 & 0.095238 & 0.505467 & N(O) \\
    0.238095 & 0.809524 & 0.505467 & N(O) \\
    0.809524 & 0.952381 & 0.505467 & O    \\
    0.523809 & 0.380952 & 0.505467 & O    \\
    0.666667 & 0.666667 & 0.505467 & N(O) \\
    0.190476 & 0.047619 & 0.564933 & O    \\
    0.761905 & 0.190476 & 0.564933 & V(O) \\
    0.333333 & 0.333333 & 0.564933 & N(O) \\
    0.476190 & 0.619048 & 0.564933 & O    \\
    0.047619 & 0.761905 & 0.564933 & N(O) \\
    0.904762 & 0.476190 & 0.564933 & O    \\
    0.619048 & 0.904762 & 0.564933 & O    \\
    0.714286 & 0.428571 & 0.624400 & O    \\
    0.857143 & 0.714286 & 0.624400 & O    \\
    0.571429 & 0.142857 & 0.624400 & O    \\
    0.000000 & 0.000000 & 0.624400 & O    \\
    0.142857 & 0.285714 & 0.624400 & O    \\
    0.285714 & 0.571429 & 0.624400 & O    \\
    0.428571 & 0.857143 & 0.624400 & V(O) \\
    \hline
  \end{tabular}
\end{table}
\end{suppinfo}

\end{document}